\documentclass[journal=ancac3]{achemso}   

\usepackage[version=3]{mhchem} 
\usepackage{multirow}
\usepackage{soul}
 \usepackage{array}
\usepackage[usenames,dvipsnames]{color} 
\usepackage[utf8]{inputenc}
\usepackage[T1]{fontenc}
\PassOptionsToPackage{hyphens}{url}
\usepackage{hyperref,url}
\usepackage{amsmath}
\usepackage{amssymb}
\usepackage{mathtools}
\usepackage{bm}
\usepackage{braket}
\usepackage{calrsfs}
\DeclareMathAlphabet{\pazocal}{OMS}{zplm}{m}{n}
\usepackage{multicol}
\usepackage{multirow}
\usepackage{makecell} 
\definecolor{mygreen}{rgb}{0.2,0.7,0.2}

\author{W. Dednam}
\email{dednaw@unisa.ac.za}
\affiliation{~Department of Physics, Florida Science Campus, University of South Africa, 1710 Johannesburg, South Africa}
\alsoaffiliation{~These authors contributed equally to this work}
\author{M. A. Garc\'ia-Bl\'azquez}
\affiliation{~Departamento de F\'\i sica de la Materia Condensada, Universidad Aut\'onoma de Madrid, E-28049 Madrid, Spain}
\alsoaffiliation{~These authors contributed equally to this work}

\author{Linda A. Zotti}
\affiliation{~Departamento de F\'\i sica Te\'orica de la Materia Condensada, Universidad Autonoma de Madrid, E-28049 Madrid, Spain}
\alsoaffiliation{~Condensed Matter Physics Center (IFIMAC), Universidad Autónoma de Madrid, E-28049 Madrid, Spain}

\author{E. B. Lombardi}
\affiliation{~Department of Physics, Florida Science Campus, University of South Africa, 1710 Johannesburg, South Africa}

\author{C. Sabater}
\alsoaffiliation{~Departamento de F\'\i sica Aplicada and Unidad asociada CSIC, Universidad de Alicante, E-03690 Alicante, Spain}

\author{S. Pakdel}
\affiliation{~CAMD, Department of Physics, Technical University of Denmark, 2800 Kgs. Lyngby, Denmark}

\author{J. J. Palacios}
\affiliation{~Instituto Nicolás Cabrera (INC) and Condensed Matter Physics Center (IFIMAC), Universidad Autónoma de Madrid, E-28049 Madrid, Spain}
\alsoaffiliation{~Departamento de F\'\i sica de la Materia Condensada, Universidad Aut\'onoma de Madrid, E-28049 Madrid, Spain}

\title{A group-theoretic approach to the origin of chirality-induced spin selectivity in non-magnetic molecular junctions}


\begin{document}

\begin{tocentry}
\includegraphics[width=8.25cm,height=4.45cm]{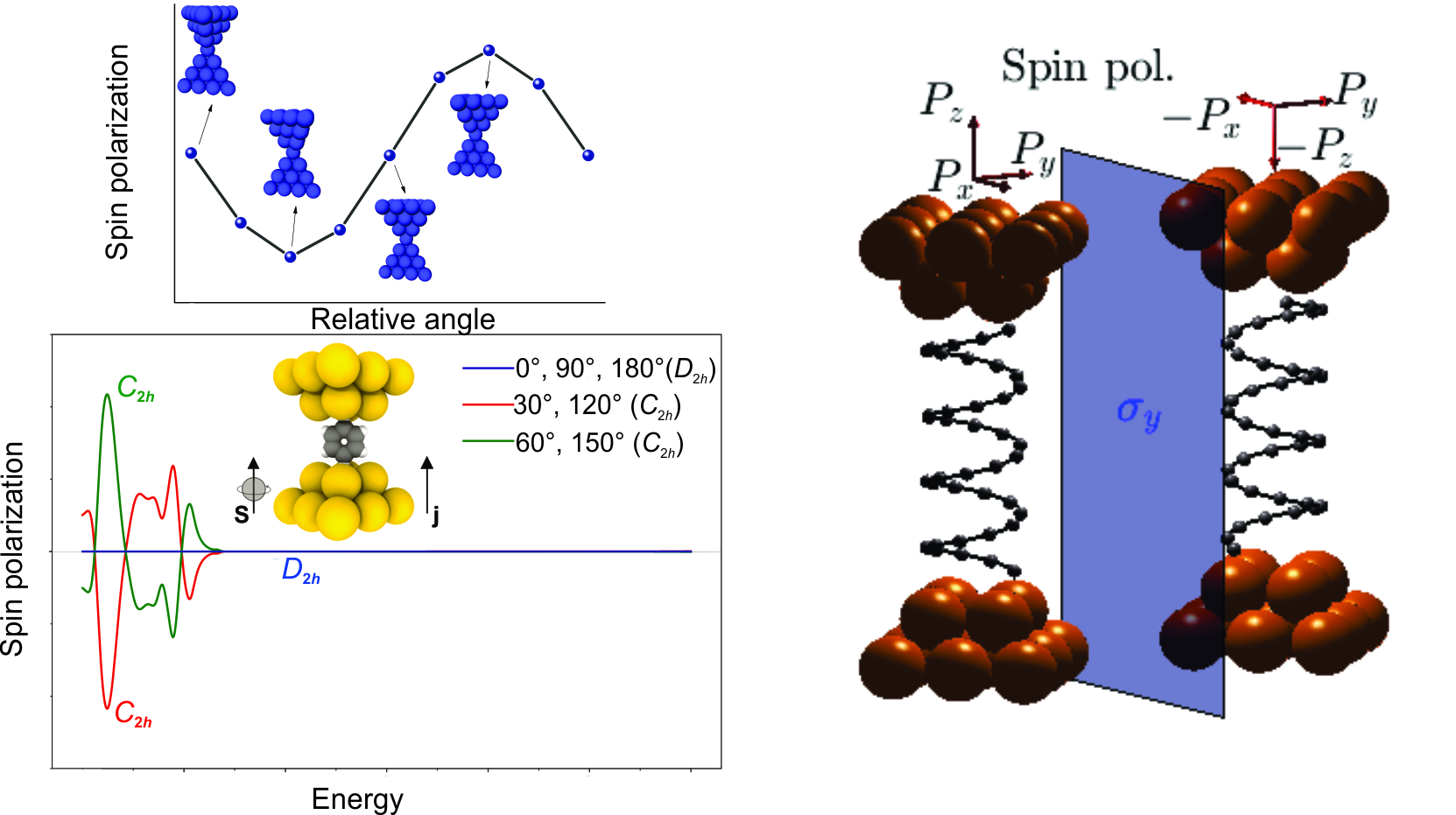}
\end{tocentry}


\begin{abstract}
Spin-orbit coupling gives rise to a range of spin-charge interconversion phenomena in non-magnetic systems where certain spatial symmetries are reduced or absent. Chirality-induced spin selectivity (CISS), a term that generically refers to a spin-dependent electron transfer in non-magnetic chiral systems, is one such case, appearing in a variety of seemingly unrelated situations ranging from inorganic materials to molecular devices. In particular, the origin of CISS in molecular junctions is a matter of an intense current debate. 
Here we derive a set of geometrical conditions for this effect to appear, hinting at the fundamental role of symmetries beyond otherwise relevant quantitative issues. Our approach, which draws on the use of point-group symmetries within the scattering formalism for transport, shows that electrode symmetries are as important as those of the molecule when it comes to the emergence of a spin-polarization and, by extension, to the possible appearance of CISS. It turns out that standalone metallic nanocontacts can exhibit spin-polarization when relative rotations which reduce the symmetry are introduced. As a corollary, molecular junctions with \emph{achiral} molecules can also exhibit spin-polarization along the direction of transport, provided that the whole junction is chiral in a specific way. This formalism also allows the prediction of qualitative changes of the spin-polarization upon substitution of a chiral molecule in the junction with its enantiomeric partner. Quantum transport calculations based on density functional theory corroborate all of our predictions and provide further quantitative insight within the single-particle framework.
\end{abstract}

\flushbottom
\maketitle

\thispagestyle{empty}

\section{Introduction}

Taking advantage of the spin degree of freedom in non-magnetic materials relies on our ability to leverage the combination of strong spin-orbit coupling (SOC) and structural asymmetries. Prototypical examples where this combination occurs include free surfaces of heavy metals\cite{LaShell.PhysRevLett.77.3419,Hoesch.PhysRevB.69.241401} and topological insulators,\cite{Qi.RevModPhys.83.1057} two-dimensional (2D) electron gases,\cite{Wunderlich.PhysRevLett.94.047204} semiconductor thin films,\cite{Kato.science.1105514} or 2D crystals with intentionally broken mirror symmetry.\cite{Safeer.acs.nanolett.9b03485,Rassekh2021,Ingla_Ayn_s_2022,benitez2020tunable}  More recently, chiral bulk systems such as Te crystals, where inversion and mirror symmetries are absent, are also being explored.\cite{Calavalle2022} In all these systems spin-related phenomena such as the spin Hall\cite{1971JETPL..13..467D,Sinova.PhysRevLett.92.126603,Valenzuela2006} or Edelstein\cite{EDELSTEIN1990233} effects (both inverse and direct) can appear and serve as a basis for exploiting the full potential of spin for spintronics applications. On the theoretical side, from basic 2D electron gas models\cite{zhai2005symmetry,nikolic2005decoherence,Pareek.PhysRevLett.92.076601} to more sophisticated models based on first-principles,\cite{Gmitra.PhysRevB.80.235431,gmitra2016trivial,Rassekh2021} many of the experimental observations can be successfully accounted for.

Molecular junctions with chiral molecules are also a playground for spin-charge interconversion phenomena, exhibiting the so-called chirality-induced spin selectivity (CISS) effect. This phenomenon, involving the spin-polarization of electrons propagating through chiral, possibly non-magnetic media (often molecules), has been the subject of a large number of experimental
\cite{Rubio.science.283.5403.814,Xie.nl2021637,Dor2013,Aragones2017,naaman2020chiral,acsnano.0c07438,acsnano.2c05601,arxiv.2209.08117}
and theoretical studies
\cite{Naaman2012,Gersten2013,Naaman2019,Dalum2019,yang2019spin,yang2020detecting,Zollner2020b,alwan2021spinterface,Liu2021,acs.jpclett.1c03925,dubi2022spinterface,Ghazaryan.acs.jpcc.0c02584,Evers.adma.202106629,acsnano.2c07088}
over the past decade. 
The CISS effect, although it may ultimately manifest in several ways, is usually identified with a finite magnetoconductance measured in transport experiments under out of equilibrium conditions, possibly also in the linear regime.\cite{acsnano.0c07438,yang2020detecting}
The microscopic origin of the CISS effect is being actively debated and could be attributed to a combination of a finite bias voltage and factors such as the chirality of the molecule, the strength of SOC in the system, electron correlations, dephasing effects, an underlying orbital polarization that is turned into spin-polarization by SOC, or even solenoid-like fields inside helical molecules.
\cite{acs.jpclett.9b02929,PhysRevB.102.035431,PhysRevB.104.L201409,acs.nanolett.1c00183,acs.jpclett.1c03925,alwan2021spinterface,Liu2021} Numerically, all of these factors should definitely play a role on the spin-polarization and the magneto-conductance, especially in the strongly non-linear regime where inelastic effects are more prominent. However, not all of them may be universally necessary features to observe measurable spin-resolved transport quantities.
At least one of these factors, the strength of SOC in the normal metal electrode component of the junction, has been identified in a recent experimental study to be an important driver of a large magneto-conductance in molecular junctions \cite{arxiv.2209.08117,acsnano.0c07438}. A strong SOC in combination with the chirality of the molecules facilitates the appearance of a significant spin-polarization (or spin current), which appears to be a necessary condition for the CISS effect to ultimately manifest in experiments. Given that chirality, a symmetry property of the system, is at the core of the phenomenon (and the terminology itself), it seems particularly appropriate to employ group theory to analyze the emergence of spin-polarization.

In this work we present a systematic and complete theoretical analysis of the electronic spin-polarization in molecular junctions (non-magnetic CISS devices), based entirely on the use of the point symmetry group of the whole system (electrodes plus, possibly, a molecule) within the scattering formulation of coherent quantum transport at the electronic single-particle level. This analysis allows us to determine the restrictions imposed by each individual spatial symmetry on the relevant quantities of the problem, the spin-resolved conductance and the spin-polarization. We list the possible symmetries that can be found in the two-terminal configurations and identify those which, when removed, allow for polarization to appear in the general case \textendash~ regardless of the system being a standalone pair of metallic nanocontacts, or a molecular junction with a chiral or achiral molecule. In particular, we show that a simple relative rotation of the electrodes, leading to the removal of certain mirror symmetries, is in general sufficient for spin-polarization to emerge, independent of the chiral nature of the molecule or even its sheer presence. 

At a quantitative level we present density functional theory (DFT) calculations for realistic systems that corroborate and quantify our theoretical predictions. We find non-zero spin-polarizations for organic molecules as long as the electrodes, not necessarily the molecule, present a strong SOC, specific mirror symmetries are removed and the orbital character near the Fermi level is not exclusively $s$-type. Chiral molecules, as expected, give rise to spin-polarization in general (as long as they are connected in suitable ways). In principle one may expect that the substitution of a chiral molecule by its enantiomeric partner in a molecular junction would result in the exact reversal of the spin-polarization. 
However, we show that this phenomenon is more subtle, requiring certain geometrical conditions on the electrodes to be met as well as on the specific anchoring of the molecules for spin-polarization to be strictly reversed.

In summary, our results grant a rather general view of the origin of the spin-polarization in transport in molecular junctions with non-magnetic electrodes, constituting a global framework with which we can predict the necessary (not in principle sufficient) conditions under which the CISS effect can be observed.

\section{Results and discussion}
\subsection{Symmetry considerations}
We consider a two-terminal device formed by two electrodes or contacts and, possibly, a molecule between them. 
The component of the spin-polarization $\bm{P}$ of the current generated at the drain electrode, along the direction of a given spin quantization axis is, for an incident unpolarized current:\cite{nikolic2005decoherence}
\begin{equation}  \label{eq1}
P=G_{\uparrow\uparrow}+G_{\uparrow\downarrow}-G_{\downarrow\uparrow}-G_{\downarrow\downarrow}
\end{equation}
where $G_{s',s}$ ($s,s'\in\set{\uparrow,\downarrow}$ referred to the spin axis) is the spin-resolved conductance at a given energy (which we omit for simplicity) measured relative to the Fermi energy. 
The three components of the vector $\bm{P}$ can then be obtained by rotating the quantization axis and applying \eqref{eq1}.

As a position-independent quantity which in the scattering formalism can be computed, at least formally, from the electronic wave functions of the electrodes and the Hamiltonian of the system, $G_{s',s}$ may in principle be subject to restrictions induced by the spatial symmetries of the whole system. These relations are obtained by recalling the invariance of the corresponding space integrals under the orthogonal coordinate transformations which form the point group $\pazocal{G}$ of the system (electrodes plus, possibly, molecule) and employing the (projective) representations according to which the spinor wave functions in the electrodes transform. The complete derivation can be found in \textit{Methods}, from which it follows that only symmetries which do not permute the electrodes can potentially impose a restriction \eqref{eqTsymAB} on the original conductance at each energy. In contrast, symmetries which do permute the electrodes may only yield a relation \eqref{eqTsymBA} between the original conductance and that corresponding to the hypothetical situation where bias polarity is reversed. The latter assertion also applies to anti-unitary symmetries, \textit{i.e.}, time-reversal and potentially particle-hole, which can be freely applied within the scalar products up to complex conjugation.

Specifically, from equation \eqref{eqTsymAB} it follows that mathematical equations relating certain spin-resolved conductance terms $G_{s',s}$ arise if the junction presents the following spatial symmetries:
\begin{itemize}
\item A $2\pi/n-$rotation $C_{n}$ which does not permute the electrodes. For a spin quantization axis directed along rotation axis, the condition on the conductance is trivial; however, for $n=2$ and any quantization axis perpendicular to the rotation axis, $G_{s',s}=G_{\overline{s}',\overline{s}}$ where $\overline{s}$ corresponds to the opposite spin state of $s$. Hence, by \eqref{eq1} the components of the spin-polarization which are normal to the rotation axis vanish. In our collinear two-terminal configuration, this rotation axis must coincide with the longitudinal direction (along which transport takes place) and we denote the operation by $C_{n,l}$.
\item A mirror plane $\sigma$ which does not permute the electrodes. Since the spinors transform as pseudo-vectors, in our formalism this operation is essentially equivalent (except for the invariance of the electrodes) to a $\pi-$rotation about the axis which is perpendicular to the mirror plane and contains the fixed point of $\pazocal{G}$. By the previous case, the components of the spin-polarization $\bm{P}$ which are parallel to the mirror plane must vanish. In our configuration, this plane must contain the longitudinal direction,  
and we denote the operation by $\sigma_{l}$ (note, however, that there will in general be more than one longitudinal plane of symmetry). We refer to any direction perpendicular to the longitudinal one as transversal, in particular the direction in which $\bm{P}$ points (unless it is the zero vector) in this case.
\end{itemize}

\begin{table}[t!]
\caption{Conditions imposed on the spin resolved conductance terms \eqref{eqT} and the resulting spin-polarization vector $\bm{P}$ by the spatial symmetries of the whole system (electrodes with possibly a molecule between them) on their own, as well as in combination with time reversal symmetry $\Theta$. Anti-unitary symmetries have been included in the last two rows. Columns: \textbf{1}, Element of the point group $g\in\pazocal{G}$ of the system. \textbf{2}, Direction of spin projection, which defines the polarization component \eqref{eq1}. \textbf{3}, Conductance term to which $G^{AB}_{s',s}$ must be equal, due to the presence of either \textbf{3.1}, the symmetry $g$ alone \eqref{eqTsymAB}, \eqref{eqTsymBA}, or  \textbf{3.2}, the symmetry $\Theta g$, where time-reversal $\Theta$ has also been applied \eqref{eqTsymTR}, \eqref{eqTsymABTR}. $\overline{s}$ corresponds to the opposite spin state of $s$, and $G^{BA}$ to the conductance from electrode $B$ to electrode $A$. \textbf{4}, Restriction imposed on the corresponding vector components of the spin-polarization; these always come from the operation $g$ alone, Column \textbf{3.1}. Entries in blank are either tautological or inconclusive (yield no compact identities). 
}
\setlength{\extrarowheight}{0.1cm}
\footnotesize
\begin{tabular}{!{\vrule width1.2pt}c!{\vrule width1.2pt}c!{\vrule width1.2pt}c|c|c!{\vrule width1.2pt}}
\Xhline{1.2pt}
\multirow{2}*{Symmetry ($g\in\pazocal{G}$)} & \multirow{2}*{Spin quantization axis}  & \multicolumn{2}{c|}{$G^{AB}_{s',s}$ identity} & \multirow{2}*{$\bm{P}$ restriction} \\[0.2cm] \cline{3-4}
& & $g$ & $\Theta g$ & 
\\[0.2cm] \Xhline{1pt}\Xhline{1pt}
\multirow{3}*{Longitudinal mirror ($\sigma_{l}$)} & Longitudinal ($l$) & $G^{AB}_{\overline{s}',\overline{s}}$ & $G^{BA}_{s,s'}$ & $P_{l}=0$  \\[0.2cm] \cline{2-5}
& Transversal, parallel to $\sigma_{l}$ ($t_{\parallel}$) & $G^{AB}_{\overline{s}',\overline{s}}$ & $G^{BA}_{s,s'}$ & $P_{t_{\parallel}}=0$  
\\[0.2cm] \cline{2-5}
& Transversal, normal to $\sigma_{l}$ ($t_{\perp}$) &  & &  \\[0.2cm] \Xhline{1.2pt}
\multirow{2}*{\makecell{\\[-0.05cm] Longitudinal $\pi-$ \\[0.1cm] rotation ($C_{2,l}$)}} & Longitudinal ($l$) &  & &  \\[0.2cm] \cline{2-5}
& Transversal ($t$) & $G^{AB}_{\overline{s}',\overline{s}}$ &$G^{BA}_{s,s'}$ &$P_{t}=0$ 
\\[0.2cm] \Xhline{1.2pt}
\multirow{2}*{\makecell{\\[-0.05cm] Longitudinal $2\pi/n-$ \\[0.1cm] rotation ($C_{n,l}$), $n\geq3$}} & Longitudinal ($l$) &  & &  \\[0.2cm] \cline{2-5}
& Transversal ($t$) & & & $P_{t}=0$  
\\[0.2cm] \Xhline{1.2pt}
\multirow{2}*{Transversal mirror ($\sigma_{t}$)} & Longitudinal ($l$) & $G^{BA}_{s',s}$ & $G^{AB}_{\overline{s},\overline{s}'}$ &  \\[0.2cm] \cline{2-5}
& Transversal ($t$) & $G^{BA}_{\overline{s}',\overline{s}}$ & $G^{AB}_{s,s'}$ &  
\\[0.2cm] \Xhline{1.2pt}
\multirow{3}*{\makecell{\\[0.02cm] Transversal $\pi-$ \\[0.1cm] rotation ($C_{2,t}$)}} & Longitudinal ($l$) & $G^{BA}_{\overline{s}',\overline{s}}$ & $G^{AB}_{s,s'}$ &  \\[0.2cm] \cline{2-5}
& Transversal, parallel to $C_{2,t}$ ($t_{\parallel}$) & $G^{BA}_{s',s}$ & $G^{AB}_{\overline{s},\overline{s}'}$ & 
\\[0.2cm] \cline{2-5}
& Transversal, normal to $C_{2,t}$ ($t_{\perp}$) & $G^{BA}_{\overline{s}',\overline{s}}$ & $G^{AB}_{s,s'}$ &  
\\[0.2cm] \Xhline{1.2pt}
Inversion ($I_{s}$) & Any &  $G^{BA}_{s',s}$ & $G^{AB}_{\overline{s},\overline{s}'}$ & 
\\[0.2cm] \hline\hline
Time-reversal ($\Theta$) & Any & $G^{BA}_{\overline{s},\overline{s}'}$& & $\bm{P}=0$ if 1 $B$ channel 
\\[0.2cm] \Xhline{1.2pt}
Particle-hole ($\pazocal{C}$) & Any & $G^{BA}_{s,s'}(-E)$ & $G^{AB}_{\overline{s}',\overline{s}}(-E)$ & $\bm{P}(E)=-\bm{P}(-E)$
\\[0.2cm] \Xhline{1.2pt}
\end{tabular}
\label{table1}
\end{table}

Consequently, we denote the operations that do not (do) permute the electrodes as \textit{longitudinal} (\textit{transversal}, resp.). An illustration can be found in Figure \ref{nanocontacts}d. A careful analysis (see the \textit{Supplementary Information} for details) reveals one further restriction on the polarization which is not associated with identities between the different conductance terms: any longitudinal rotation symmetry $C_{n,l}$ ($n\geq2$), not only $C_{2,l}$, guarantees the vanishing of all the transversal components of the spin-polarization. Only rotations with $n=2,3,4,6$ are, however, allowed in electrodes which possess a three-dimensional crystalline structure in the bulk. In contrast, chain-like electrodes may present other $C_{n,l}$ symmetries, with one-atom chains presenting any of them. 

We thus conclude that there can be no spin-polarization in two-terminal systems whose point group $\pazocal{G}$ is one of the following: $C_{nv}$, $D_{nh}$, $D_{nd}$, $\forall ~n\geq2$ (where the principal rotation axis is oriented along the longitudinal direction), due to the simultaneous presence of $C_{n,l}$ and $\sigma_{l}$, which forces all the vector components of the spin-polarization to vanish. Note that the polyhedral groups are not compatible with a two-terminal configuration, hence are automatically excluded. Several examples are shown in Figure \ref{nanocontacts} and discussed in the next section. The remaining groups which may 
admit a finite polarization vector, $C_{i}$, $C_{2n,h}$, $S_{4n-2}$, $\forall~n\geq1$, contain inversion symmetry, the geometrical breaking of which is thus not a necessary condition to observe spin-polarization in the transmitted current.

The results for all symmetry operations that are compatible with the two-terminal configuration are summarized in Table \ref{table1}, where the identities between spin-resolved conductance terms for which the polarity is reversed (induced by anti-unitary and electrodes-permuting unitary symmetries) have also been included. 
As can be observed, time-reversal symmetry $\Theta$ forces the oddness \eqref{eqTsymTRC} of the polarization \eqref{eq1} in combination with particle-hole symmetry (in the exceptional cases in which the latter holds\cite{Chico2015,Guo2016}) with respect to the zero of energy that the latter defines. And, perhaps more importantly, time-reversal symmetry guarantees the vanishing of the whole polarization vector, at all energies at which the final electrode has only one mode\cite{zhai2005symmetry} (see \textit{Methods}). There are no further restrictions on the spin-polarization induced by anti-unitary symmetries, including the ones of the form $\Theta g$ with $g$ an unitary symmetry, as can be concluded from column 3.2 in Table \ref{table1}.

\begin{figure}[t!]
\centering 
\includegraphics[width=0.45\textwidth]{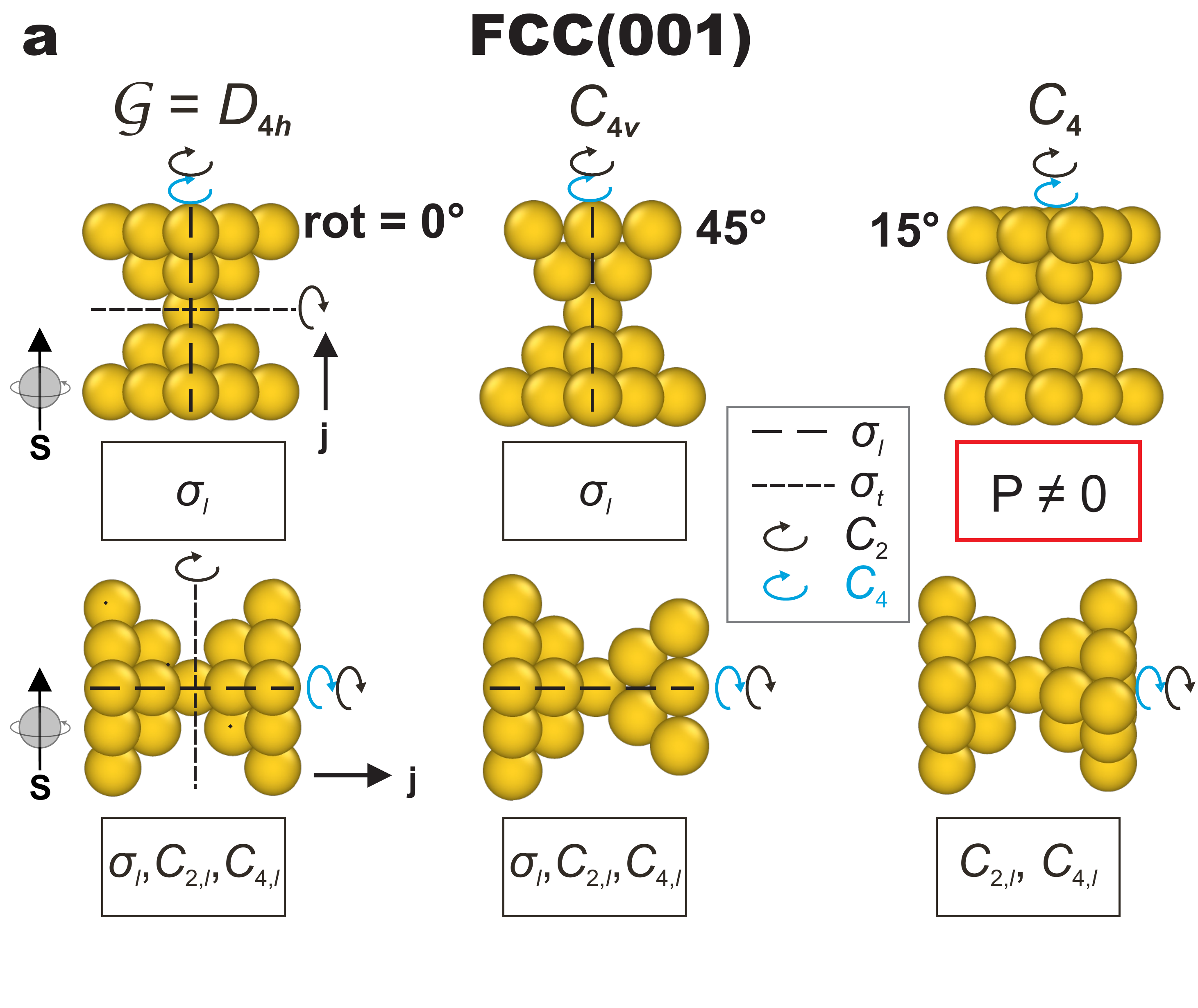}
\hspace{0.5cm}
\includegraphics[width=0.45\textwidth]{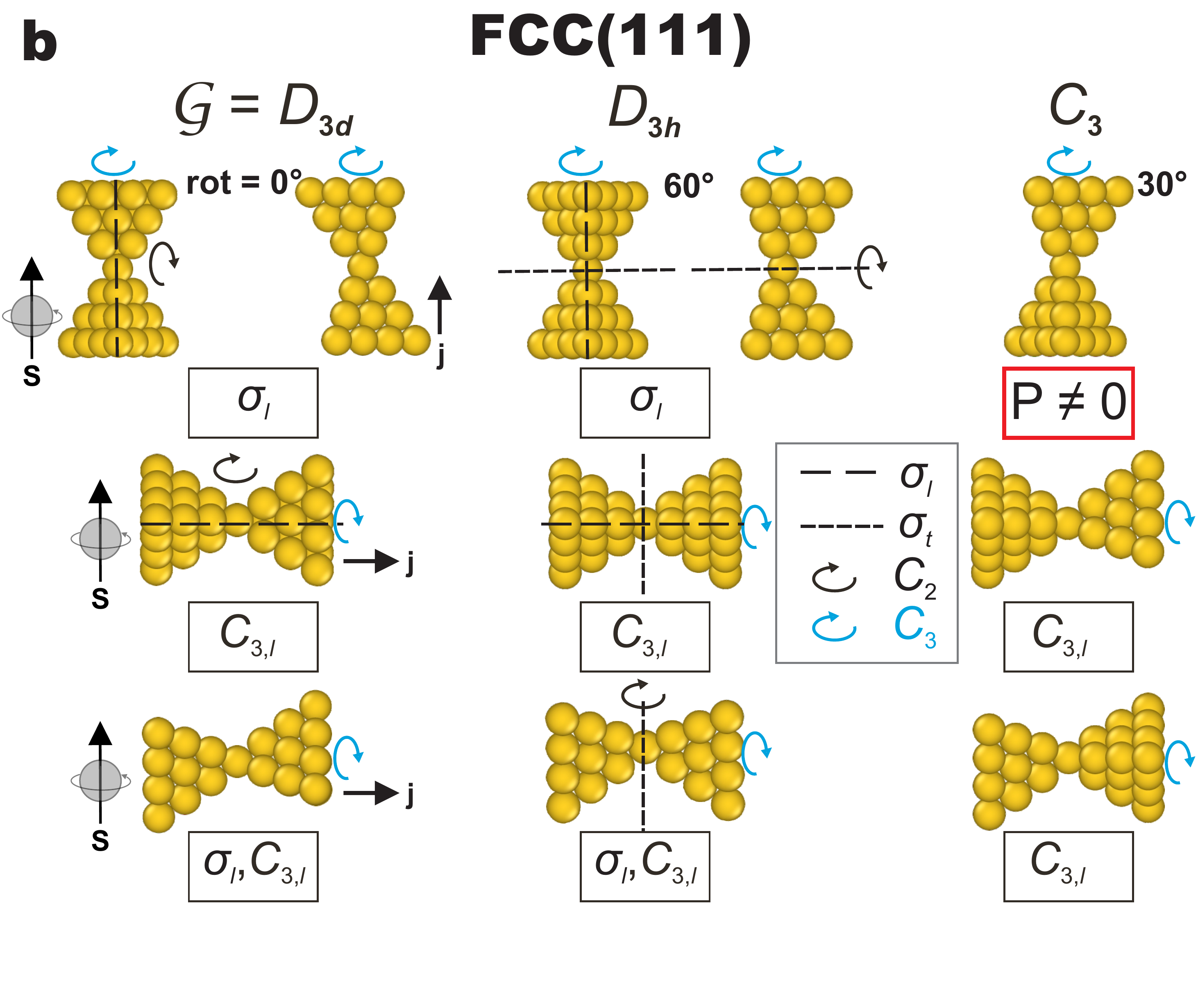} \\[0.5cm]
\includegraphics[width=0.45\textwidth]{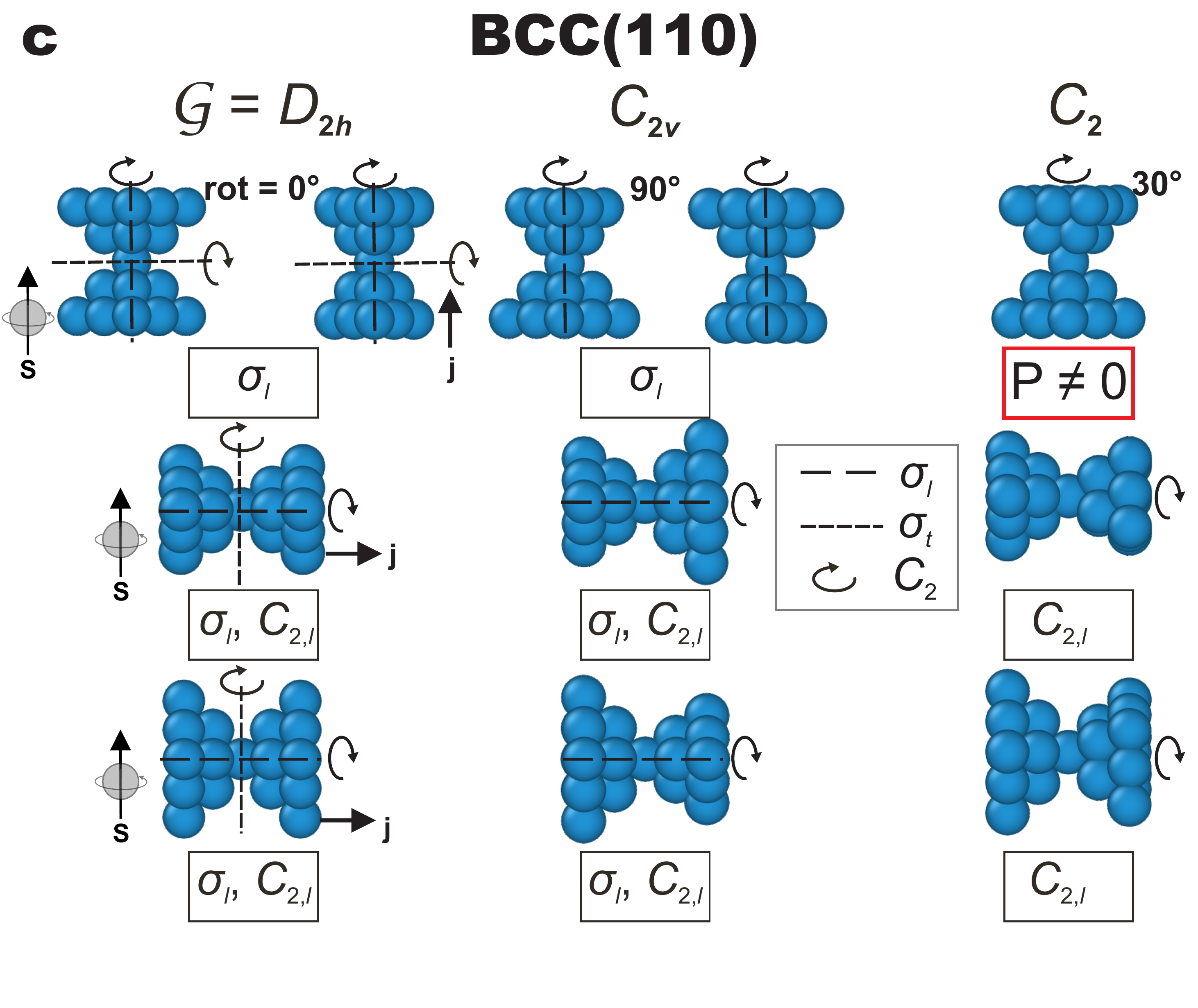}
\hspace{0.5cm}
\includegraphics[width=0.45\textwidth]{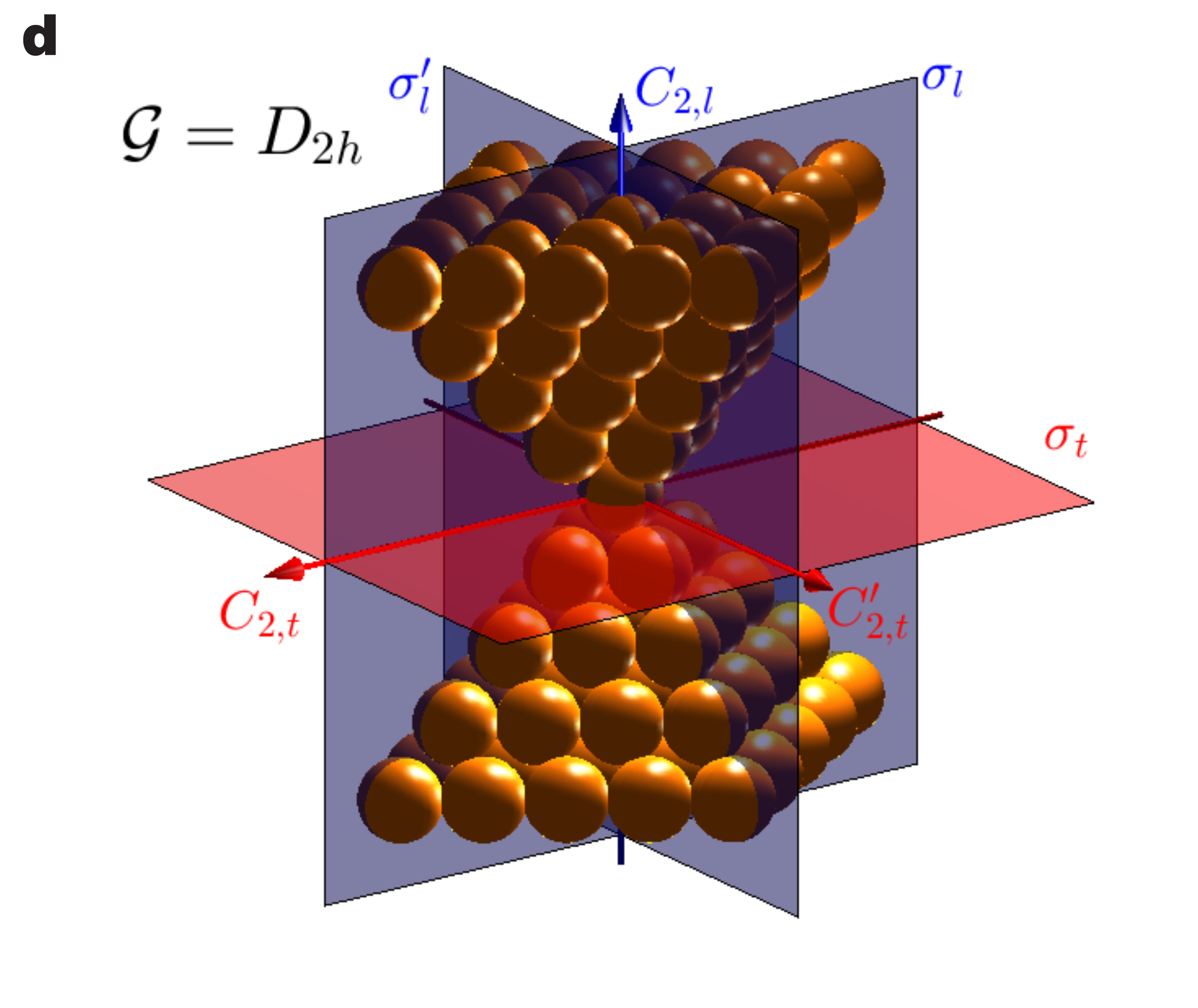}
\caption{
Representative examples of 
metallic nanocontacts. \textbf{a}, FCC(001) nanocontacts (\textit{e.g.} Au) with $0$\textdegree, $45$\textdegree\space and $15$\textdegree\space electrode relative rotations. \textbf{b}, FCC(111) nanocontacts (\textit{e.g.} Au) with $0$\textdegree, $60$\textdegree\space and $30$\textdegree\space relative rotations. \textbf{c}, BCC(-110) nanocontacts (\textit{e.g.} W) with $0$\textdegree, $90$\textdegree\space and $30$\textdegree\space relative rotations. Rotation angles of the drain electrode (with respect to the source one, ordering indicated by the current direction $\bm{j}$) vary across columns, and the orientation of the electrodes with respect to the spin quantization axis, indicated by $\bm{s}$ and with fixed direction here, vary across rows. The point groups $\pazocal{G}$ of the corresponding systems are indicated above each column. Relevant symmetry operations (including a set of generators of the corresponding groups) are explicitly indicated, and those that force the vanishing of the spin-polarization component along $\bm{s}$ are indicated in the boxes below each figure, following the notation of Table \ref{table1}. 
For the polarization in transversal directions, $\sigma_{l}$ is the longitudinal symmetry plane which is parallel to the page.
Red boxes with no symmetry operations indicate a finite polarization component instead. \textbf{d}, Detailed three-dimensional view of the symmetry planes and axes with the group $\pazocal{G}=D_{2h}$, as in the first column of \textbf{c}. Operations depicted in blue (red) are longitudinal (transversal, resp.), \textit{i.e.}, they do not  (do, resp.) permute the electrodes.
\label{nanocontacts}}
\end{figure}

Nanocontacts  are the simplest possible systems in which the previous discussion on spin-polarization can be applied. As shown in Figure \ref{nanocontacts}, these typically consists of two crystalline fragments or electrodes, source and drain, in contact and placed so that their principal symmetry axes are coincident\footnote{The relations in Table \ref{table1} would hold for an arbitrary arrangement of the two electrodes, but then all rotation symmetries would necessarily permute them. The case we consider here allows one to obtain the largest possible point groups $\pazocal{G}$ for a two-electrode system.}. We consider three different pairs of identical electrodes, all of them presenting a crystallographic cubic system in the bulk, and with the main symmetry axis coincident with a $\langle001\rangle$ ($4-$fold axis), $\langle111\rangle$ ($3-$fold axis), $\langle-110\rangle$ ($2-$fold axis) direction of the cubic structure. These are respectively shown in Figure \ref{nanocontacts}a,b,c. Metallic, non-magnetic structures of such kind can experimentally be made of Au (FCC), Pb (FCC) or W (BCC), among other elements, including the perfect crystallographic atomic arrangement.\cite{Sabater.PhysRevLett.108.205502}

Once the point group $\pazocal{G}$ of the structure is known, one can immediately foresee whether spin-polarization of the transmitted current is possible. According to our previous discussion, the resulting current in systems with groups $C_{nv}$, $D_{nh}$, $D_{nd}$ ($n\geq2$) must be spin-unpolarized. As can be observed in Figure \ref{nanocontacts}, a simple way of reducing the otherwise high symmetry of a system (for aligned electrodes) is to rotate one electrode while keeping the other fixed. This action preserves all longitudinal rotation symmetries, but in general removes the longitudinal mirror planes (except if the rotation is by an integer multiple of the dihedral angle $\pi/n$, with $n$ corresponding to one of the previously mentioned groups); thereby reducing $\pazocal{G}$ to a subgroup\footnote{Up to isomorphism, as between $D_{3d}$ and $D_{3h}$ in Figure \ref{nanocontacts}b.} and allowing for a finite polarization at least in the longitudinal direction. Of course, it is not necessary to identify the whole point group of the system in order to rule out spin-polarization; it suffices to look for rotational and mirror symmetries that do not permute the electrodes. 

In complete analogy with the relative rotation of the electrodes, the placement of a molecule (or, in general, a piece of material) between the contacts either leaves $\pazocal{G}$ invariant or turns it into one of its subgroups, since it obviously cannot add any symmetry that was not already present in the standalone pair of electrodes. As a result, the qualitative effect of adding the molecule to the system is a potential lifting of symmetry-induced restrictions on the spin-resolved conductance and spin-polarization. In particular it may allow for an otherwise forbidden finite polarization, but it cannot (strictly) cancel it if the bare electrodes already exhibited a non-vanishing polarization. 

The spin-resolved conductance, and hence the spin-polarization, can be related with those of an alternative system obtained by application of an orthogonal transformation. If such an operation is a symmetry of the system, then the previous analysis applies and one may find restrictions for these quantities. If, however, the transformation does not leave the system invariant, then by a similar procedure one may relate the conductance terms and polarization of the two systems by equation \eqref{eqTnosym}. In particular, the spin-polarization of the transmitted current across a chiral molecule and across its enantiomeric partner may differ only in sign \eqref{eqTenantlong}, as long as the connection of the enantiomer molecule with the electrodes is done in such a way that the mirror plane which relates both molecules is separately a symmetry of the two electrodes. This topic is elaborated further in the \textit{Enantiomeric partners and polarization reversal} section.

\subsection{DFT-based quantum transport calculations}

\begin{figure*}[t!]
\centering 
\includegraphics[width=0.45\textwidth]{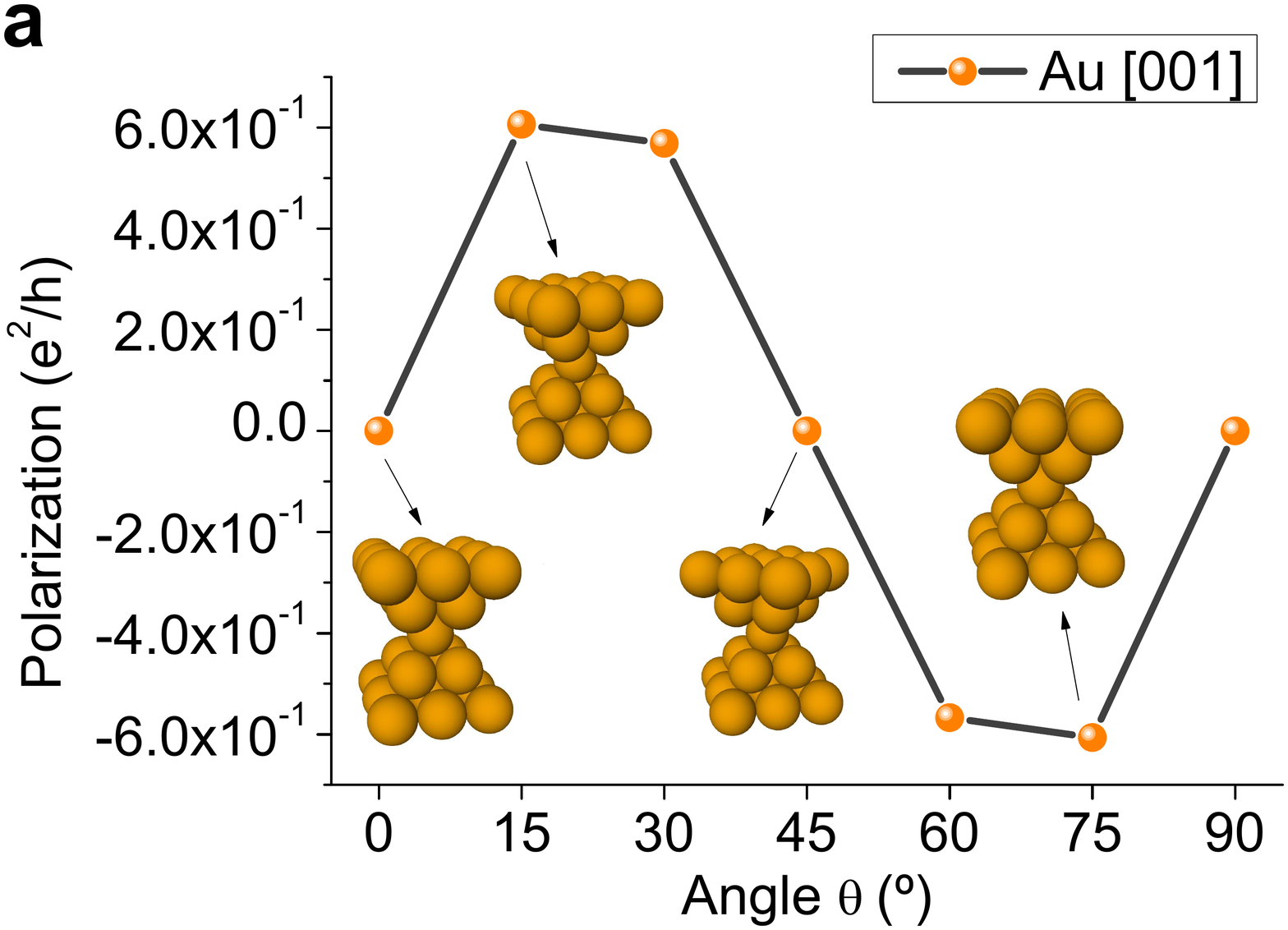}
\includegraphics[width=0.45\textwidth]{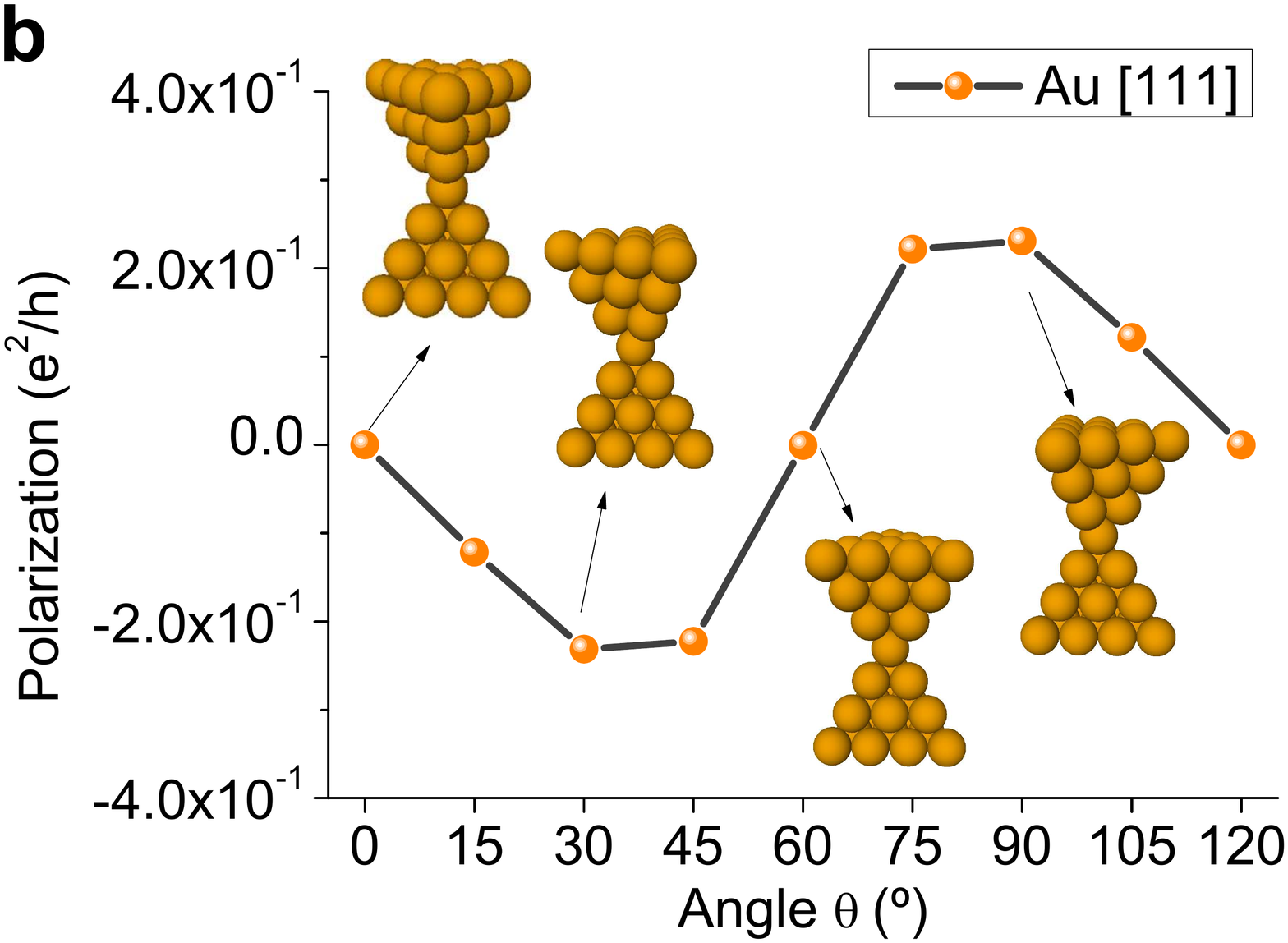}
\includegraphics[width=0.45\textwidth]{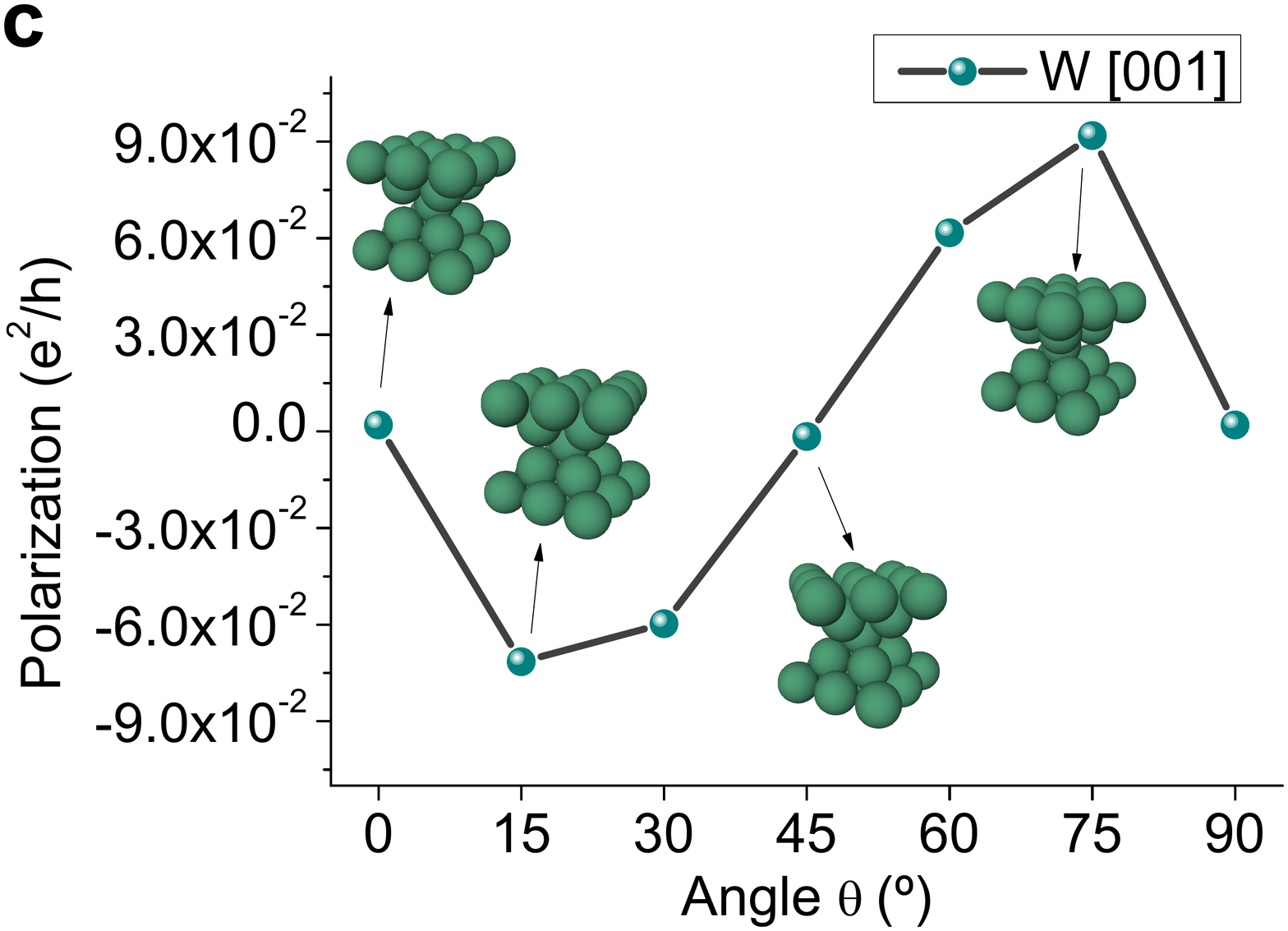}
\includegraphics[width=0.45\textwidth]{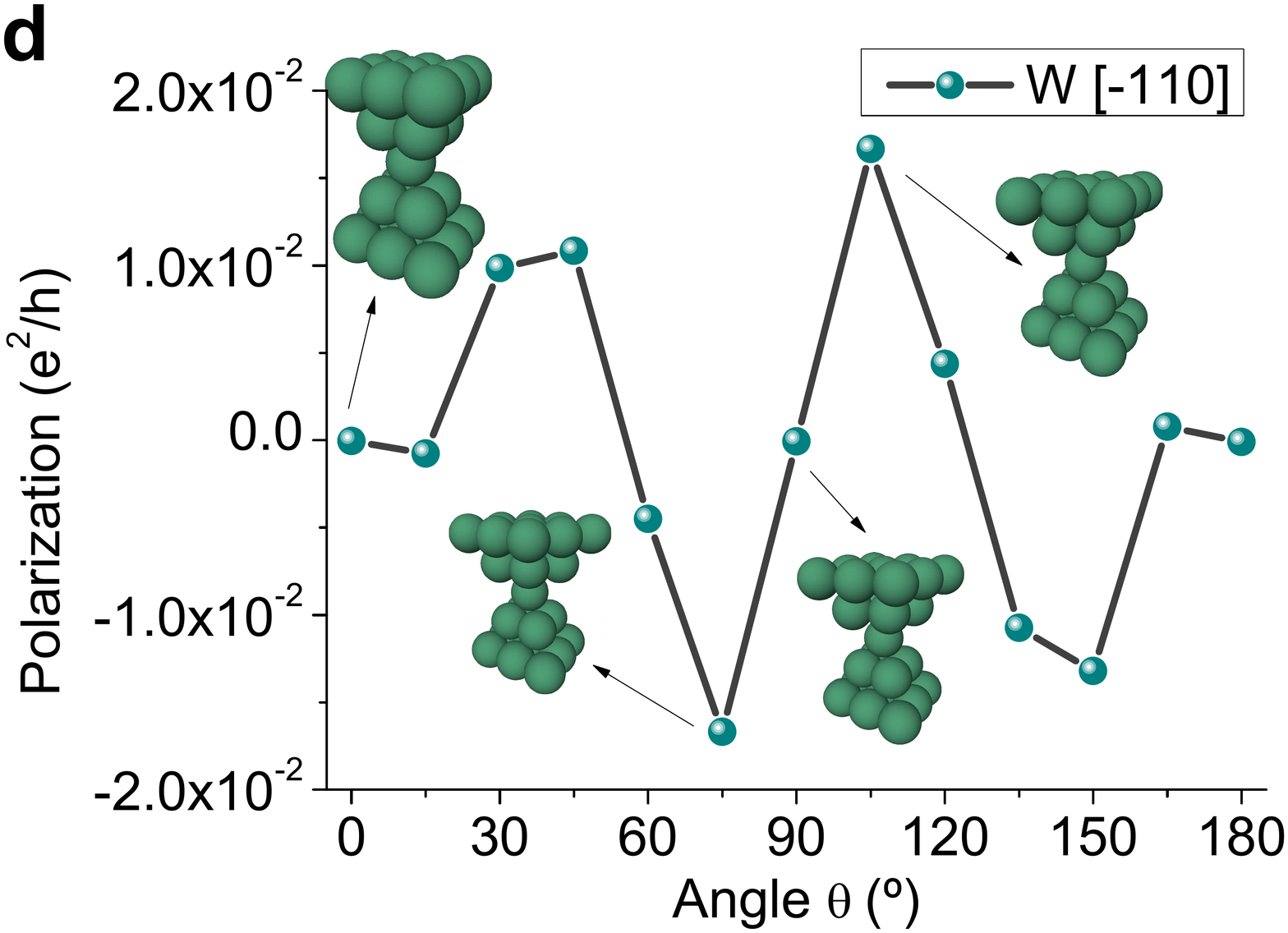}
\includegraphics[width=0.45\textwidth]{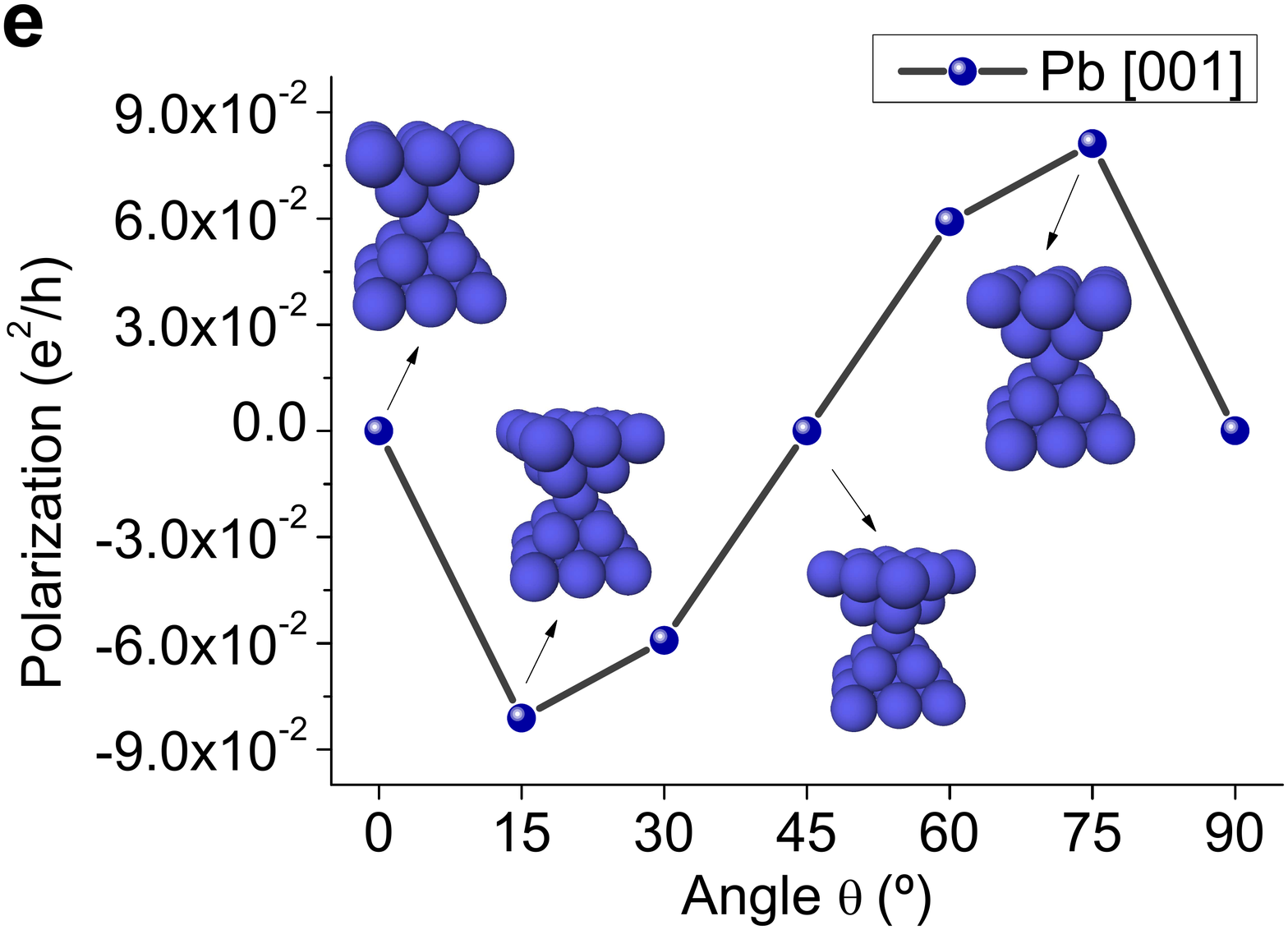}
\includegraphics[width=0.45\textwidth]{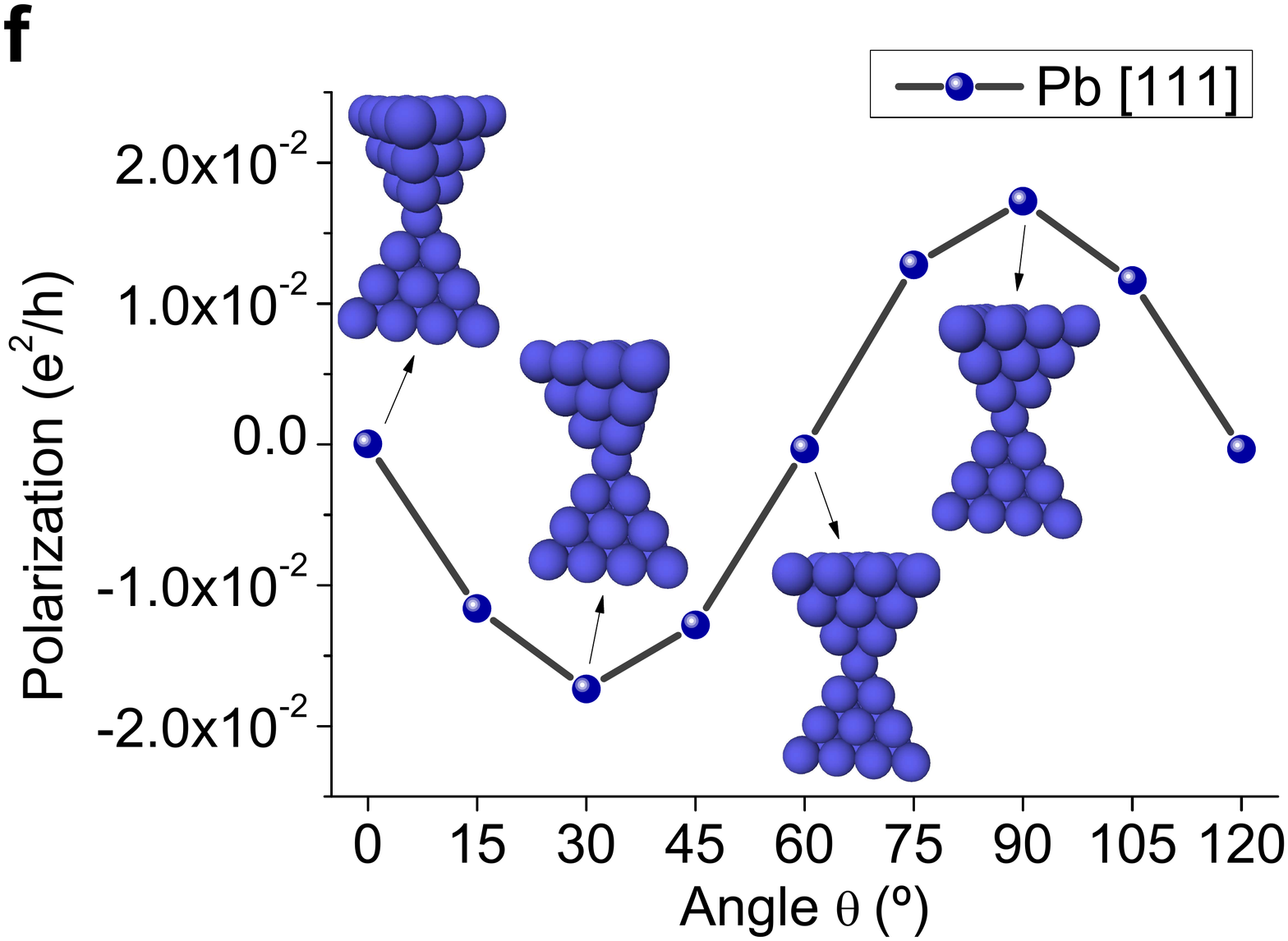}
\caption{
DFT-computed zero-bias spin-polarization \eqref{eq1} along the direction of transport as a function of the angle of relative rotation between the electrodes (only the top half is rotated), at the fixed energy that maximizes the maximum absolute value of spin-polarization across both energies and angles. The oddness of the polarization with respect to the dihedral angles fits nicely into the CISS phenomenology, each pair of systems with opposite polarization being related by a longitudinal mirror plane. \textbf{a}, Au(001), point group $\pazocal{G}=D_{4h}$ at $\theta=0$. \textbf{b}, Au(111), $D_{3d}$ at $\theta=0$. \textbf{c}, W(001), $D_{4h}$ at $\theta=0$. \textbf{d}, W(-110), $D_{2h}$ at $\theta=0$. \textbf{e}, Pb(001), $D_{4h}$ at $\theta=0$. \textbf{f}, Pb(111), $D_{3d}$ at $\theta=0$.}
\label{fig2}
\end{figure*}

The predictions in Figure \ref{nanocontacts} can be verified by means of SOC-corrected DFT quantum transport calculations as implemented in our code Atomistic NanoTransport~(\texttt{ANT.Gaussian}).
\cite{palacios2001fullerene,palacios2002transport,ANTG}
See the \textit{Supplementary Information} for further details. Specifically, we show in Figure \ref{fig2} that a (longitudinal) mirror symmetry breaking is sufficient to produce a rather large spin-polarization in strong-SOC metal nanocontacts, with Au, Pb and W as our representative elements. To this end we rotate the drain electrode of the nanocontact by an angle $\theta$ in increments of $15$\textdegree\space and compute the polarization along the longitudinal direction according to equation \eqref{eq1}. The energy is fixed at the value in the $(-3,3)$ eV range that maximizes the maximum absolute value of polarization across the selected angles\footnote{This maximum occurs at different energy values for the different elements and configurations, depending on the orbital nature of the conductance channels. In particular for Au the maximum occurs below the Fermi energy, where the contribution of the $d$ orbitals becomes significant. In contrast, a finite polarization (not the maximum one) at the Fermi energy is obtained for Pb and W. See the \textit{Supplementary Information}.}. The maximum $\theta$ corresponds to the symmetry operation $C_{n,l}\in\pazocal{G}$ ($\theta=2\pi/n$), with $n=4$ in Figure \ref{fig2}a,c,e, $n=3$ in \ref{fig2}b,f and $n=2$ in \ref{fig2}d. The $C_{n,l}$ symmetry guarantees that the displayed curves are $2\pi/n-$periodic.

\begin{figure*}[t!]
\centering 
\includegraphics[width=0.47\textwidth]{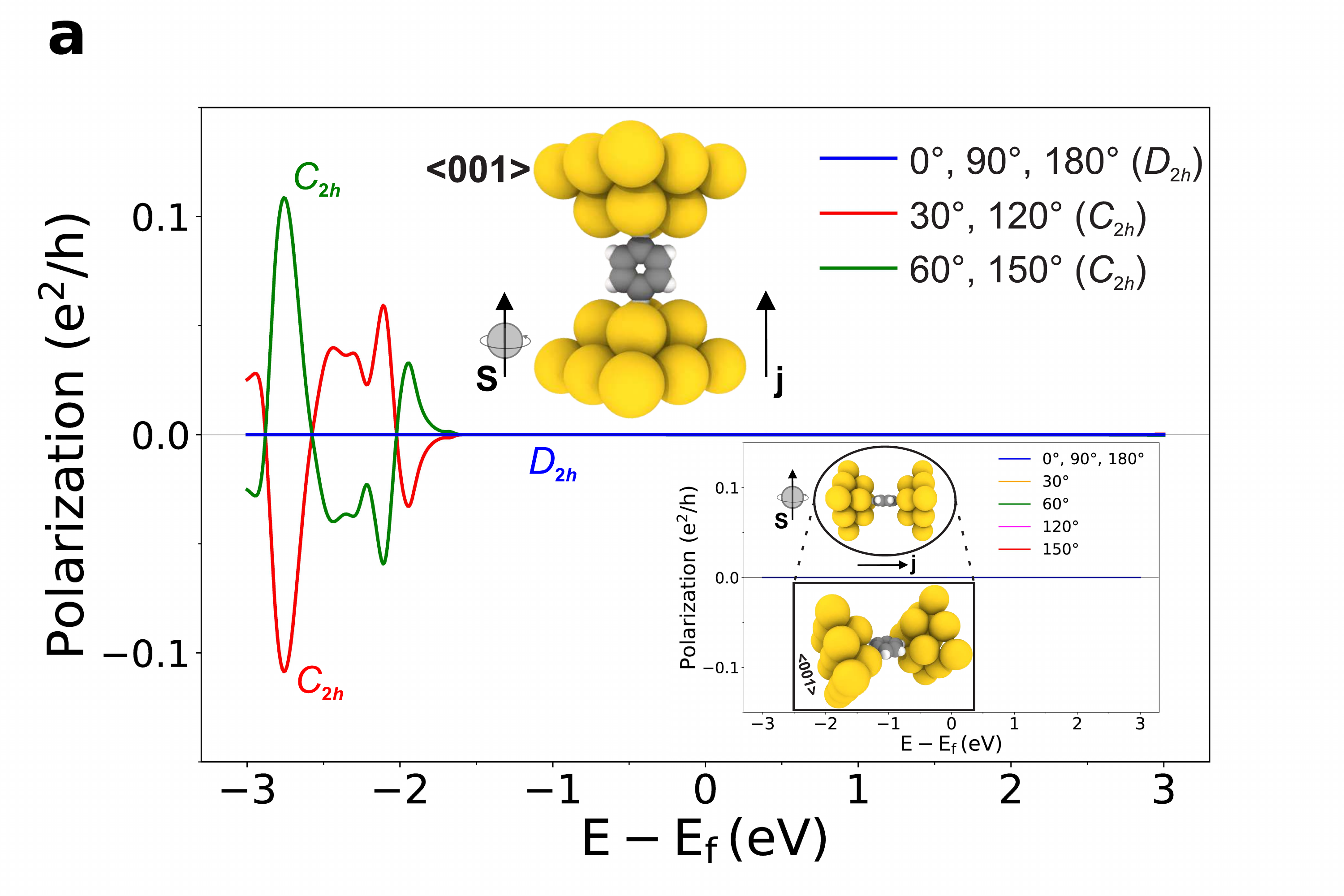}
\includegraphics[width=0.47\textwidth]{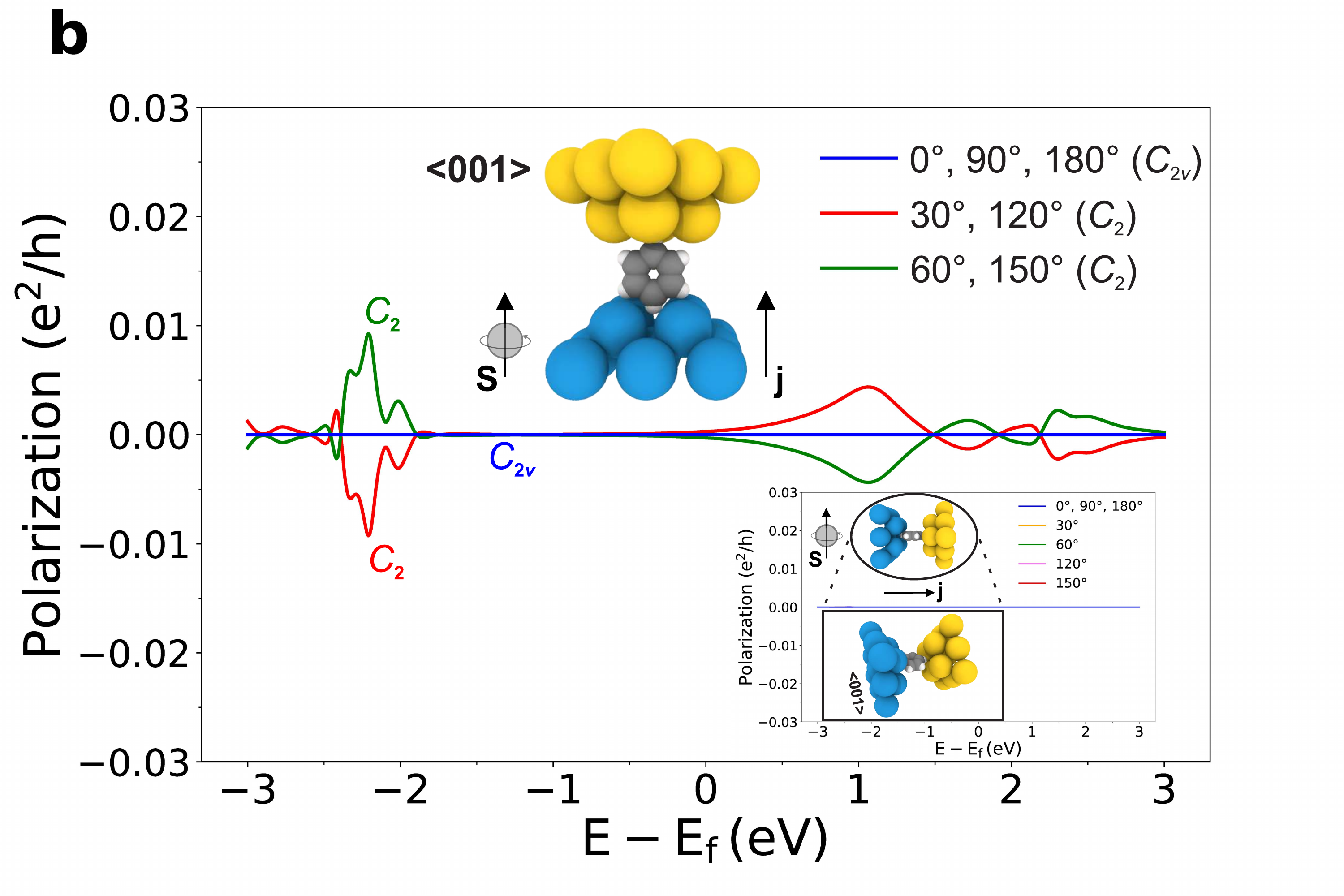}
\includegraphics[width=0.47\textwidth]{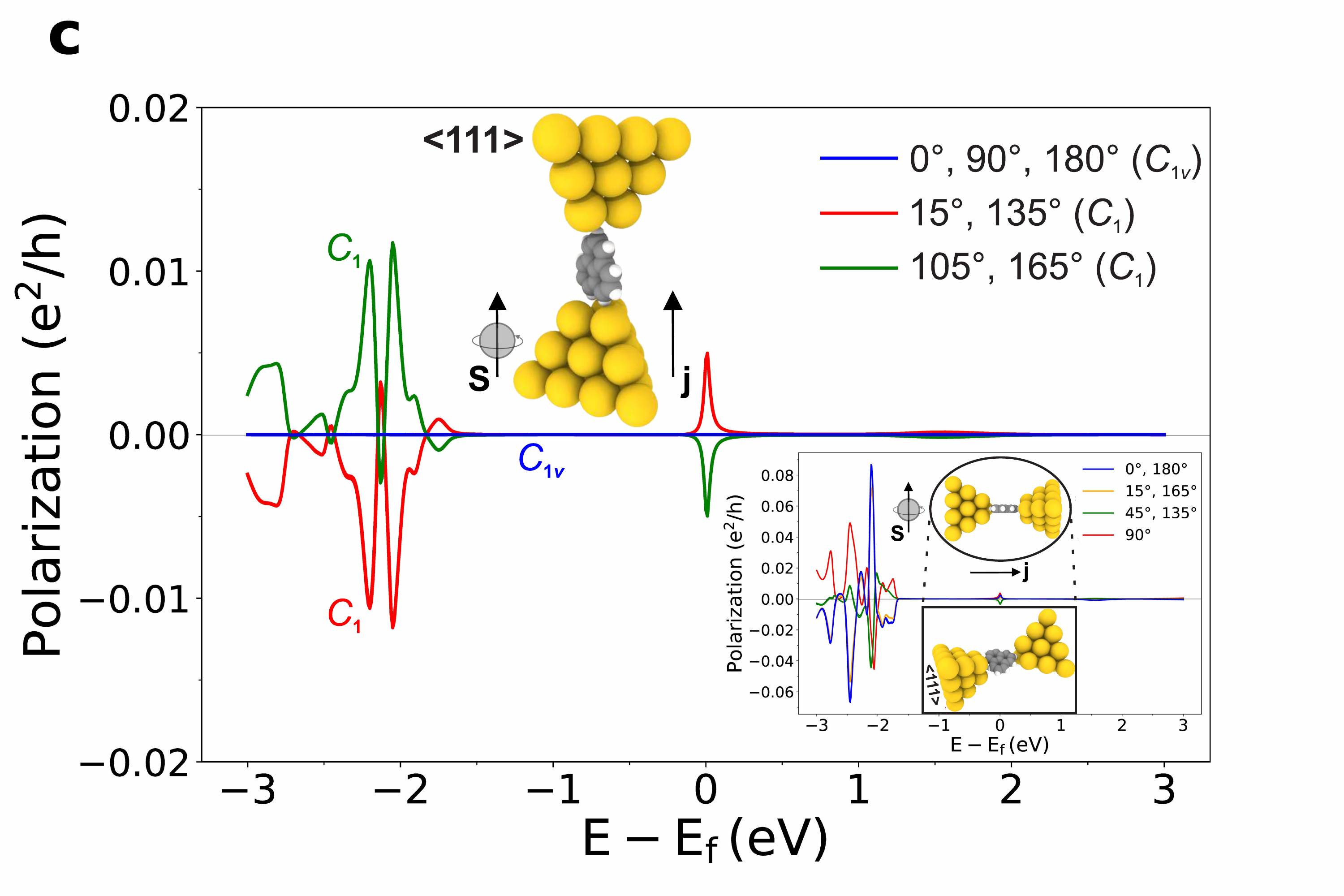}
\includegraphics[width=0.47\textwidth]{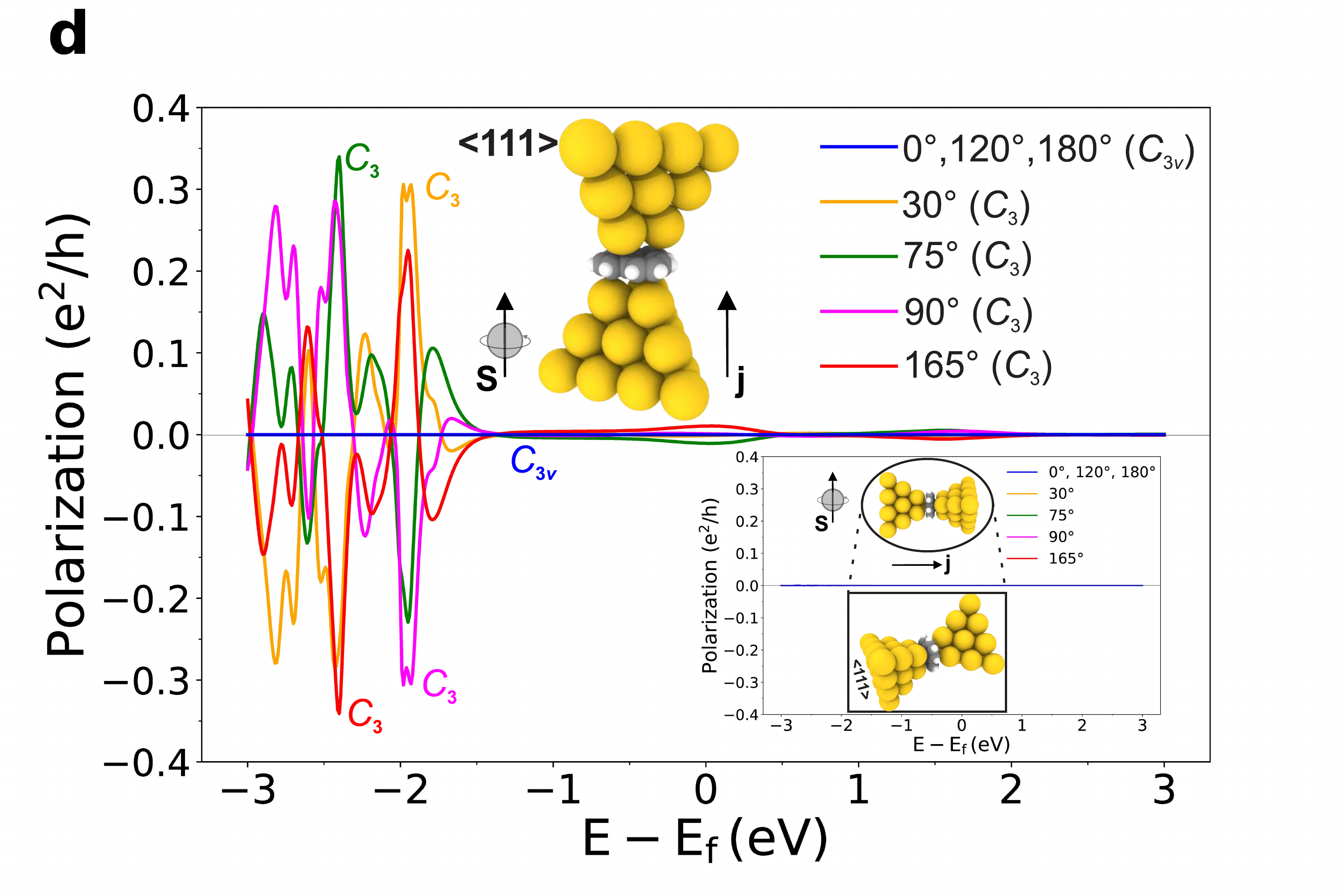}
\caption{
DFT-computed zero-bias spin-polarization \eqref{eq1} along the longitudinal and a transversal direction (the latter shown in the inset panels), as a function of energy and for different arrangements involving achiral molecules. In all cases the direction of propagation and the polarization component are respectively indicated by $\bm{j}$ and $\bm{s}$ (inset panels thus corresponding to transversal polarization of the original structures of which orthographic and perspective views are also shown). The angles correspond to rotations of the molecule alone, along the longitudinal direction. SOC is considered only on the metallic electrodes. \textbf{a}, Au(001) electrodes with a parallel benzene (main rotation axis perpendicular to the current direction). 
\textbf{b}, Source W(001) and drain Au(001) electrodes with a parallel benzene. 
\textbf{c}, Au(111) electrodes with a parallel triangulene. 
\textbf{d}, Au(111) electrodes with a triangulene perpendicular to the current direction. 
The corresponding point groups of the junctions are indicated in the figures along with the general crystallographic directions of each electrode fragment in the inset structures. In \textbf{a}, \textbf{b} and \textbf{d}, the vanishing of the transversal polarization for all energies at the angles with lower symmetry is respectively induced by $C_{2,l}$, $C_{2,l}$ and $C_{3,l}$ symmetries. Note that the longitudinal polarization in \textbf{a} is non-vanishing even in the presence of inversion symmetry $I_{s}\in C_{2h}$ (the benzene molecules themselves are also centrosymmetric).
\label{fig5}}
\end{figure*}

The point group $\pazocal{G}$ at each rotation angle can be inferred directly from Figure \ref{nanocontacts}. At $\theta=\pi/n$, that is, the dihedral angle of the corresponding $\pazocal{G}$ at $\theta=0$, the system recovers the longitudinal mirror planes and so the polarization is again vanishing. 
Importantly, for any two angles $\theta=\pi/n\pm\alpha$ (or more generally, $\theta=m\pi/n\pm\alpha$, $\forall~ m\in\mathbb{Z}$) the corresponding systems are related by a longitudinal mirror plane, that is, each arrangement of electrodes transforms into the other by application of a reflection $\sigma_{l}$ to the whole system. In this case, for $\alpha$ not an integer multiple of $\pi/n$ the systems are chiral, one being the enantiomorph of the other. This is a simplified case of the situations with achiral and chiral molecules which are treated below, see Figure \ref{fig5},\ref{fig8}, but the results \eqref{eqTenantlong} are the same: \textit{any two systems that can be obtained from one another by a reflection across a plane that does not permute the electrodes present opposite spin-polarization in the direction of propagation}. In fact, the assertion is true for any direction contained in the plane. This phenomenon can be thought of as a generalization of the CISS effect (in regards to spin-polarization), in the sense that the spin-polarization along the direction of transport vanishes if the system presents such a plane of symmetry, which is forbidden by chirality. In the context of Figure \ref{fig2}, the polarization$-$angle curves are hence $\pi/n-$antiperiodic when for some $\theta$ there is at least one longitudinal mirror plane in $\pazocal{G}$.
 
From Figure \ref{fig2} it follows that Au exhibits the greatest spin-polarization among the chosen materials, albeit the values with W and Pb nanocontacts are also significant. For all three metals, the $4-$fold nanocontacts noticeably exhibit the largest polarization (across all rotation angles), as compared to the $3-$fold ones. 

Therefore, standalone metal nanocontacts with strong SOC can exhibit significant spin-polarization along the direction of transport due to the symmetry reduction induced by a simple continuous transformation: a rotation of one electrode with respect to the other. This result is reminiscent of the Rashba-Edelstein effect, \textit{e.g.}, as reported in graphene\cite{Chico2015,Rassekh2021} and other 2D crystals,\cite{Safeer.acs.nanolett.9b03485}.

Achiral molecules, such as benzene or polycyclic aromatic hydrocarbons (in our case a three-ring polybenzenoid, or ``triangulene'', with a $3-$fold rotation axis), can still give rise to significant spin-polarization when the entire molecular junction, electrodes plus molecule, breaks longitudinal spatial symmetries. This is demonstrated for particularly interesting configurations in Figure \ref{fig5} \textit{via} DFT quantum transport calculations, in analogy with Figure \ref{fig2} but as a function of energy and for several rotation angles of the molecule (keeping both electrodes fixed). Unlike previous studies,\cite{Zollner2020b} here SOC is only considered in the metallic electrodes and ignored in the molecules, which emphasizes that the qualitative role of the molecule 
is purely geometrical.

The introduction of a molecule between the electrodes potentially removes longitudinal rotation and mirror symmetries from the point group $\pazocal{G}$ of the system, the latter depending on the rotation angle of the molecule with respect to the electrodes. The presence of these symmetries forced the spin-polarization to vanish along the corresponding axes (see Table \ref{table1}) in the systems of standalone electrodes (see Figure \ref{nanocontacts},\ref{fig2}). Hence the qualitative effect of the addition of the molecule in the polarization$-$energy curves is similar to that of the relative rotation of electrodes\footnote{The quantitative effect, however, is in general completely different. Specific finite values of spin-polarization can vary greatly depending on the molecular energy levels and their orbital character. This can be seen in Figure \ref{fig5}c where a finite polarization appears at the Fermi energy despite of the use of Au electrodes. The presence of molecular states at the Fermi energy is due to the zero-gap degenerate nature of the triangulene spectrum, since open-shell calculations have not been considered here for simplicity\cite{palacios.PhysRevLett.99.177204}.}, further allowing to break longitudinal rotation symmetries. 

The importance of the exact position of the molecule on the spin-polarization is apparent from all cases in Figure \ref{fig5}. For the discrete set of rotation angles that make the longitudinal (regarding the connection to the contacts) symmetry planes of the achiral molecule coincident with the longitudinal symmetry planes of both electrodes, $\pazocal{G}$ contains some $\sigma_{l}$ and so the longitudinal spin-polarization is vanishing at all energies, see the blue curve in each subfigure. For any other rotation angle of the molecule, the separately achiral components constitute a chiral system. The longitudinal polarization at any energy is thus reversed between any pair of angles $m\pi/\text{lcm}(n_{\text{el}},n_{\text{mol}})\pm\alpha$, with $m\in\mathbb{Z}$, $\text{lcm}$ denoting the least common multiple of two integers, and $n_{\text{el}},n_{\text{mol}}\geq1$ the number of longitudinal planes of the electrodes and molecule, respectively. This is because the resulting systems are enantiomorph pairs, in analogy with the bare nanocontacts configurations in Figure \ref{fig2}. In particular, $\text{lcm}(n_{\text{el}},n_{\text{mol}})=\text{lcm}(4,2)=4$ in Figure \ref{fig5}a,b, $\text{lcm}(3,2)=6$ in Figure \ref{fig5}c and $\text{lcm}(3,3)=3$ in Figure \ref{fig5}d. More compactly, the longitudinal polarization at any energy will be $\pi/\text{lcm}(n_{\text{el}},n_{\text{mol}})-$antiperiodic in the rotation of the molecule alone, since the latter is achiral (assuming that the bare pair of electrodes presents any longitudinal mirror symmetry and that it extends to the whole junction for some connection of the molecule); hence the corresponding function is $2\pi/\text{lcm}(n_{\text{el}},n_{\text{mol}})-$periodic.

The insets in Figure \ref{fig5}a,b,d demonstrate the vanishing of the transversal polarization at all energies due to $C_{2,l}$, $C_{2,l}$, $C_{3,l}$ symmetries (at least for the angles that yield the smaller point groups, as specified in the caption, which lack longitudinal mirror planes), respectively. The numerical values obtained in these cases were below the computational error threshold.

As stated above, inversion symmetry is compatible with a finite spin-polarization in the two-terminal device. This is numerically exemplified in Figure \ref{fig5}a for the angles at which $\pazocal{G}=C_{2h}$, a group that allows for a non-null longitudinal polarization. The point groups containing inversion symmetry that do not force a vanishing polarization, listed in the \textit{Symmetry considerations} section, are somewhat elusive with standard electrode choices unless an appropriate symmetry-breaking
molecule (removing all transverse rotation symmetries or the transverse mirror plane while keeping inversion symmetry), is introduced into the system.
This may disguise the fact that inversion symmetry (or the geometrical breaking thereof) is qualitatively irrelevant for spin-polarized transport.

These general rules are consistent with the results of Guo \textit{et al}. \cite{Guo2016}, see \textit{Supplementary Information} for further details.

It is worth noting that the vanishing of the spin-polarization in Figure \ref{fig5}a above approximately $-2.5$ eV is due to the exclusive $s-$orbital character of the bulk Au(001) bands in that energy range, see the \textit{Supplementary Information}. Being proportional to the $\bm{L}\cdot\bm{S}$ operator, SOC is therefore not present in the system at these energies and so the current must be spin-unpolarized, see \textit{Methods}. In contrast, the W(001) electrode in Figure \ref{fig5}b does not share this peculiarity, hence enabling spin-polarization in the previous energy range.
Nevertheless, depending on where the energy levels of the molecule lie relative to the Fermi level of the junction, it may  be possible to observe spin-polarized current at accessible energies (bias voltages) in experimental molecular junctions even with Au contacts (see Figure \ref{fig5}c). In the case of W or Pb electrodes, the chance  of always detecting a finite signal at bias voltages on the order of a few hundred mV is greatly enhanced.

\subsection{Enantiomeric partners and polarization reversal}

\begin{figure}[H]
\centering 
\includegraphics[width=0.32\textwidth]{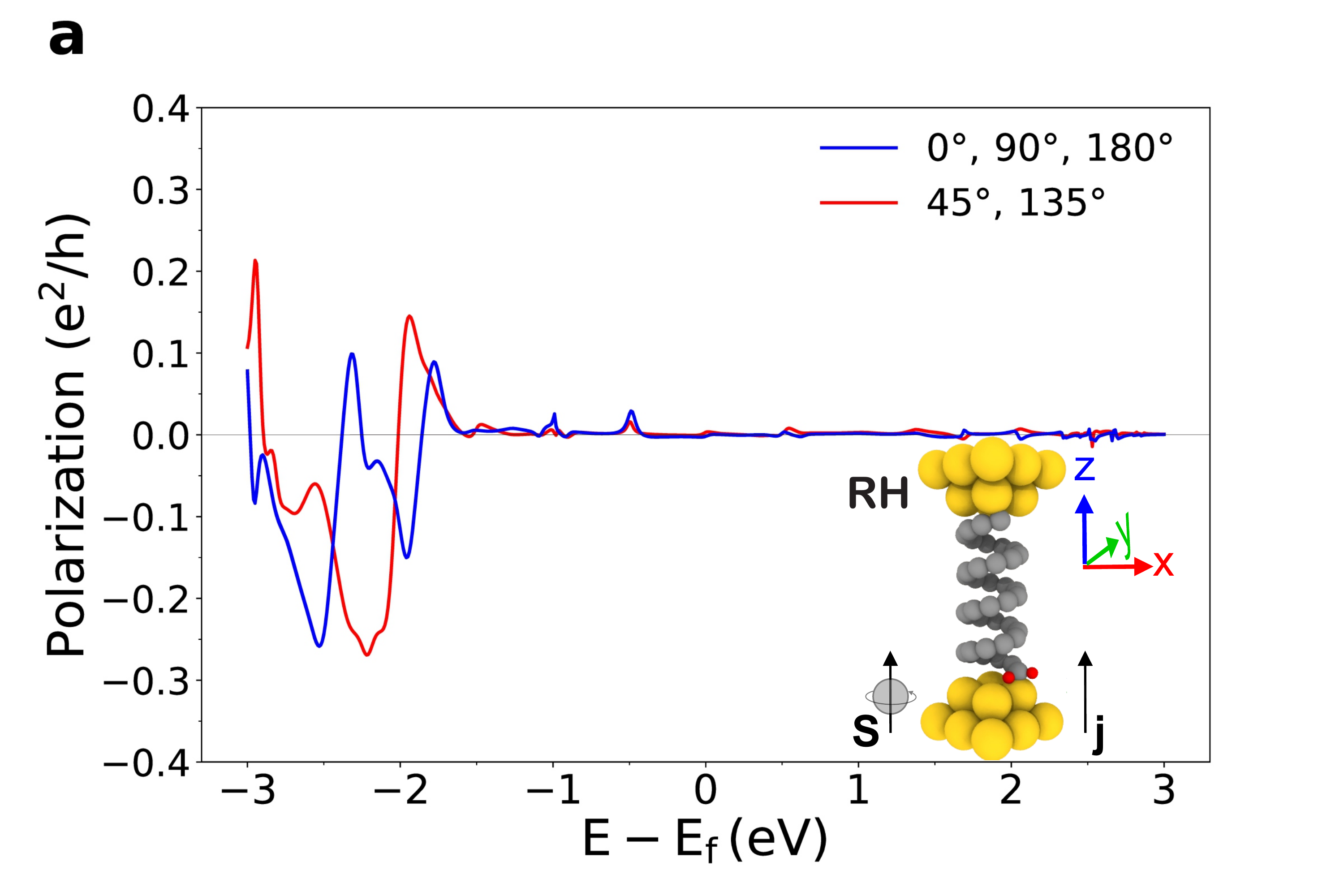}
\includegraphics[width=0.32\textwidth]{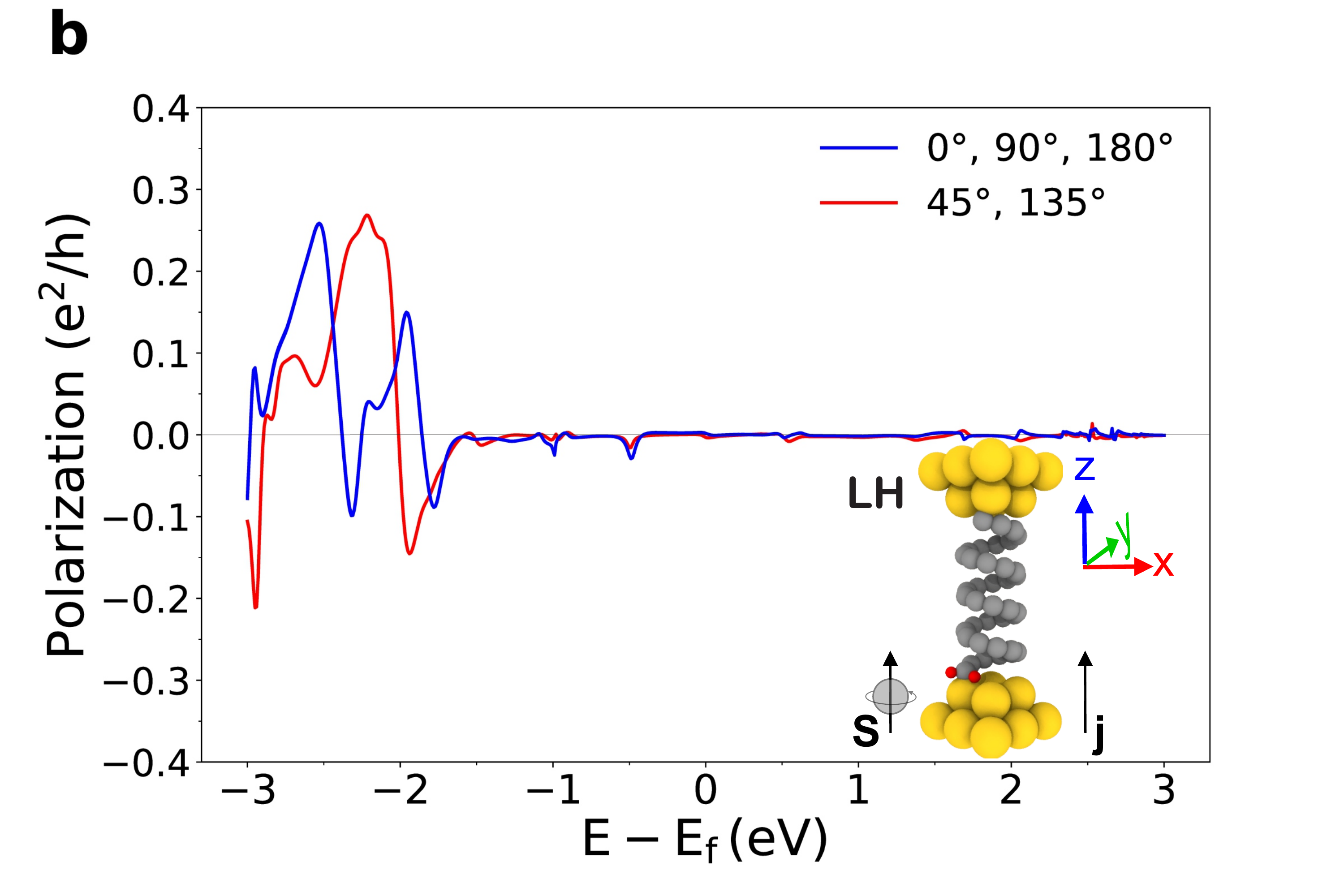}
\includegraphics[width=0.32\textwidth]{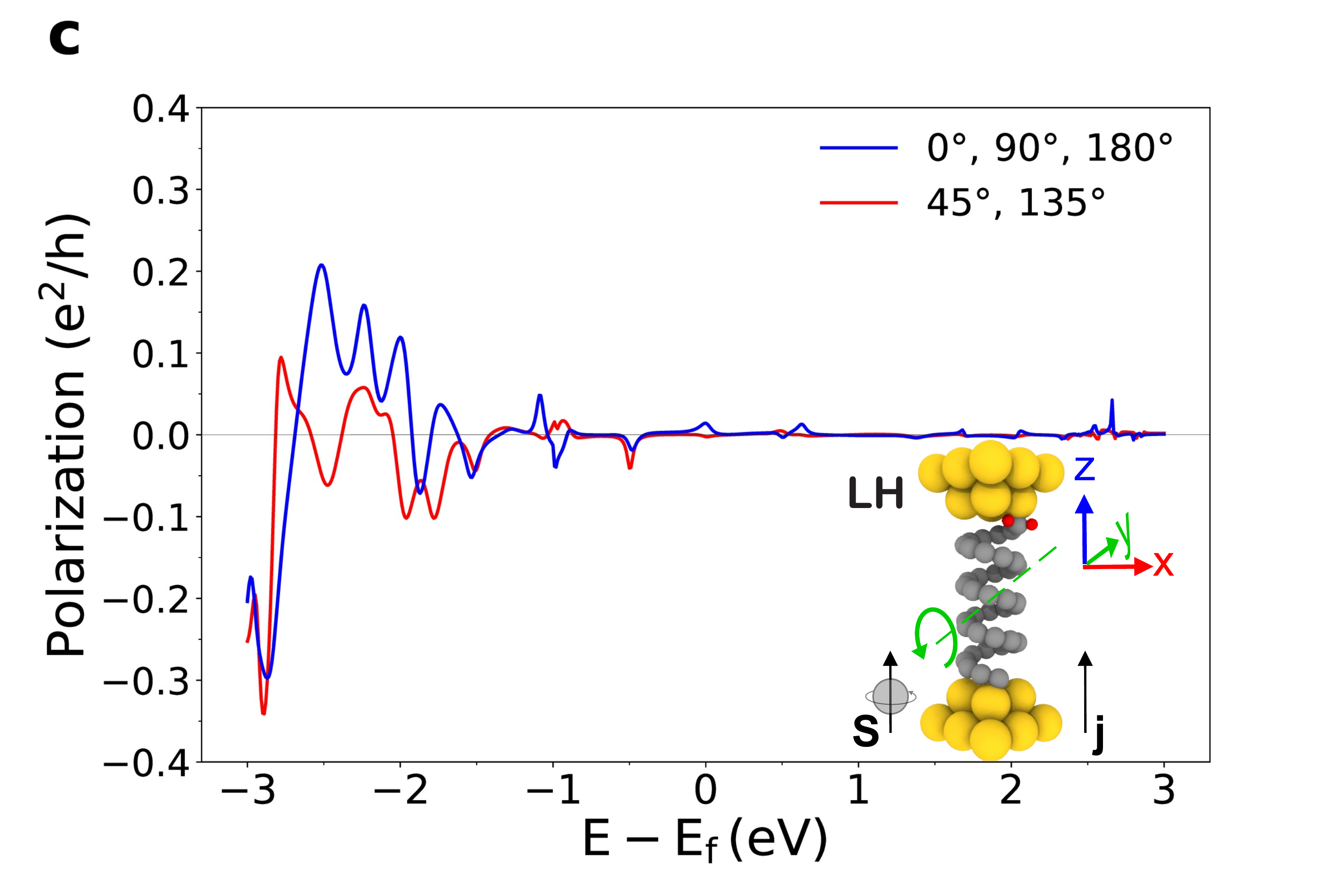}
\includegraphics[width=0.32\textwidth]{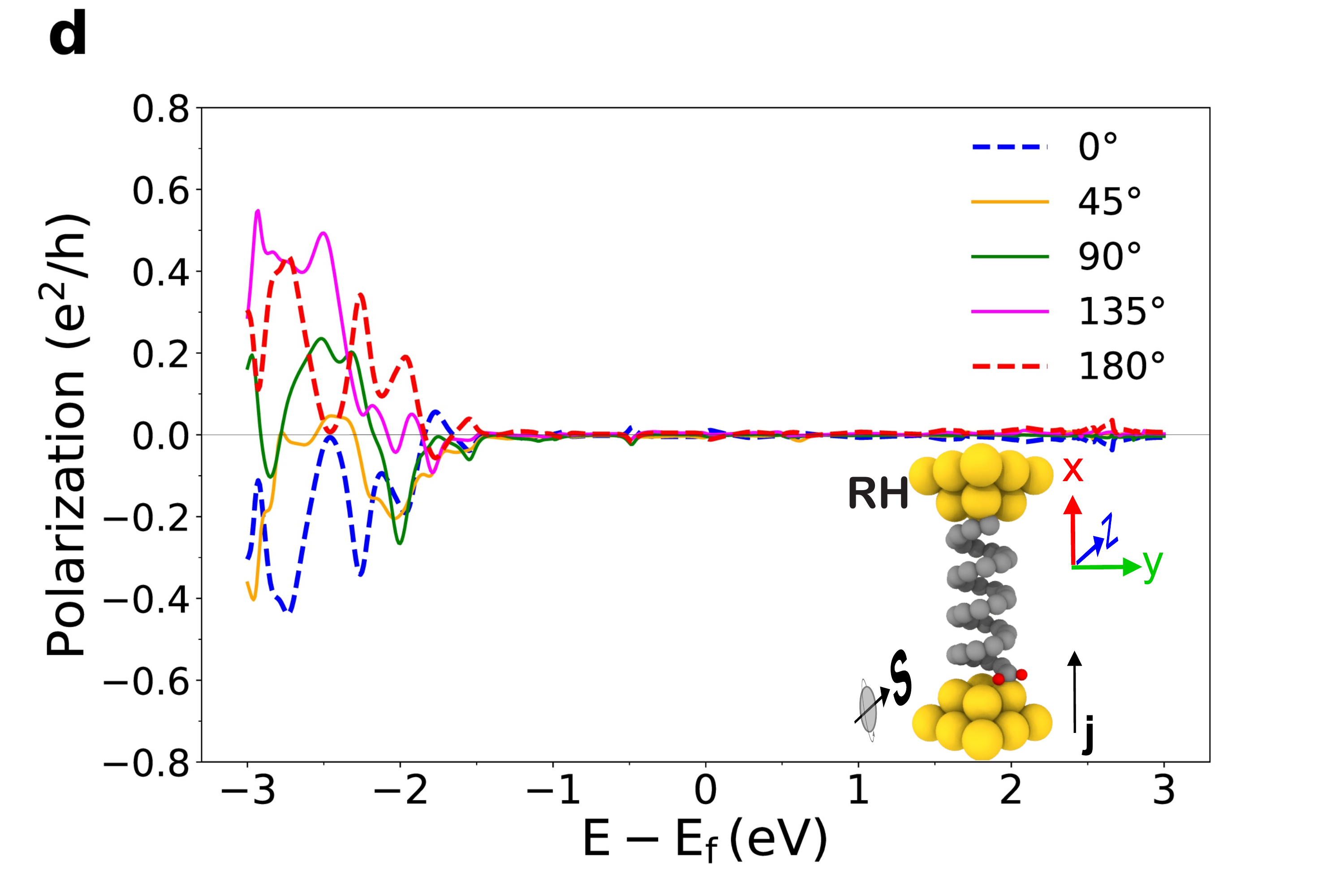}
\includegraphics[width=0.32\textwidth]{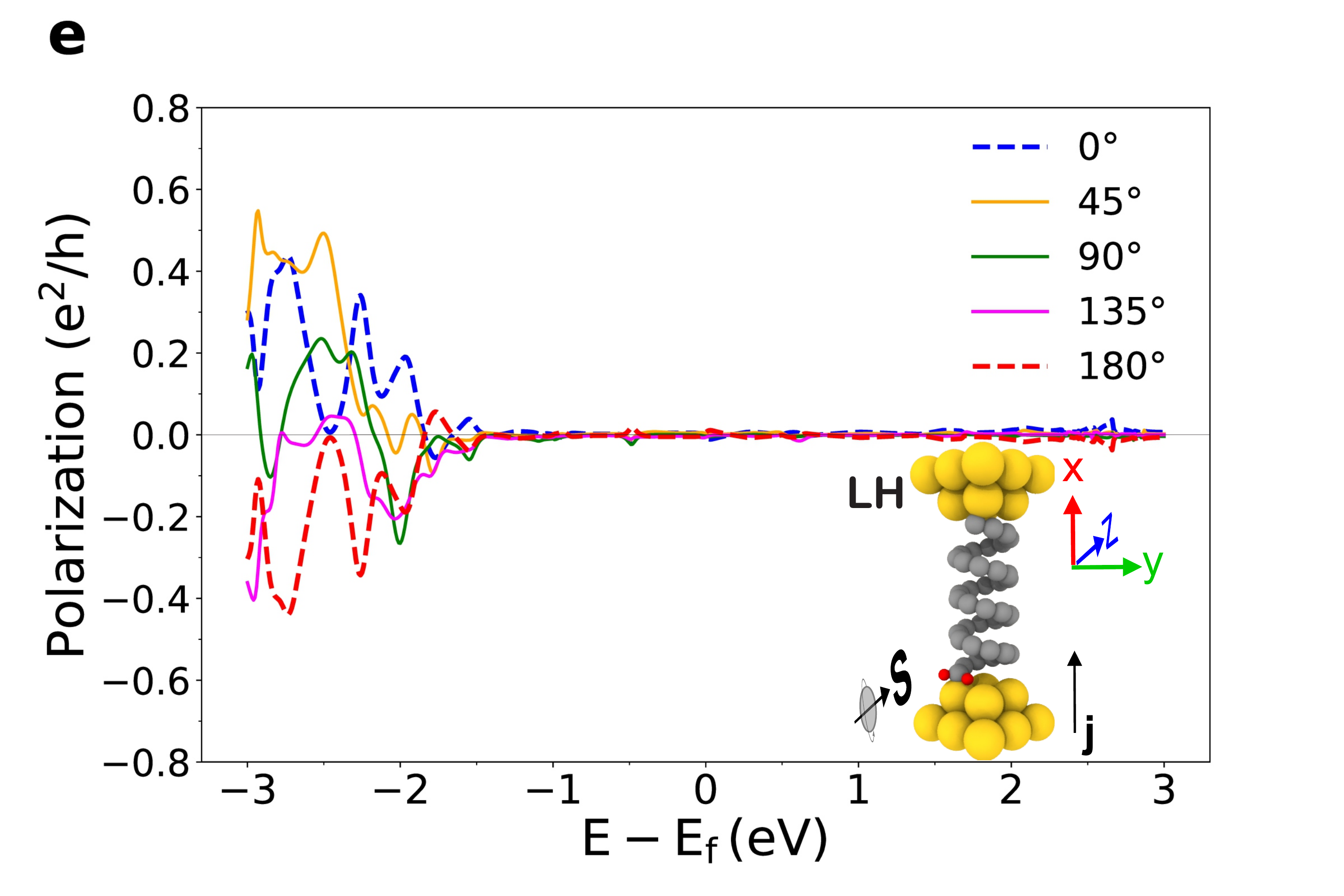}
\includegraphics[width=0.32\textwidth]{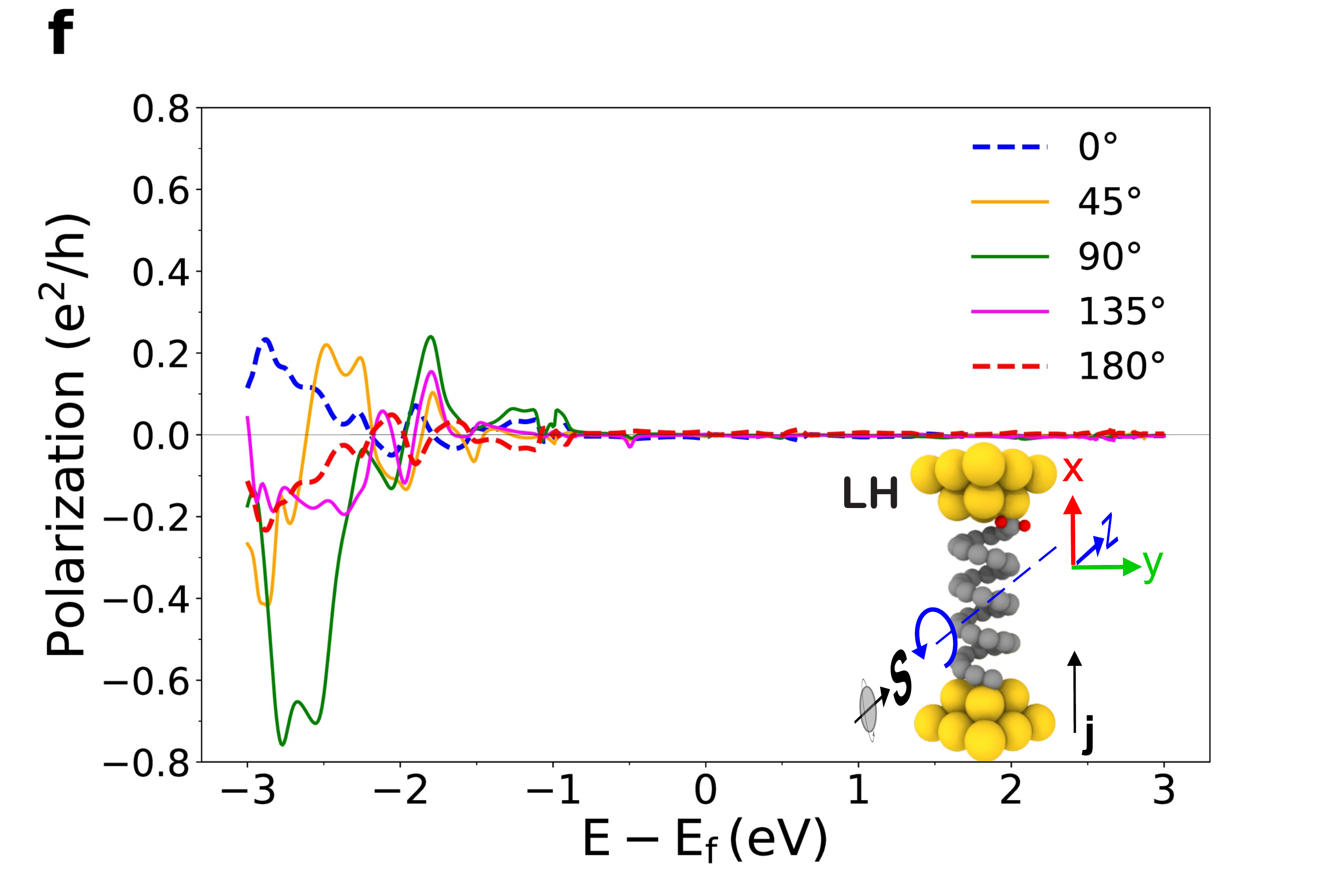}
\includegraphics[width=0.32\textwidth]{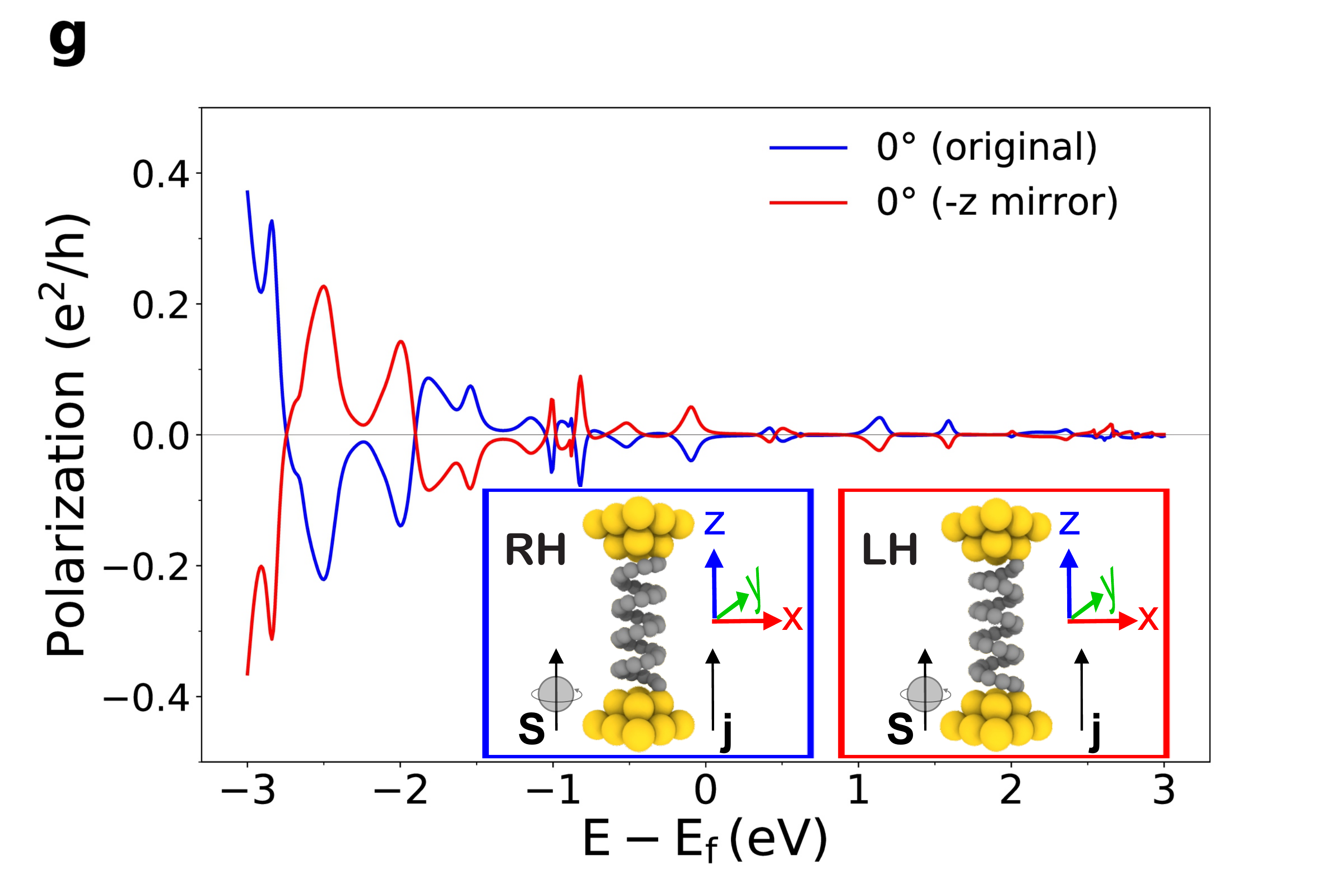}
\includegraphics[width=0.32\textwidth]{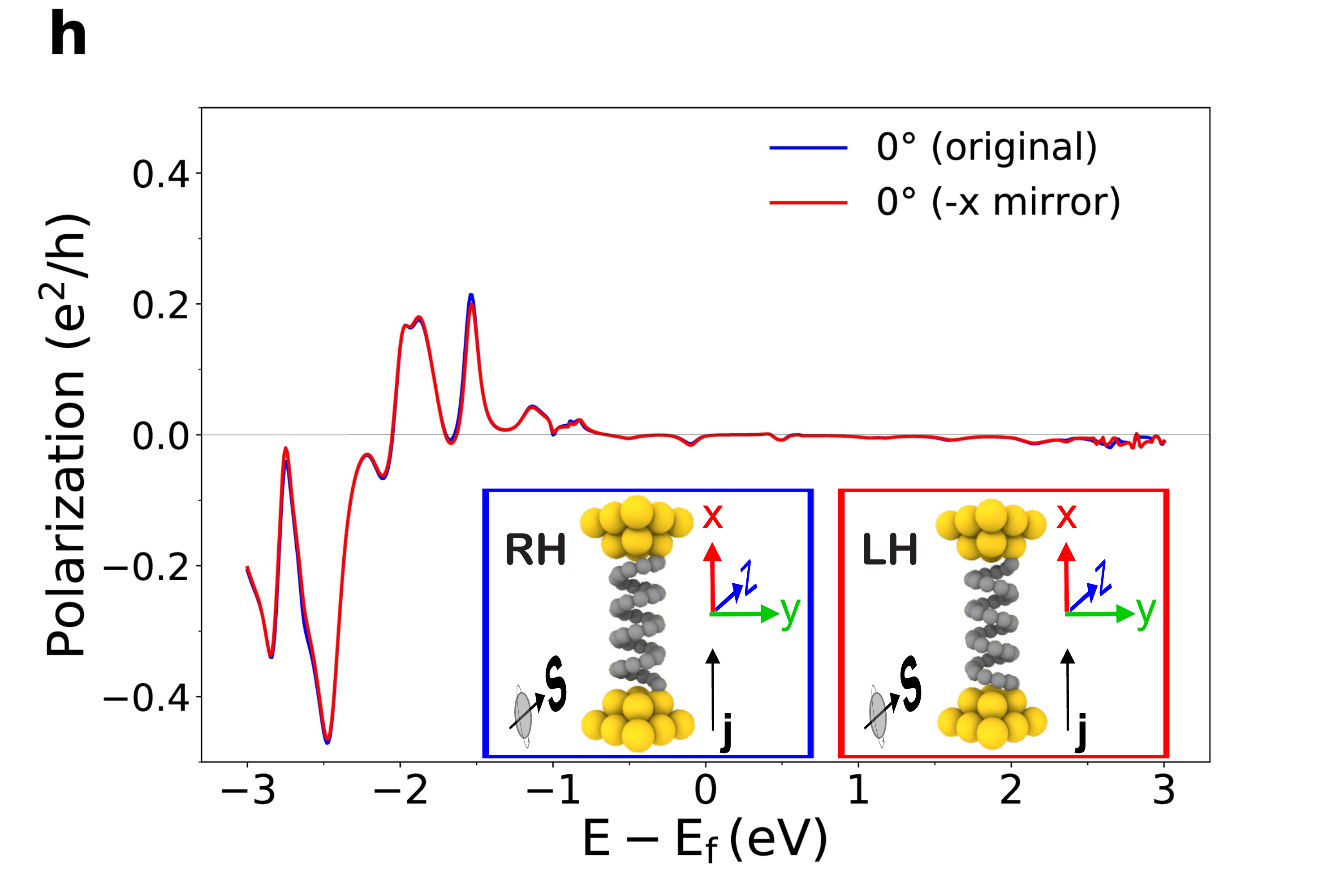}
\includegraphics[width=0.32\textwidth]{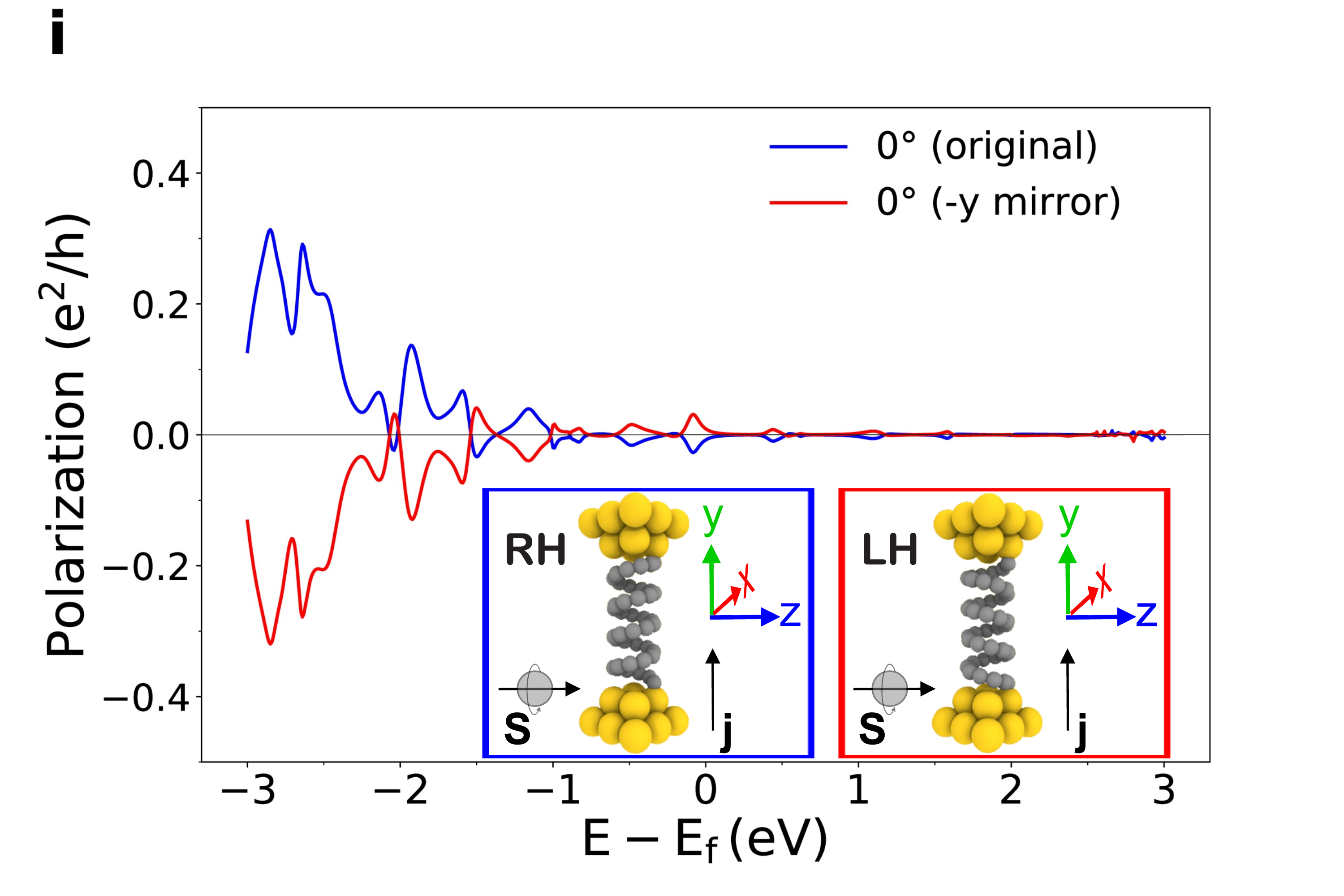}
\caption{
DFT-computed zero-bias spin-polarization \eqref{eq1} along the longitudinal and transversal directions, as a function of energy and for different arrangements involving chiral molecules, in particular two types of carbon helices (see the main text). In all cases, $\bm{z}$ and $\bm{s}$ indicate the corresponding polarization component. The angles correspond to rotations of the molecule alone, along the longitudinal direction. SOC is considered only on the metallic electrodes, which are Au(001) in all cases. \textbf{a}, Asymmetric carbon helix; longitudinal polarization. \textbf{b}, Its enantiomeric partner connected such that the system is the specular reflection of \textbf{a} through a (centred) longitudinal plane; longitudinal polarization. \textbf{c}, Analogous to \textbf{b}, but with the connection of the enantiomer realized through a reflection on the (centred) transversal plane; longitudinal polarization. \textbf{d}, \textbf{e}, \textbf{f}, Analogous to \textbf{a}, \textbf{b}, \textbf{c} respectively, but for transversal polarization (spins point into the page as the perspective in the insets suggests). Point group of the junctions in \textbf{a}-\textbf{f}: $\pazocal{G}=C_{1}$ (trivial). \textbf{g}, Symmetric carbon helix with the transversal rotation $C_{2,x}$. Original (blue) and enantiomeric partner (red) connected such that the system is the specular reflection of the original one through the transversal plane; longitudinal polarization. \textbf{h}, Analogous to \textbf{g}, but for transversal polarization in a direction perpendicular to the $C_{2}$ symmetry axis. \textbf{i}, Analogous to \textbf{g}, but for transversal polarization in a direction parallel to the $C_{2}$ symmetry axis. Point group of the junctions in \textbf{g}-\textbf{i}: $\pazocal{G}=C_{2}$, with transversal orientation. LH and RH refer to left- and right-handed enantiomers, respectively. The longitudinal spin-polarization is reversed between enantiomeric pairs as long as both systems as a whole (including the nanocontacts) are related by reflection through a plane that does not permute the electrodes (\textbf{a},\textbf{b}). This is also true for the transversal polarization component contained in such plane (\textbf{d},\textbf{e}), while the transversal component normal to it remains invariant. If the enantiomeric partner is connected according to a transversal plane, the polarization is altered unpredictably (\textbf{c},\textbf{f}) unless the molecule has a $2-$fold transversal rotation symmetry, in which case both connections of the enantiomer are identical (\textbf{g}-\textbf{i}).  
\label{fig8}}
\end{figure}

In the following we consider left- and right-handed chiral molecules which make up enantiomeric pairs. Specifically, these molecules are helices made out of a carbon chain and we explicitly refer to them as a carbon helix. They have been employed in previous theoretical studies of the CISS effect\cite{Zollner2020b} and we consider here two variants (see the \textit{Supplementary Information}) along with their respective enantiomeric partners. The first one, which we refer to as asymmetric, has no spatial symmetries and is depicted in Figure \ref{fig8}a-f. The second, which we refer to as symmetric, presents in contrast a single spatial symmetry, namely a $2-$fold transversal rotation symmetry through its center, and appears in Figure \ref{fig8}g-i. The difference between them is the removal of two $C$ atoms and the presence of $H$ atoms (depicted in red) on a single end of the former. 

As shown in Figure \ref{fig5},\ref{fig8}a,d and proved in equation \eqref{eqTenantperp}, a longitudinal rotation of the molecule changes the spin-polarization along any direction in an unpredictable way, except for a discrete set of angles determined by the symmetries of the molecule and electrodes (see equation \eqref{eqTnosym} and the discussion thereafter). It is then to be expected that a variation in the anchoring of the enantiomeric partner between the contacts also has an unpredictable effect on the polarization. By equation \eqref{eqTenantlong}, however, the two sets of polarization$-$energy curves (along a given direction) obtained from the pair of molecules have a one-to-one correspondence, given by a reversal of sign at least along the longitudinal direction according to equation \eqref{eqTenantlong}. This reversal of the spin-polarization, which accounts for the CISS effect, occurs when the enantiomeric partner junction is realized through the application of a mirror symmetry of both electrodes, turning one system into another (enant$_{1}$ in \textit{Methods}). These two junctions are shown in Figure \ref{fig8}a,b.

This way of realizing the enantiomeric partner junction (up to longitudinal rotations), namely, through longitudinal mirror symmetries of the electrodes, yields, to an extent, a familiar result. In contrast, if the connection of the enantiomeric partner is realized through a reflection on the transversal plane (enant$_{2}$ in \textit{Methods}), by equation \eqref{eqTenantperp} the spin-polarization will in general not be related to that of the original molecule, even if longitudinal rotations are considered. This is displayed in Figure \ref{fig8}c,f. Note that it has been implicitly assumed in these general discussions that the standalone electrodes present the corresponding planes of symmetry, for which Figure \ref{nanocontacts},\ref{fig2} may be used as a guide.

Both ways of realizing the enantiomeric partner connection (up to longitudinal rotations) are qualitatively different due to the absence of a $2-$fold transversal rotation of the molecule. The spin-polarization of the enantiomeric partner junction obtained through the transversal plane $\sigma_{t}$ will be equal to that obtained through a longitudinal plane $\sigma_{l}$ if the molecule presents the aforementioned rotation symmetry and it is placed with the appropriate longitudinal angle such that the rotation symmetry is shared by the electrodes, \textit{i.e.}, $C_{2,t}=\sigma_{l}\sigma_{t}\in\pazocal{G}$. This is exemplified in Figure \ref{fig8}g-i, where the symmetries are respectively $C_{2,x},C_{2,y},C_{2,z}\in\pazocal{G}$ (shared by the electrodes and chiral molecule). The ``transversal'' enantiomeric partner (analogous to that in Figure \ref{fig8}c,f and displayed in the red panels) is identical to that obtained through a reflection on a longitudinal plane, as it should be since $\sigma_{t}=\sigma_{l}C_{2,t}$ and $C_{2,t}\in\pazocal{G}$ is a symmetry of the system; and so is the resulting polarization. The invariance (reversal) of the polarization of the enantiomeric partner in Figure \ref{fig8}h (\ref{fig8}i) occurs due to the geometrical equivalence between inverting the longitudinal coordinate and the spin direction $\hat{\bm{z}}$ (the transversal direction normal to $\hat{\bm{z}}$, respectively), see equation \eqref{eqTenantC2}.

\subsection{Experimental perspective}
We have not addressed in the above how the predicted spin-polarization, as defined in \eqref{eq1}, may be actually verified in the laboratory. This would certainly require 
the application of a finite bias between the electrodes, establishing a charge current $I$ which is then accompanied by preferred spin orientations, constituting the spin current $\bm{\pazocal{I}}\propto\bm{P}I$\cite{nikolic2005decoherence}. In principle, the spin current could be detected in two-terminal devices in the equilibrium limit as long as the unitary scattering formalism does not strictly hold, which in particular would be ensured in the presence of significant mode-selective dissipation.\cite{acsnano.2c07088}. However, one should also consider that the polarization of the current may be greatly suppressed not far from the scattering region if dissipative processes are abundant in the drain electrode. 

Currently, CISS experiments typically involve magneto-conductance measurements, where finite values have been reported with two-terminals in the linear regime, exploiting the spin-valve effect.\cite{acsnano.0c07438} Ferromagnetic components, either metals or doped semiconductors, are employed for such detections and the magneto-conductance can be identified with $\Delta G(\bm{M},V)=G(\bm{M},V)-G(-\bm{M},V)$, where $\bm{M}$ is the total magnetization, $V$ is the bias voltage (fixing the drain/source nature of the electrodes) and $G$ is the total conductance. Onsager's relation fundamentally forbids a finite equilibrium value $\Delta G(\bm{M},0)$, relying only on the applicability of the unitary scattering formalism as well as, of course, the perfect reversal of the magnetization in the two separate conductance measurements (see \textit{Methods}). The detection of a finite value could then indicate the presence of significant inelastic effects.
A more revealing feature, also applicable at finite bias, can be obtained by application of the symmetry analysis to the magnetic system. The Hamiltonian of the junction is then no longer time-reversal invariant and $\Theta$ is thus not a symmetry, but more importantly, a finite magnetization along the longitudinal (a transversal, resp.) direction will remove all longitudinal mirror (rotation, resp.) symmetries. This will induce a finite spin-polarization, if it was not already present, along the corresponding directions according to Table \ref{table1}. Note, however,  that its origin is magnetic (it also perfectly changes sign upon magnetization reversal, as shown below) and should not be attributed to CISS. In the case of magnetization along the direction of transport, if the underlying structure disregarding magnetism presents a longitudinal symmetry plane $\sigma_{l}$, \textit{i.e.}, $\Theta\sigma_{l}$ is a symmetry of the magnetic point group, a straight-forward application of \eqref{eqTnosym} with $\hat{S}(\sigma_{l}\bm{r},\bm{M})=\hat{S}(\bm{r},-\bm{M})$ yields $G_{s',s}(\bm{M})=G_{\overline{s}',\overline{s}}(-\bm{M})$, hence $\Delta G=0$. Likewise, in the case of magnetization along any direction perpendicular to transport, if $\Theta C_{n,l}$ (recall that $C_{n,l}$ is a rotation of $2\pi/n$ along the transport direction) is a symmetry, then $\Delta G=0$ by an analogous argument. Therefore, the geometrical selection rules for the spin-polarization are the same as for the magneto-conductance, establishing the analogy between the vector components of $\bm{P}$ and $\bm{M}$. Note that the relevant set of symmetries is that of the purely spatial or unitary ones, not the magnetic point group introduced by the ferromagnet. 


It may be worth noting that, in practice, the symmetry of the system may be reduced by defects, often in an uncontrolled way. As is customary in solid-state physics, group-theoretical results may then be understood as a limiting case. In particular, the spin-polarization would be expected to be small, but not strictly vanishing, if the deviation from the suitable, perfectly symmetrical configuration was also small, \textit{i.e.}, the symmetry of the system is preserved on a course scale. The relevant defects, however, would be those close to the tips of the electrodes, in the scattering region (on which our analysis is based). Furthermore, the detection of a vanishingly small spin-polarization along a given direction may allow to determine the presence of certain symmetries according to Table \ref{table1}. Another important topic in connection with practical applications and experiments, especially at room temperature, is the presence of molecular vibrations and their coupling to the electronic spin \cite{acs.nanolett.1c00183,das2022temperature}. In principle, we would not expect qualitative changes to our steady-state results in the limit in which the unitary scattering formalism is still applicable and one should stick to it, since the group $\pazocal{G}$ is unchanged for the perturbed Hamiltonian around the equilibrium positions (see, for example, equation 2 in reference \cite{PhysRevB.102.235416}); although a further study would be convenient in this regard. It is worth noting that temperature-induced atomic vibrations will most likely not have an appreciable effect in systems with bare contacts of heavy atoms, while they will generally enhance the effect of SOC in the molecule (here disregarded compared to that of the leads), constituting in principle an important correction to numerical results for the latter systems at room temperature. In any case, we have shown that the reduction of symmetry will, if anything, favor the emergence of a finite spin-polarization and magneto-conductance.

Regarding the possible reversal of the spin-polarization upon substitution of a chiral molecule with its enantiomeric partner, it should be noted that even though the relative longitudinal rotation angle between the molecule and its enantiomeric partner may not be controllable in practice (so that the polarization reversal could not be ensured), still the averaged spin-polarizations over multiple different connections (with varying relative rotational angles) of the two enantiomers will tend to be the opposite of each other. Furthermore, while for the longitudinal direction the average spin-polarization will in general be a finite value, for any transversal direction the average will tend to zero since the polarization at any given energy is $\pi-$antiperiodic due to equation \eqref{eqTenantlong}\footnote{A longitudinal $\pi-$rotation of the molecule is equivalent to the reversal of the transversal direction from which the spin-polarization is measured, if that is a symmetry of the electrodes.}. This is illustrated in Figure \ref{fig8}d,e. Meanwhile for achiral molecules, if the junction presents a longitudinal mirror symmetry for any relative rotation angle, then the angle-average longitudinal spin-polarization will always tend to zero.

At a quantitative level, our DFT results show that, as expected, Au nanocontacts do not exhibit finite spin-polarization near the Fermi level (due to predominant $s$-orbitals) and, by extension, will not show a measurable magneto-conductance at small bias voltages, unless an organic molecule with frontier $p$-orbitals near the Fermi level is included in the junction. This is most likely the case in the recent experimental study by Liu \textit{et al.} \cite{acsnano.0c07438}, where a measurable magneto-conductance at a bias voltage as low as $100$ mV is reported for a two-terminal molecular junction with Au as one of the electrodes. In general, however, we note that FCC Pb and BCC W electrodes exhibit non-negligible spin-polarization near the Fermi level and thus, are more likely to show experimental traces of magnetoconductance at low bias voltages. In any case, even if the presence of a molecule is not needed to generate a finite spin-polarization, they can still be crucial to achieve the rather large values that have been measured.\cite{Kulkarni.adma.201904965,naaman2020chiral}

As a final remark, we note that the phenomenon displayed in Figures \ref{fig2}, \ref{fig5}, namely the emergence of a finite spin-polarization along the transport direction upon rotation of a single component of the junction, indicates a potential mechanism to mechanically switch between polarized and unpolarized currents within the same device, something that cannot be accomplished with permanent ferromagnetic elements.

\section{Conclusions}
The lack of certain symmetry planes is the distinctive feature of chiral molecules regarding their potential to induce spin-polarized transport in molecular junctions. This is so irrespective of the specific shape of the molecule (whether they are helix-like or not) or the specific dominant electronic orbitals in the junction (as long as they are not all isotropic), which may nevertheless play an important quantitative role. Here we have extended the concept of chirality in the CISS effect to the junction as a whole (or extended molecule) and provide the following generalization: any two systems obtained from one another by the reflection through a plane that does not permute the electrodes will present opposite spin-polarizations along the direction of transport. In particular, systems that are left invariant under such a reflection cannot induce a finite polarization along that direction, in agreement with the traditional CISS concept. The inevitable breaking of these symmetry planes in junctions with chiral molecules provides a sufficient condition to obtain a finite spin-polarization, which is enhanced if the electrodes present a strong SOC. 

This therefore extends the class of molecules that can be considered for the CISS effect to not only include chiral molecules which have been widely considered in the literature up to now, but to any molecule - and even nano-contacts without any intermediate molecule - which meet the aforementioned symmetry criteria. In particular, a simple rotation of any component of the junction, molecule (if present), drain, or source electrode will generally induce spin-polarization along the direction of transport due to the breaking of the mirror symmetry planes that do not permute the electrodes.

We finally note that the results presented here are strictly valid within the unitary scattering formalism, in particular at equilibrium. Nevertheless, we postulate that a vanishing spin-polarization in equilibrium remains so when applying a finite bias (and by extension also the magneto-conductance, as long as it does not vanish exclusively due to Onsager's reciprocity), since the longitudinal symmetries are still present in the Hamiltonian. The arguments, however, have to be adapted to the Green's function formalism to account for non-equilibrium conditions, and will be presented in a subsequent work.

\section{Methods}
\textit{Note}: in this work we assume the existence of a single longitudinal direction which acts as a symmetry axis for both electrodes separately, and along which electronic propagation takes place. Most of the analysis in this section would still hold for an arbitrary arrangement of the electrode pair, particularly equations \eqref{eqT}-\eqref{eqTsymABTR}, \eqref{eqTnosym}; but care should be taken when cataloging the possible symmetries of the system and the directions of spin quantization axis for which the spin-polarization is qualitatively affected by them. \\\\
Electronic transport in a system with two leads can be formulated as a scattering problem between Bloch states of the isolated electrodes. \cite{chico1996quantum,zhai2005symmetry,beenakker1997random,paulsson2007transmission,utsumi2020spin} At a given energy $E$, let $\alpha_{i}$, $\alpha'_{i}$ ($i=1,\dots,M_{A}$) be the incoming and outgoing (with respect to the scattering region) modes of electrode $A$, and $\beta_{j}$, $\beta'_{j}$ ($j=1,\dots,M_{B}$) the outgoing and incoming modes of electrode $B$. These modes label the eigenstates $\psi^{\alpha},\psi^{\beta}$, which are in general spinors, of the corresponding isolated electrode which obey the flux (or energy) normalization\cite{paulsson2007transmission}\footnote{We have implicitly assumed that both electrodes present time-reversal or inversion symmetry, so that at any energy in each electrode the number of incoming modes coincides with the number of outgoing modes (for crystalline electrodes, this is due to the evenness of the energy spectrum in the Brillouin zone). Nevertheless, only the analysis of the anti-unitary symmetries requires that condition; the rest would be analogous. More generally, there is a one-to-one correspondence between the incoming (outgoing) eigenstates of a ferromagnetic electrode and the outgoing (incoming, resp.) eigenstates of the electrode with opposite magnetization, given by the time-reversal operation.}. The spin-resolved conductance between the leads $A$ (initial) and $B$ (final) can then be defined in terms of the scattering matrix elements as: \begin{equation} \label{eqT}
G^{AB}_{s',s}(E)=\frac{e^{2}}{h}\sum_{i=1}^{M_{A}}\sum_{j=1}^{M_{B}}\left|S^{\beta_{j}\alpha_{i}}_{s',s}\right|^{2}=\frac{e^{2}}{h}\sum_{i=1}^{M_{A}}\sum_{j=1}^{M_{B}}\left| \bra{\psi^{\beta_{j}}_{s'}} \hat{S} \ket{\psi^{\alpha_{i}}_{s}} \right|^{2}=\frac{e^{2}}{h}\sum_{i=1}^{M_{A}}\sum_{j=1}^{M_{B}}\left|\int_{\mathbb{R}^{3}}\psi^{\beta_{j}}_{s'}(\bm{r})^{*}\hat{S}(\bm{r})\psi^{\alpha_{i}}_{s}(\bm{r})d^{3}\bm{r}\right|^{2}
\end{equation} 
where $s,s'\in\set{\uparrow,\downarrow}$ respectively label the initial and final spin states, or spinor components, referred to a given direction of quantization and $\hat{S}$ is the unitary scattering operator.\cite{merzbacher1998quantum,belkic2020principles} 
The energy and electrodes labels will be dropped for simplicity, except when mandatory. $\hat{S}$ depends on the spatial coordinates $\bm{r}$ exclusively \textit{via} the Hamiltonian of the complete system, which includes the scattering region formed by the contacts (or part of the leads which is sufficiently close to the scattering region) plus, possibly, a molecule or piece of material between them. For an incident unpolarized current, the component of the spin-polarization (of the outgoing current) along an arbitrary spin quantization axis reads,\cite{nikolic2005decoherence} up to a normalization factor:
\begin{equation*} 
P=G_{\uparrow\uparrow}+G_{\uparrow\downarrow}-G_{\downarrow\uparrow}-G_{\downarrow\downarrow}\;\;, \end{equation*}
where the $\uparrow,\downarrow$ spin states are again referred to the spin quantization axis. The whole polarization vector $\bm{P}$ may then be evaluated by rotating the spin axis and applying \eqref{eq1}, which is the method here employed in numerical calculations with \texttt{ANT.Gaussian}. Analogously, for a fixed quantization axis along $\hat{\bm{z}}$ one may compute the perpendicular components as:\cite{nikolic2005decoherence}
\begin{equation*}
P_{x}=\frac{2e^{2}}{h}\sum_{i=1}^{M_{A}}\sum_{j=1}^{M_{B}}\operatorname{Re}\tilde{P}\:,\;\;\;\;P_{y}=\frac{2e^{2}}{h}\sum_{i=1}^{M_{A}}\sum_{j=1}^{M_{B}}\operatorname{Im}\tilde{P}\:,\;\;\;\;\tilde{P}\equiv S^{\beta_{j}\alpha_{i}*}_{\uparrow\uparrow}S^{\beta_{j}\alpha_{i}}_{\downarrow\uparrow} + S^{\beta_{j}\alpha_{i}*}_{\uparrow\downarrow}S^{\beta_{j}\alpha_{i}}_{\downarrow\downarrow}
\end{equation*} 
with $\hat{\bm{x}},\hat{\bm{y}}$ the directions along which the Pauli vector has components $\sigma_{x},\sigma_{y}$ (in the usual representation) respectively.

In the case of no spin-dependent terms (originating from SOC or non-collinear magnetism), anywhere in the contacts or molecules, then it must be $S^{\beta_{j}\alpha_{i}}_{\uparrow\uparrow}=S^{\beta_{j}\alpha_{i}}_{\downarrow\downarrow}\:$, $\:S^{\beta_{j}\alpha_{i}}_{\uparrow\downarrow}=S^{\beta_{j}\alpha_{i}}_{\downarrow\uparrow}=0$, so that all the polarization components are vanishing. In our case, the metallic contacts present strong SOC and we ignore it in the molecules, as in practice it is negligibly small in light elements. 

Before proceeding, it may be worth noting that we are here dealing with steady-state conditions, as is usual in CISS studies. While one could also perform time-dependent DFT calculations \cite{bushong2005approach}, as long as the potential preserves the original symmetry during the evolution and the density matrix is consistent with it, our analysis would in principle remain valid for both formalisms.

\subsection{Spatial (unitary) symmetries}
Let $\pazocal{G}$ $\subset\:O(3)$ be the point group of the complete system. Then $\hat{S}(g^{-1}\bm{r})=\hat{S}(\bm{r})$ for any spatial symmetry operation $g\in\pazocal{G}$, and owing to the invariance of the integral under this coordinate transformation: \cite{bir1974symmetry}
\begin{equation} \label{eqTg}
S^{\beta_{j}\alpha_{i}}_{s',s}=\sum_{s_{1},s'_{1}}\pazocal{D}^{1/2}(g)_{s',s'_{1}}\pazocal{D}^{1/2}(g)^{*}_{s,s_{1}}\int_{\mathbb{R}^{3}}(\overline{g}\psi^{\beta_{j}}_{s'_{1}}(\bm{r})^{*})\hat{S}(\bm{r})(\overline{g}\psi^{\alpha_{i}}_{s_{1}}(\bm{r}))d^{3}\bm{r}\;\:,\;\:\forall g\in\pazocal{G}
\end{equation}
where $\overline{g}$ denotes the action of the coordinate transformation $g$ on the spinor functions which does not affect its (spin) components, and $\pazocal{D}^{1/2}$ is the (projective) representation of $O(3)$ of angular momentum $1/2$, given by \begin{equation}\label{eqD1/2} \begin{aligned}
\pazocal{D}^{1/2}(C_{\theta})=\pm\pazocal{D}^{1/2}(I_{s}C_{\theta})=e^{i\bm{\sigma}\hat{\bm{e}}\theta/2}=\begin{pmatrix}a&b\\-b^{*}&a^{*}\end{pmatrix}\;,\\ a=\cos(\theta/2)+ie_{z}\sin(\theta/2)\:,\;b=(e_{y}+ie_{x})\sin(\theta/2)
\end{aligned}\end{equation}
for a $\theta-$rotation along direction $\hat{\bm{e}}$, where $\bm{\sigma}$ is the Pauli vector and the parity under inversion $I_{s}$ has been dropped from the notation since it does not affect equation \eqref{eqT} or \eqref{eq1}. This representation carries the transformation of the spinor components under an operation $g\in\pazocal{G}$, and the particular form we have given is valid under the assumption that the spin is projected along $\hat{\bm{z}}$; otherwise an orientation-preserving change of coordinates is needed to determine $\hat{\bm{e}}$.

At this point we must distinguish two cases among the symmetry operations $g\in\pazocal{G}$. Let $\pazocal{G}_{A}$ ($\pazocal{G}_{B}$) be the point group of the isolated electrode $A$ ($B$, resp.) with the same fixed point as $\pazocal{G}$:
\begin{itemize}
\item $g$ is a symmetry for each electrode in isolation, that is, $g\in\pazocal{G}_{A}\cap\pazocal{G}_{B}$. Then $g$ cannot affect the longitudinal coordinate and thus $\overline{g}$ does not alter the incoming/outgoing nature of the modes, nor their energy. Therefore $\overline{g}\psi^{\alpha_{i}}=\sum_{k}\pazocal{D}^{\alpha_{i}}(g)_{k,i}\psi^{\alpha_{k}}$ where $\pazocal{D}^{\alpha_{i}}$ is the representation of $\pazocal{G}_{A}$ according to which $\psi^{\alpha_{i}}$ transforms, and an analogous result follows for electrode $B$. Inserting this in equation \eqref{eqTg}, it is straightforward to prove (invoking the unitarity of such representations) that $G^{AB}_{s',s}(E)$ is actually independent on all $\pazocal{D}^{\alpha_{i}},\pazocal{D}^{\beta_{j}}$ regardless of their dimensions, hence in this case: 
\begin{equation} \label{eqTsymAB}
G^{AB}_{s',s}(E)=\frac{e^{2}}{h}\sum_{i=1}^{M_{A}}\sum_{j=1}^{M_{B}}\left|\sum_{s_{1},s'_{1}}\pazocal{D}^{1/2}(g)_{s',s'_{1}}\pazocal{D}^{1/2}(g)^{*}_{s,s_{1}}\bra{\psi^{\beta_{j}}_{s'_{1}}}\hat{S}\ket{\psi^{\alpha_{i}}_{s_{1}}}\right|^{2}\;\:,\;\:\forall g\in\pazocal{G}\cap\pazocal{G}_{A}\cap\pazocal{G}_{B}
\end{equation}
\item $g$ is not a symmetry for each electrode in isolation, that is, $g\notin\pazocal{G}_{A}\cap\pazocal{G}_{B}$. Then $g$, if it exists, necessarily permutes the electrodes, which must thus have identical composition. In this case the longitudinal coordinate is inverted, hence the existence of such a symmetry establishes a one-to-one correspondence ($\overline{g}$) between the incoming modes $\set{\alpha_{i}}\leftrightarrow\set{\beta'_{j}}$ and also between the outgoing modes $\set{\beta_{j}}\leftrightarrow\set{\alpha'_{j}}$ ($i,j=1,\dots,M_{A}=M_{B}=M$), including the group velocities in the flux normalization, which must then coincide for both modes in the pair; so that: 
\begin{equation} \label{eqTsymBA}
G^{AB}_{s',s}(E)=\frac{e^{2}}{h}\sum_{i,j=1}^{M}\left|\sum_{s_{1},s'_{1}}\pazocal{D}^{1/2}(g)_{s',s'_{1}}\pazocal{D}^{1/2}(g)^{*}_{s,s_{1}}\bra{\psi^{\alpha'_{i}}_{s'_{1}}}\hat{S}\ket{\psi^{\beta'_{j}}_{s_{1}}}\right|^{2}\;\:,\;\:\forall g\in\pazocal{G}\backslash(\pazocal{G}_{A}\cap\pazocal{G}_{B})
\end{equation}
\end{itemize}
Therefore we may treat symmetry operations individually irrespective of the groups $\pazocal{G}_{A}$, $\pazocal{G}_{B}$ they belong to. From equations \eqref{eqTsymAB}, \eqref{eqTsymBA} it follows that there will be a relation between the spin elements of $G^{AB}$ or between those of $G^{AB}$, $G^{BA}$ (respectively) whenever $\pazocal{D}^{1/2}(g)$ has exactly two out of four non-null entries, which occurs for rotations along the spin quantization direction ($b=0$ in equation \eqref{eqD1/2}) and for $\pi-$rotations perpendicular to it ($a=0$ in equation \eqref{eqD1/2}), or equivalently for reflections through planes perpendicular to these axes.

\subsection{Anti-unitary symmetries}
For non-magnetic systems the group of symmetries is a grey point group, which is constructed from the previous $\pazocal{G}$ by allowing for the time-reversal operation $\Theta=\sigma_{y}K$, where $\sigma_{y}$ acts on the spinor components only and $K$ denotes complex conjugation. Noting that $\Theta$ is an anti-unitary operation, that $\Theta\hat{S}\Theta^{-1}=\hat{S}^{\dagger}$ (in analogy with the time evolution operator) if time-reversal is indeed a symmetry of the Hamiltonian ($\sigma_{y}H^{*}\sigma_{y}=H$, in the time-independent case) and that $K$ inverts the propagation direction of the modes within each electrode (thus making a correspondence $\set{\alpha_{i}}\leftrightarrow\set{\alpha'_{i}}$ and $\set{\beta_{j}}\leftrightarrow\set{\beta'_{j}}$): 
\begin{equation} \label{eqTsymTR} \begin{aligned}
G^{AB}_{s',s}(E)=\frac{e^{2}}{h}\sum_{i=1}^{M_{A}}\sum_{j=1}^{M_{B}}\left| 
\bra{\Theta\psi^{\beta_{j}}_{s'}}\Theta\hat{S}\Theta^{-1}\ket{\Theta\psi^{\alpha_{i}}_{s}}^{*}\right|^{2}= \\
\frac{e^{2}}{h}\sum_{i=1}^{M_{A}}\sum_{j=1}^{M_{B}}\left| \sum_{s_{1},s'_{1}}(\sigma^{*}_{y})_{s,s_{1}}(\sigma_{y})_{s',s'_{1}}
\bra{\psi^{\alpha'_{i}}_{s_{1}}}\hat{S}\ket{\psi^{\beta'_{j}}_{s'_{1}}}\right|^{2}=
G^{BA}_{\overline{s},\overline{s}'}(E)
\end{aligned} \end{equation} 
where $\overline{s},\overline{s}'$ denote the opposite spin states of $s,s'$, respectively; and we have introduced the conductance $G^{BA}$ from electrode $B$ to $A$: 
\begin{equation*}
G^{BA}_{s',s}(E)=\frac{e^{2}}{h}\sum_{i=1}^{M_{A}}\sum_{j=1}^{M_{B}}\left| \bra{\psi^{\alpha'_{i}}_{s'}}\hat{S}\ket{\psi^{\beta'_{j}}_{s}}  \right|^{2}
\end{equation*}
Combining equations \eqref{eqTsymBA} and \eqref{eqTsymTR} we obtain the action of the symmetries $g\Theta$ and $\Theta g$ on the conductance for any spatial operation $g$ that permutes the electrodes. The result in both cases is: 
\begin{equation} \label{eqTsymABTR}
G^{AB}_{s',s}(E)=\frac{e^{2}}{h}\sum_{i,j=1}^{M}\left| \sum_{s_{1},s'_{1}}\pazocal{D}^{1/2}(g)_{\overline{s},s'_{1}}\pazocal{D}^{1/2}(g)_{\overline{s}',s_{1}}^{*}\bra{\psi^{\beta_{j}}_{s'_{1}}}\hat{S}\ket{\psi^{\alpha_{i}}_{s_{1}}}\right|^{2}\;\:,\;\:\forall g\in\pazocal{G}\backslash(\pazocal{G}_{A}\cap\pazocal{G}_{B})
\end{equation}
which readily allows us to obtain the restrictions imposed on $G^{AB}$ by these operations. 

Another important consequence of time-reversal symmetry, which was first stated by Zhai and Xu \cite{zhai2005symmetry} and latter expanded by Utsumi \textit{et al}.,\cite{utsumi2020spin} is the fact that all spin-polarization components vanish at any energy at which the final electrode has only one mode, \textit{i.e.}, $M_{B}=1$ in our notation. The proof is as follows. Setting $M_{B}=1$, by the unitarity of the $S$ matrix: 
\begin{equation*}
\sum_{i=1}^{M_{A}}\sum_{s}S^{\beta\alpha_{i}}_{s_{1},s}S^{\beta\alpha_{i}*}_{s_{2},s}+S^{\beta\beta'}_{s_{1},\uparrow}S^{\beta\beta'*}_{s_{2}\uparrow}+S^{\beta\beta'}_{s_{1}\downarrow}S^{\beta\beta'*}_{s_{2}\downarrow}=\delta_{s_{1},s_{2}}
\end{equation*}
and the condition imposed by time-reversal symmetry (note that in this case $\Theta\psi^{\beta}_{s}=i(\delta_{s,\downarrow}-\delta_{s,\uparrow})\psi^{\beta'}_{\overline{s}}$): 
\begin{equation*}
S^{\beta\beta'}_{s_{1},s_{2}}=(2\delta_{s_{1},s_{2}}-1)S^{\beta\beta'}_{\overline{s}_{2},\overline{s}_{1}}\;\;\Rightarrow\;\;S^{\beta\beta'}_{\uparrow\uparrow}=S^{\beta\beta'}_{\downarrow\downarrow}\;\;,\;\;S^{\beta\beta'}_{\uparrow\downarrow}=S^{\beta\beta'}_{\downarrow\uparrow}=0
\end{equation*}
setting $s_{1}=s_{2}=s'$, one obtains $G_{s',\uparrow}+G_{s',\downarrow}=G_{\overline{s}',\uparrow}+G_{\overline{s}',\downarrow}$, which implies that $P=0$ in \eqref{eq1} for every spin quantization axis, hence $\bm{P}=0$ at the corresponding energy. 

An extension of the previous procedure allows to prove Onsager's relation, which states that $G(\bm{M})=G(-\bm{M})$, $\bm{M}$ being the total magnetization of the junction and $G=G_{\uparrow\uparrow}+G_{\uparrow\downarrow}+G_{\downarrow\uparrow}+G_{\downarrow\downarrow}$ the total conductance. Although we do not treat magnetic junctions numerically in the present work, it may be worthwhile to explicitly derive this result both for its practical importance (in magneto-conductance experiments) and to illustrate how this analysis can be readily generalized to account for magnetic elements. For an arbitrary $M_{B}$, employing the unitarity of the $S$ matrix and taking the sum over the $M_{B}$ $\beta$ modes, one obtains:
\begin{equation*}
\sum_{s}\sum_{j=1}^{M_{B}}\left[\sum_{i=1}^{M_{A}}S^{\beta_{j}\alpha_{i}}_{s_{1},s}(\bm{M})S^{\beta_{j}\alpha_{i}*}_{s_{2},s}(\bm{M}) +\sum_{j'=1}^{M'_{B}}S^{\beta_{j}\beta'_{j'}}_{s_{1},s}(\bm{M})S^{\beta_{j}\beta'_{j'}*}_{s_{2},s}(\bm{M})\right]=M_{B}\delta_{s_{1},s_{2}}
\end{equation*}
where we have allowed for a possibly different number of outgoing ($M_{B}$) and incoming ($M'_{B}$) modes in the $B$ electrode (same for $A$ also) due to the potential absence of both inversion and time-reversal symmetry. Application of the time-reversal operation (not symmetry) now yields the more general relation $S^{\beta_{j}\beta'_{j'}}_{s_{1},s_{2}}(\bm{M})=(2\delta_{s_{1},s_{2}}-1)S^{\beta_{j'}\beta'_{j}}_{\overline{s}_{2},\overline{s}_{1}}(-\bm{M})$, since $\Theta\hat{S}(\bm{M})\Theta^{-1}=\hat{S}(-\bm{M})^{\dagger}$ (inherited from $\Theta\hat{H}(\bm{M})\Theta^{-1}=\hat{H}(-\bm{M})$ ). Combining these two expressions, one concludes that $G(\bm{M})=G(-\bm{M})$. Furthermore, in the presence of a single outgoing channel, $\bm{P}(\bm{M})=-\bm{P}(-\bm{M})$; but not otherwise because the vanishing of the spin-flipping terms is necessary. Note that Onsager's relation holds in equilibrium, implying the vanishing of the magneto-conductance $\Delta G(\bm{M})=G(\bm{M})-G(-\bm{M})$ also in presence of SOC.

For completeness we also comment on particle-hole (or charge conjugation) symmetry $\pazocal{C}$, although it is not present in the physical systems of this work. In our basis, this operator is represented as $\pazocal{C}=IK$, where $I$ is the identity operator acting on the spinor components only. The condition on the Hamiltonian $\pazocal{C}H\pazocal{C}^{-1}=-H$ implies $\pazocal{C}\psi^{\alpha_{i}}(E)=\psi^{\alpha'_{i}}(-E)$, where we have explicitly included the energy (measured from a Fermi level obtained from the previous condition on the Hamiltonian) corresponding to each eigenfunction, and omitted the arbitrary phase factor since it is cancelled in our calculations. Therefore: 
\begin{equation} \label{eqTsymC} \begin{aligned}
G^{AB}_{s',s}(E)=\frac{e^{2}}{h}\sum_{i=1}^{M_{A}}\sum_{j=1}^{M_{B}}\left| \bra{\pazocal{C}\psi^{\beta_{j}}_{s'}}\pazocal{C}\hat{S}\pazocal{C}^{-1}\ket{\pazocal{C}\psi^{\alpha_{i}}_{s}}^{*}\right|^{2}=\\
\frac{e^{2}}{h}\sum_{i=1}^{M_{A}}\sum_{j=1}^{M_{B}}\left| \bra{\psi^{\beta'_{j}}_{s'}(-E)}\hat{S}^{\dagger}(-H)\ket{\psi^{\alpha'_{i}}_{s}(-E)} \right|^{2} = G^{BA}_{s,s'}(-E)
\end{aligned} \end{equation}
By equations \eqref{eqTsymTR}, \eqref{eqTsymC}, the combination $\Theta\pazocal{C}$ would yield the condition: 
\begin{equation} \label{eqTsymTRC}
G^{AB}_{s',s}(E)=G^{AB}_{\overline{s}',\overline{s}}(-E)\;\;\Rightarrow\;\; \bm{P}(E)=-\bm{P}(-E)
\end{equation}
However, particle-hole symmetry will rarely be present in realistic systems beyond simplified models. A family of materials which may reasonably exhibit this symmetry are the carbon allotropes.\cite{Chico2015,Guo2016}

This exhausts the set of anti-unitary symmetries. 
Note that for \eqref{eqTsymTR}-\eqref{eqTsymTRC} to be applicable, the whole system needs to have the corresponding symmetry. In particular, placing a non-magnetic molecule or material between magnetic electrodes (or vice-versa) would break the time reversal symmetry of the system.

\subsection{Rotated systems and enantiomers}
Consider now an orthogonal spatial operation which is not a symmetry of the system; $g\notin\pazocal{G}$. Performing the corresponding change of coordinates in equation \eqref{eqT} we obtain a new scattering operator $\hat{S}(\bm{r})'=\hat{S}(g^{-1}\bm{r})\neq\hat{S}(\bm{r})$ and eigenfunctions $\pazocal{D}^{1/2}(g)^{*}\psi(\bm{r})'=\psi(g^{-1}\bm{r})$ (which have the opposite incoming/outgoing nature if and only if $g$ inverts the longitudinal direction) of the transformed system, but the integral is still invariant. In defining $\psi^{\prime}=\overline{g}\psi$ we are keeping the spin projection along the same, untransformed direction. Therefore the spin-polarization of the transformed system could in principle be related to that of the original system along any fixed projection direction, since the spin-resolved conductance satisfy a similar equation to \eqref{eqTg}: 
\begin{equation} \label{eqTnosym}
G^{AB}_{s',s}(E)=\frac{e^{2}}{h}\sum_{i=1}^{M_{A}}\sum_{j=1}^{M_{B}}\left|
\sum_{s_{1},s'_{1}}\pazocal{D}^{1/2}(g)_{s',s'_{1}}\pazocal{D}^{1/2}(g)^{*}_{s,s_{1}}
\int_{\mathbb{R}^{3}}\psi^{\beta_{j}}_{s'_{1}}(\bm{r})^{\prime*}\hat{S}(\bm{r})'\psi^{\alpha_{i}}_{s_{1}}(\bm{r})^{\prime}d^{3}\bm{r}
\right|^{2}\;\;,\;\;\forall g\in O(3)\backslash\pazocal{G}
\end{equation}
If the electrodes are fixed and the molecule is rotated around the longitudinal direction by an operation which is not a symmetry of the molecule (otherwise equation \eqref{eqTsymAB} would apply) or the electrodes (otherwise equation \eqref{eqTnosym} would apply, since it would be equivalent to rotating the whole system), then the integral in equation \eqref{eqT} is not invariant under this transformation and the new components (in the rotated system, but along the original direction) of the spin-polarization are in general unrelated to the old ones, as can be observed in each subfigure of Figure \ref{fig5},\ref{fig8}. The exception being if the rotation is geometrically equivalent to a longitudinal reflection, in which case equation \eqref{eqTenantlong} applies. 

If a chiral molecule is placed between the electrodes, then the system cannot have planes of symmetry (more precisely and assuming that the molecule also lacks inversion symmetry, $\pazocal{G}\subset SO(3)$). It follows from equation  \eqref{eqTnosym} that the substitution, while keeping the electrodes fixed, of a chiral molecule by its enantiomeric partner may in principle yield a spin-polarization that is related to the original in a deterministic way (that is, independent on the symmetry-compatible details of the system). For equation \eqref{eqTnosym} to be applicable, it is necessary that the system of electrodes possesses a symmetry plane, and to connect the molecule in such a way that the whole system is obtained from the original by reflection through that mirror plane, as done in Figure \ref{fig8}. Two cases can then be distinguished, corresponding to the two essentially different ways to connect the enantiomer (both of them related by a $\pi-$rotation of the molecule around an axis perpendicular to the longitudinal direction, which swaps the anchoring to the electrodes), which we label by enant$_{1}$, enant$_{2}$: 
\begin{itemize}
\item The system of electrodes has a longitudinal symmetry plane $\sigma_{l}$, containing the longitudinal direction. Let $t_{\parallel},t_{\perp}$ be transversal directions parallel and perpendicular, respectively, to $\sigma_{l}$. Then following the discussion of equation \eqref{eqTsymAB} and employing equation \eqref{eqTnosym} with $g=\sigma_{l}$, we obtain:
\begin{equation}  \label{eqTenantlong}
\left\{ \begin{aligned} 
& \text{Longitudinal spin projection ($l$): } \;&G^{AB}_{s',s}(E)=G^{AB}_{\overline{s}',\overline{s}}(E)^{\text{enant}_{1}}\;\;\Rightarrow\;\;P_{l}^{\text{\:enant}_{1}}=-P_{l} \\ 
& \text{Transversal, $\parallel$ to $\sigma_{l}$ ($t_{\parallel}$): } \;&G^{AB}_{s',s}(E)=G^{AB}_{\overline{s}',\overline{s}}(E)^{\text{enant}_{1}}\;\;\Rightarrow\;\;P_{t_{\parallel}}^{\text{\:enant}_{1}}=-P_{t_{\parallel}}  \\
& \text{Transversal, $\perp$ to $\sigma_{l}$ ($t_{\perp}$): } \;&G^{AB}_{s',s}(E)=G^{AB}_{s',s}(E)^{\text{enant}_{1}}\;\;\Rightarrow\;\;P_{t_{\perp}}^{\text{\:enant}_{1}}=P_{t_{\perp}}
\end{aligned} \right.
\end{equation}
where $G^{AB}_{s',s}(E)^{\text{enant}_{1}}=\frac{e^{2}}{h}\sum_{i=1}^{M_{A}}\sum_{j=1}^{M_{B}}\left|\int_{\mathbb{R}^{3}}
(\overline{\sigma}_{l}\psi^{\beta_{j}}_{s'_{1}}(\bm{r})^{*})S(\sigma_{l}^{-1}\bm{r})(\overline{\sigma}_{l}\psi^{\alpha_{i}}_{s_{1}}(\bm{r}))
d^{3}\bm{r} \right|^{2}$ is the conductance in the system with the present connection of the enantiomeric partner of the original molecule.

Usually there will be more than one such plane of symmetry for the system of electrodes, each of them determining a position of the enantiomer (all of them related by a rotation of the molecule alone around the longitudinal direction) for which the spin-polarization is related to that of the original molecule.
\item The system of electrodes has a transversal symmetry plane $\sigma_{t}$, perpendicular to the longitudinal direction. This operation is uniquely determined in our configuration and it permutes the electrodes, so that following the discussion of equation \eqref{eqTsymABTR} (in particular, invoking time reversal symmetry) and employing equation \eqref{eqTnosym} with $g=\sigma_{t}$, we obtain for the following spin projections: 
\begin{equation}  \label{eqTenantperp}
\left\{ \begin{aligned} 
& \text{Longitudinal ($l$): } \;&G^{AB}_{s',s}(E)=G^{AB}_{\overline{s},\overline{s}'}(E)^{\text{enant}_{2}} \\ 
& \text{Transversal ($t$): } \;&G^{AB}_{s',s}(E)=G^{AB}_{s,s'}(E)^{\text{enant}_{2}}
\end{aligned} \right.
\end{equation}
where $G^{AB}_{s',s}(E)^{\text{enant}_{2}}$ is defined in analogy with $G^{AB}_{s',s}(E)^{\text{enant}_{1}}$, only changing $\sigma_{l}$ by $\sigma_{t}$. There is thus no conclusive relation between $\bm{P}$ and $\bm{P}^{\text{\:enant}_{2}}$.
\end{itemize}
There is then one connection of the enantiomer, enant$_{2}$, that may yield a spin-polarization which is unrelated, neither equal nor opposite, to that of the original molecule. This is no longer true if (and only if, in our configuration) the system of electrodes has both symmetries $\sigma_{l},\sigma_{t}$ and the original system, including the chiral molecule, has a transversal $C_{2,t_{\parallel}}$ rotation symmetry whose axis is parallel to the plane $\sigma_{l}$, so that $\sigma_{t}=\sigma_{l}C_{2,t_{\parallel}}$. In this case, successively applying equations \eqref{eqTsymABTR}, \eqref{eqTenantlong} in the right hand side, and equations \eqref{eqTsymTR}, \eqref{eqTenantperp} in the left hand side ($\sigma_{t}\Theta=\sigma_{l}\Theta C_{2,t_{\parallel}}$), it follows that: 
\begin{equation}  \label{eqTenantC2}
\left\{ \begin{aligned} 
& \text{Longitudinal ($l$):} &G^{AB}_{\overline{s},\overline{s}'}(E)^{\text{enant}_{2}}=G^{AB}_{\overline{s},\overline{s}'}(E)^{\text{enant}_{1}}\;\;\Rightarrow\;\; P_{l}^{\text{enant}_{2}}=P_{l}^{\text{enant}_{1}}=-P_{l}   \\ 
& \text{Transversal, $\parallel$ to $\sigma_{l}$ and $C_{2}$ ($t_{\parallel}$):} &G^{AB}_{s,s'}(E)^{\text{enant}_{2}}=G^{AB}_{s,s'}(E)^{\text{enant}_{1}}\;\;\Rightarrow\;\; P_{t_{\parallel}}^{\text{enant}_{2}}=P_{t_{\parallel}}^{\text{enant}_{1}}=-P_{t_{\parallel}}   \\ 
& \text{Transversal, $\perp$ to $\sigma_{l}$ and $C_{2}$ ($t_{\perp}$):}
&G^{AB}_{s,s'}(E)^{\text{enant}_{2}}=G^{AB}_{s,s'}(E)^{\text{enant}_{1}}\;\;\Rightarrow\;\; P_{t_{\perp}}^{\text{enant}_{2}}=P_{t_{\perp}}^{\text{enant}_{1}}=P_{t_{\perp}}
\end{aligned} \right.
\end{equation}
which effectively makes the polarization equal for the two ways of connecting the enantiomer. This was to be expected, since both systems are identical due to the $C_{2,t_{\parallel}}$ symmetry. 

\section{Author contributions}
W.D and M.A.G.B contributed equally to this work, the former carrying out the DFT calculations and a basic theoretical analysis and the latter the full theoretical symmetry analysis. Both also contributed to the writing of the manuscript. L.Z. brought to our attention the problem and carried out preliminary calculations. E.B.L. provided computational support, aided with guidance on presentation of numerical results and figures, discussions on the theoretical implications and contributed to the editing of the manuscript. S.P. provided the original code to implement SOC in ANT.Gaussian. C.S. aided with calculations and discussions on the experimental implications. J.J.P. had the original idea, coordinated the team, and contributed to the writing. 

The authors declare no competing financial interests.


\begin{acknowledgement}
J.J.P. and M.A.G.B acknowledge financial support from Spanish MICIN through Grant No. PID2019-109539GB-C43/AEI/ 10.13039/501100011033, the María de Maeztu Program for Units of Excellence in R\&D (Grant No. CEX2018-000805-M), the Comunidad Autónoma de Madrid through the Nanomag COST-CM Program (Grant No. S2018/NMT-4321), the Generalitat Valenciana through Programa Prometeo/2021/017, the Centro de Computación Científica of the Universidad Autónoma de Madrid, and the computer resources of the Red Española de Supercomputación. L.A.Z. thanks financial support from MCIN/AEI/ 10.13039/501100011033 (grant PID2021-125604NB-I00) and from the Universidad Autónoma de Madrid/Comunidad de Madrid (grant No. SI3/PJI/2021-00191). C.S. thanks the financial support from the Generalitat Valenciana through CIDEXG/2022/45, CDEIGENT/2018/028 and PROMETEO/2021/017. S.P. acknowledges the grant from Erasmus+ 2018 programme (Collaboration between University of Tehran, Iran and Autonomous University of Madrid, Spain). W. D. thanks A. E. Botha for sharing python scripts used to make some of the figures in this work. The computational results contained in this work would also not have been possible without access to the High Performance Computing (HPC) facility at Unisa and the supercomputing facility in the Department of Applied Physics at the University of Alicante.
\end{acknowledgement}

\section{Supplementary information}

The \textit{Supplementary Information} contains further details of the numerical calculations performed in this work; the mathematical verification of the vanishing of the transversal spin-polarization under any longitudinal rotation symmetry; some comments on the carbon helices employed in Figure \ref{fig5}; some comments on how the results of reference \cite{Guo2016} fit our analysis with achiral molecules; the full polarization-energy curves corresponding to Figure \ref{fig2}; a spin-polarization plot along directions other than the longitudinal and transversal; and a numerical example with random defects.



\bibliography{Symmetry}

\end{document}









\section{SOC-corrected DFT transport calculations}

Pakdel et al.\cite{Pakdel2018} developed a convenient method to correct the band structure of solid materials for spin-orbit coupling (SOC). The core feature of this approach is that quality SOC-corrected bands can be obtained if an existing (or optimized) Gaussian-type orbital (GTO) basis set already reproduces the non-SOC band structure of a material to a good degree, so that SOC can be treated perturbatively. Once that condition is met, adding SOC as a correction in a post self-consistent-field (SCF) step gives good bands in comparison to reference methods which are known to generate very high quality SOC bands by fully-relativistic self-consistent approaches, e.g., \texttt{VASP}\cite{Kresse1996}, \texttt{Wien2k}\cite{Blaha2001}, \texttt{Quantum Espresso}\cite{giannozzi2009} or \texttt{OpenMX}\cite{Ozaki2005,Ozaki2017}. The latter is used as the reference method in this work.

Here we very briefly outline the method developed by Pakdel et al. \cite{Pakdel2018} in addition to a few recent improvements we have made to it. We use the standard atomic SOC matrix to express the lowest-order Dirac-Kohn-Sham Hamiltonian as follows:
\begin{equation}
		\xi \left (r \right){\rm \bf L}\cdot{\rm \bf S}=
		\left[ \xi_{\rm ij} \left< l_{\rm i};m_{l_{\rm i}};s | {\rm \bf L}\cdot{\rm \bf S} | l_{\rm j};m_{l_{\rm j}};s' \right> \right],
		\label{eqS1}
\end{equation}
where:
\begin{equation}
		\xi_{\rm ij} = \frac{e^{2}}{2m_e c^{2}}\int_{0}^{\infty}\frac{1}{r}\frac{{\rm d}V_{\rm eff}\left (r \right)}{{\rm d}r}R_{\rm i}\left (r \right)R_{\rm j}^{*}\left (r \right)r^{2}{\rm d}r.
		\label{eqS2}
\end{equation}

The above simplification relies upon the orthogonality of the radial and angular parts of wave functions in atomic-orbital based DFT, e.g., the Gaussian-type orbitals (GTOs) used by  \texttt{CRYSTAL14} \cite{CRYSTAL14} or \texttt{GAUSSIAN09} \cite{GAUSSIAN09}.

The effective nuclear potential in Eq. ~(\ref{eqS2}) is defined as $V_{\rm eff}\left (r \right)=-\frac{Z}{r}$ ~\cite{Pakdel2018}, with \textit{Z} the atomic number. The radial wave functions $R_{\rm i}\left (r \right)$
are (un)contracted GTOs (CGTOs). The fact that SOC is an intra-atomic phenomenon means that only CGTOs on the same atom and of the same shell type ($L=$ 1, 2 or 3, referring to $P$, $D$, or $F$ shells, respectively) contribute to the integral in Eq. ~(\ref{eqS2})  \cite{Pakdel2018}. Note that, due to the lack of adequate nodal structure near the nucleus in pseudopotentials, CGTO basis sets with pseudopotentials require a multiplicative correction to $\xi_{\rm ij}$ if $V_{\rm eff}\left (r \right)$ is to reproduce the correct effective charge of a given atom.

We describe two minor modifications that improve upon the method described in Ref. \cite{Pakdel2018}. First, we have replaced the original effective nuclear potential by a Yukawa screening potential:

\begin{equation}
{V}_{\rm eff} \left (r \right) = 
\left\{
        \begin{array}{ll} \frac{- \left (Z - 1 \right ) \left [\exp \left (- \frac{ln{Z}}{{r}_{c}} {r} \right ) + 1 \right ]}{r} & {r} \leq {r}_{c} \\
        \frac{-1}{r} &  {r} > {r}_{c} 
        \end{array}
    \right.
\label{eqS3}
\end{equation}

where $r_c$ is a cutoff, typically the size of an atomic radius ($\sim 2.5-3.0$ a.u.).

\begin{figure*}[t!]
\centering 
\includegraphics[width=0.47\textwidth]{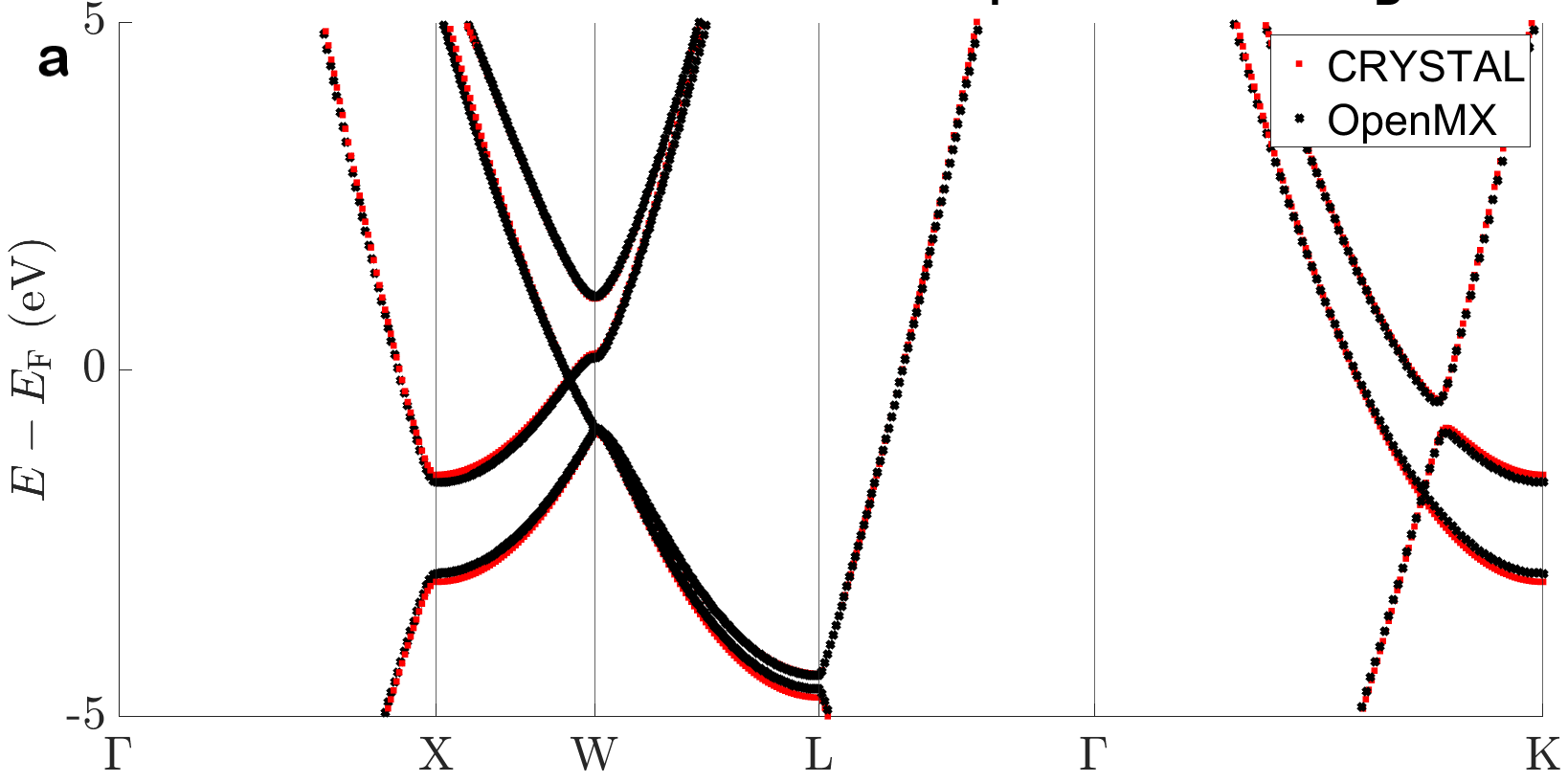}
\includegraphics[width=0.47\textwidth]{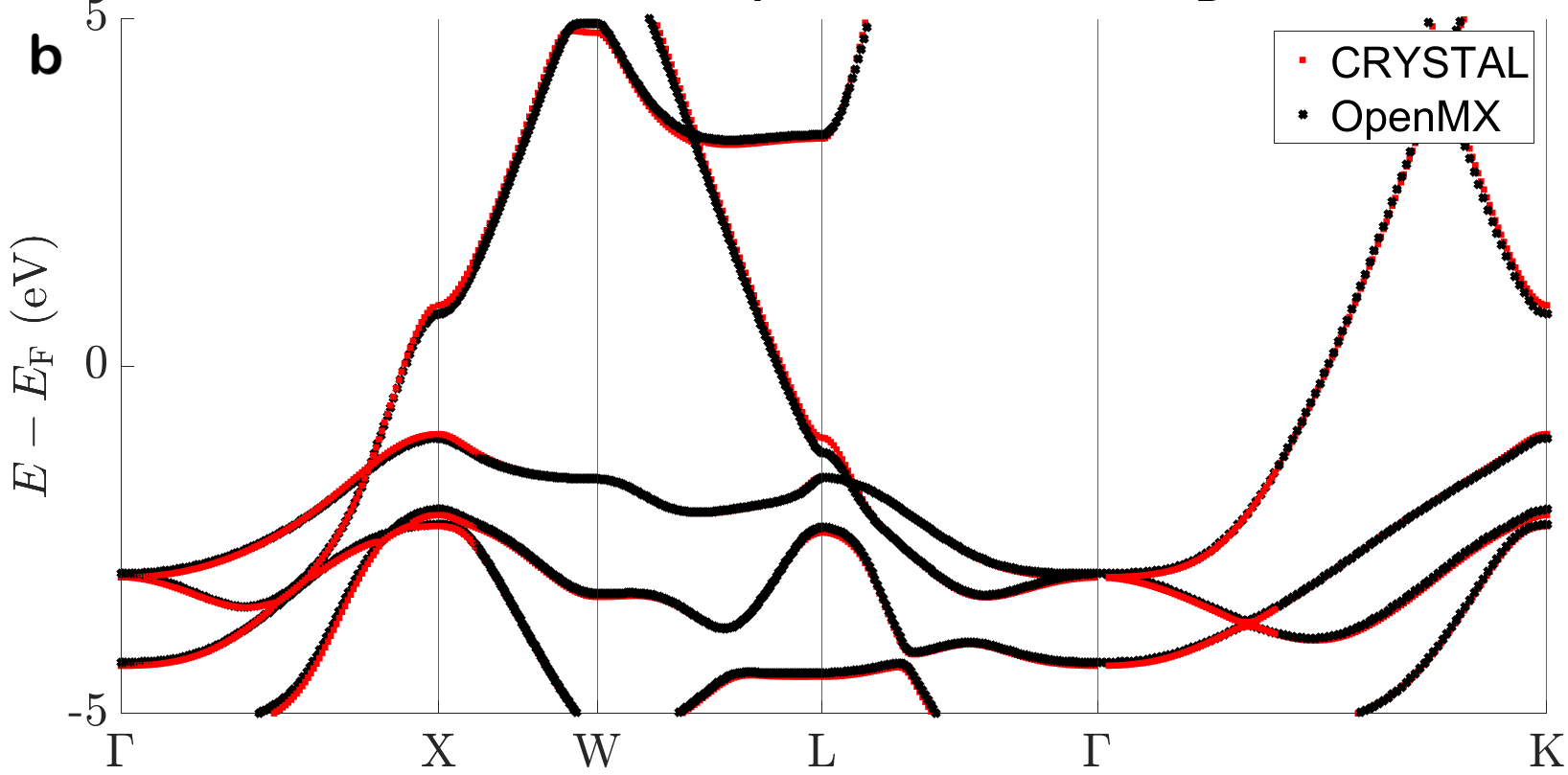}
\includegraphics[width=0.47\textwidth]{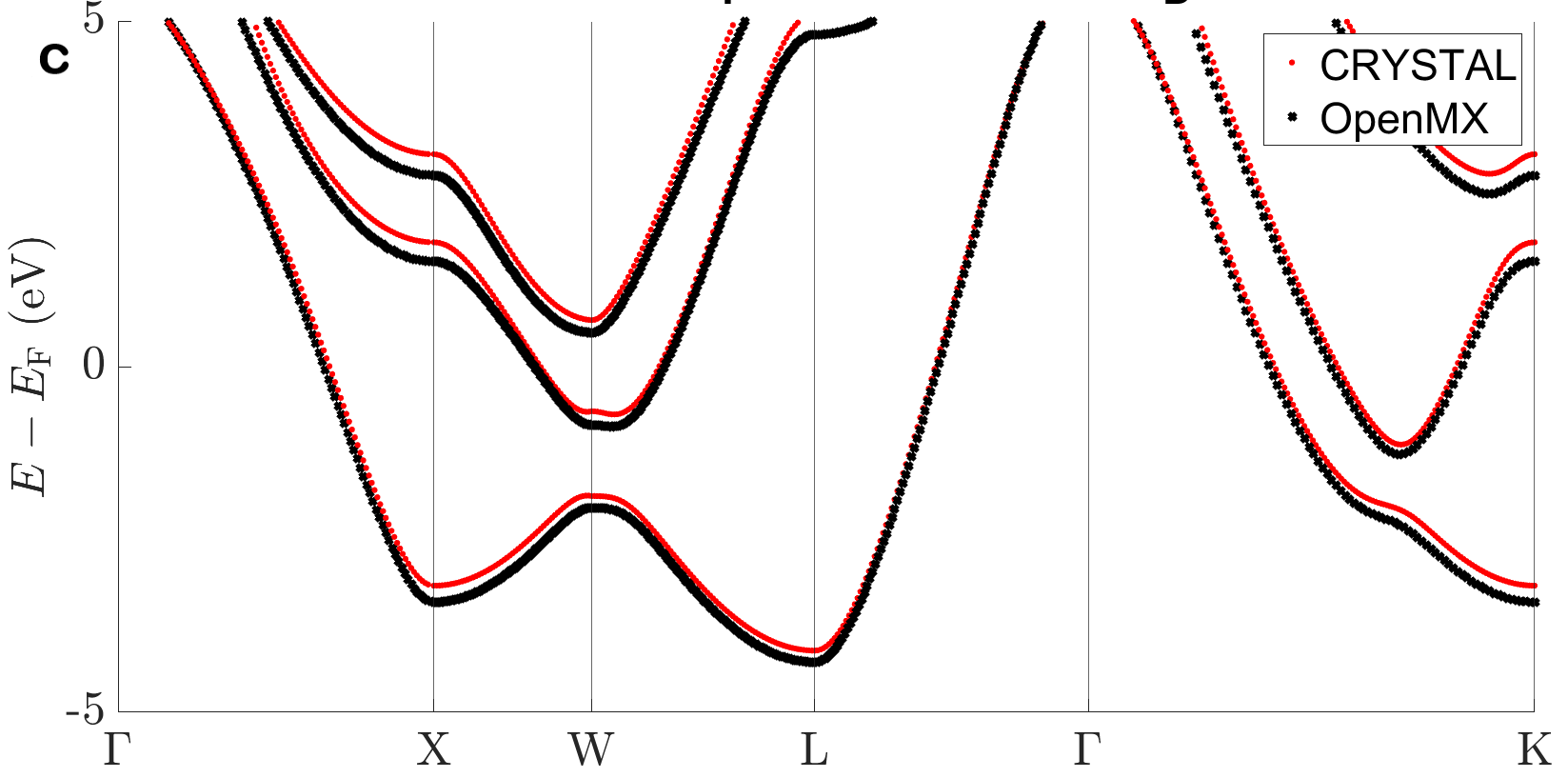}
\includegraphics[width=0.47\textwidth]{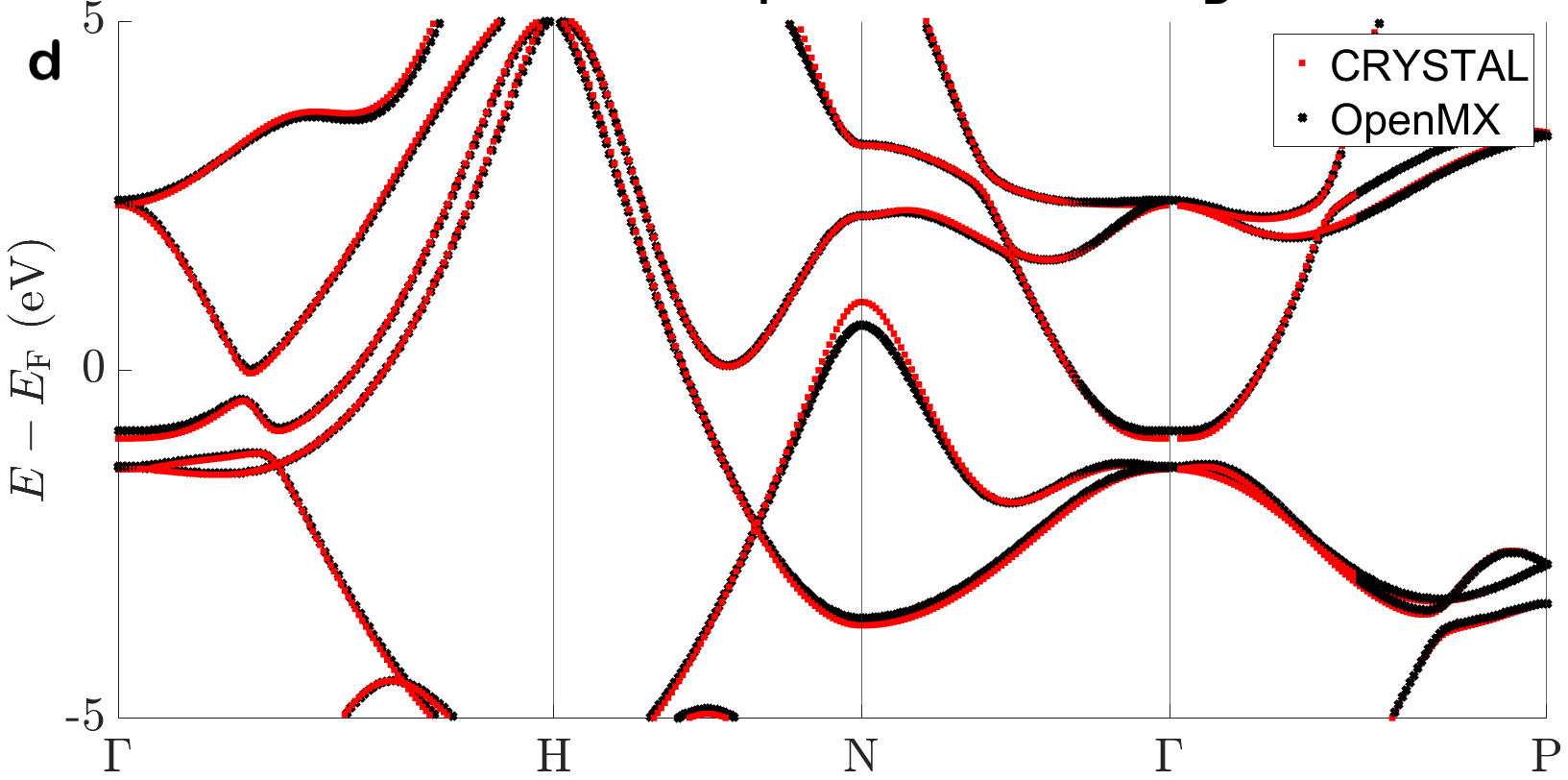}
\caption{SOC-corrected bands obtained from previous non-SOC calculations in \texttt{CRYSTAL14} using our method (solid red markers) and the reference method \texttt{OpenMX} (black crosses) for FCC \textbf{a} Al, \textbf{b} Au, \textbf{c} Pb and \textbf{d} BCC W. 
\label{figS1}}
\end{figure*}

\begin{figure*}[t!]
\centering 
\includegraphics[width=0.47\textwidth]{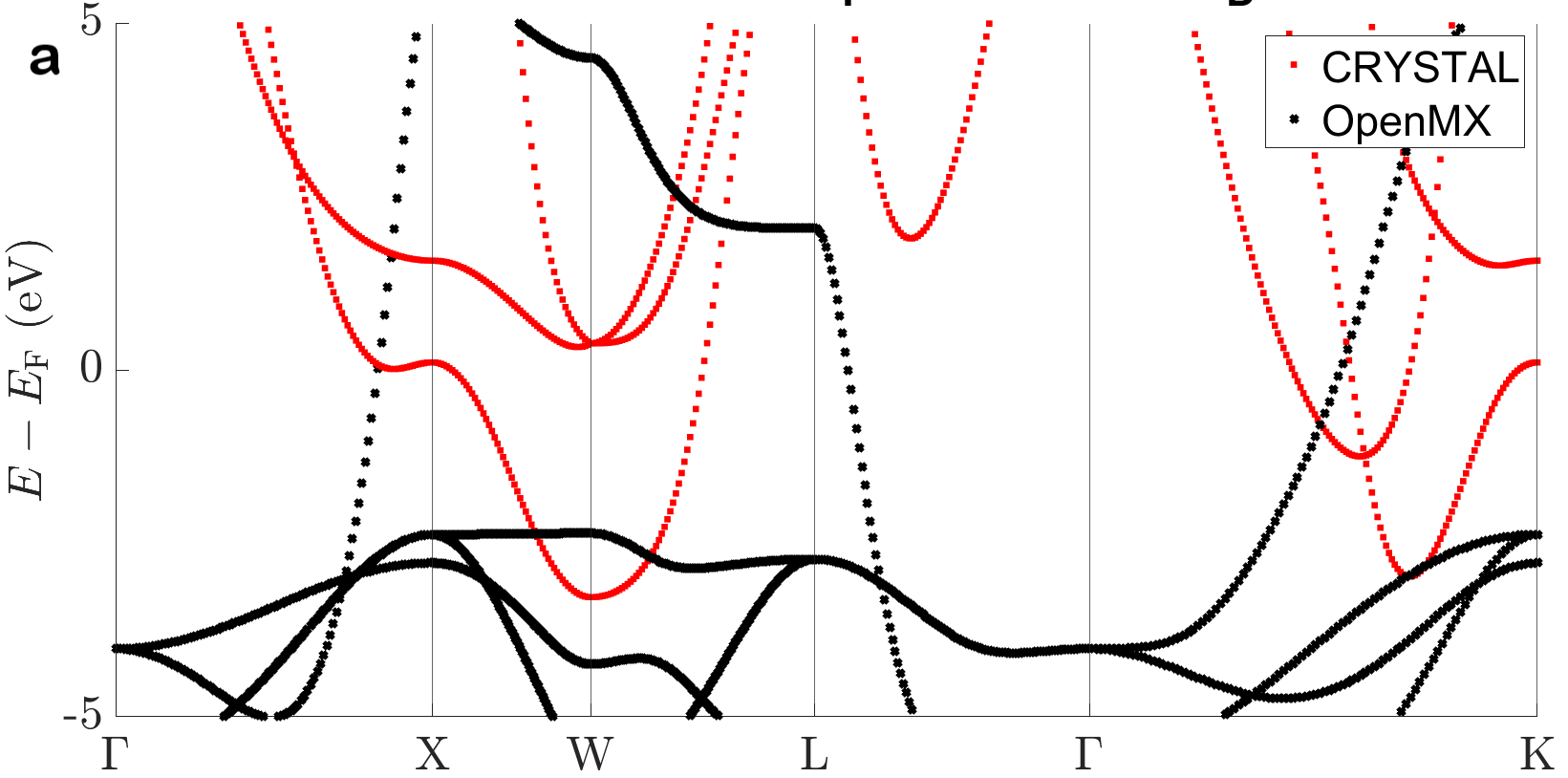}
\includegraphics[width=0.47\textwidth]{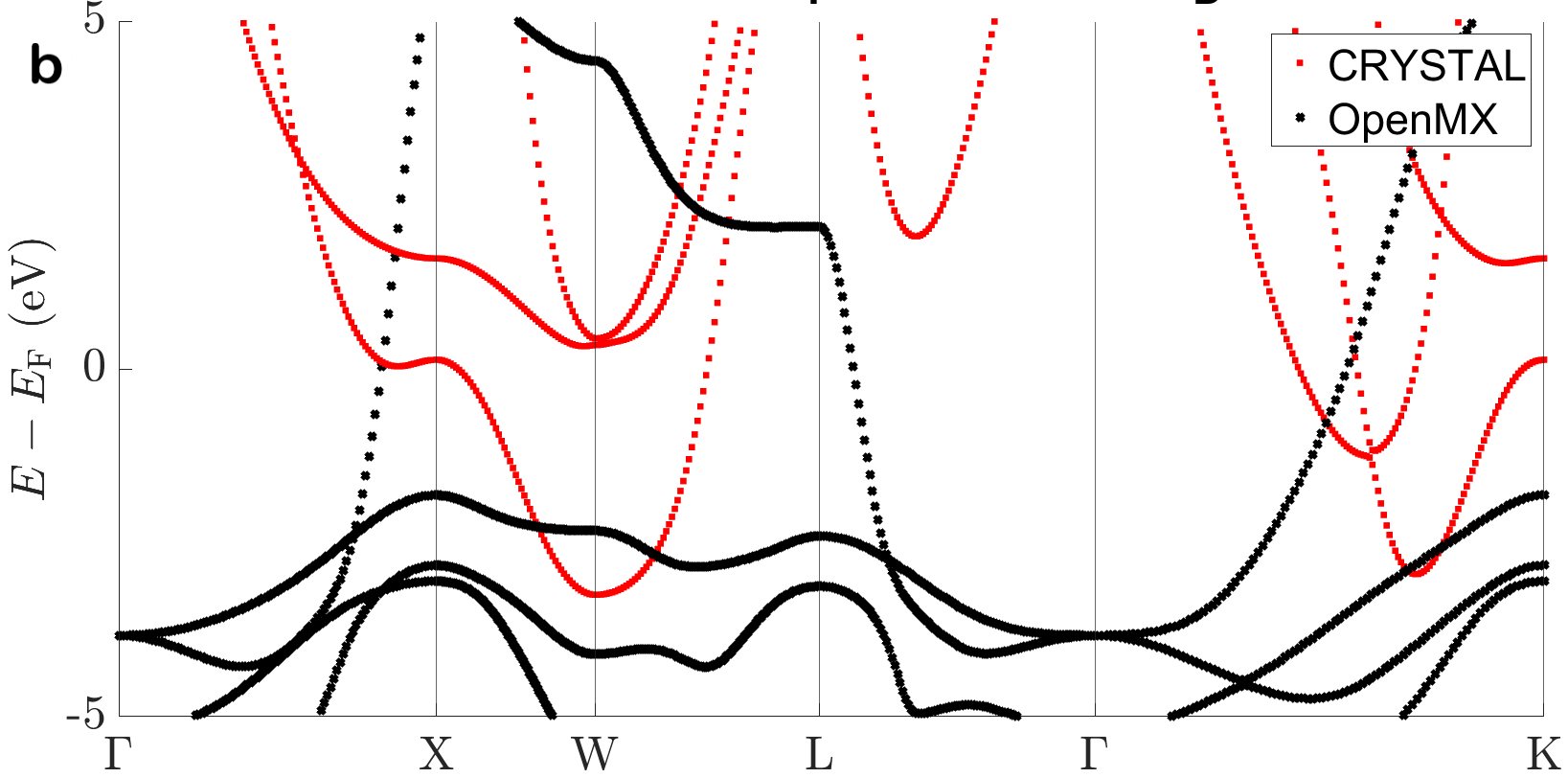}
\caption{\textbf{a} Non-SOC and \textbf{b} SOC-corrected bands obtained in \texttt{CRYSTAL14} calculations using the unoptimized \texttt{x2c-SVPall-2c} all-electron basis set  for FCC Au in our method (solid red markers) and a large pseudo-potential basis set in the reference method \texttt{OpenMX} (black crosses). The fit is very poor and the addition of SOC does not seem to affect the electronic structure about the Fermi level in our method while the reference method clearly exhibits the lifting of degeneracies expected for a strong SOC metal such as Au and reproduced by our method too in Figure \ref{figS1}. The reference bands in \textbf{b} differ slightly from those of Figure \ref{figS1}b because a larger lattice parameter was required to converge the calculation using the all-electron basis set. 
\label{figS2}}
\end{figure*}

Secondly, we do not use a single global multiplicative factor to account for the lack of nodal structure when basis sets with pseudopotentials are used. Instead, we have implemented a multiplicative factor for each shell type ($P$, $D$, or $F$) since it is known that the radial SOC coefficients of different angular momentum shells are multiples of each other \cite{Barreteau2016}.

The ultimate goal of fitting SOC corrected bands to a reference method is to choose high quality basis sets that can be used in DFT quantum transport calculations where transition metal elements, sometimes with strong SOC, are used as the electrodes. We therefore need such GTO basis sets for metals like Al, Au, Pb, and W, which are used as electrode and/or substrate materials in scanning tunneling microscopy (STM) or mechanically controllable break junction (MCBJ) experiments\cite{Agrait2003}. In this work, the basis sets used in our electronic transport calculations in the main manuscript have also been employed in band structure calculations in \texttt{CRYSTAL14} to make Figure \ref{figS1}, which shows an empirical fit of SOC-corrected bands to \texttt{OpenMX} band structure calculations for Al, Au, Pb and W. In all electronic structure and transport calculations we use the \texttt{PBE} exchange-correlation functional \cite{Perdew1996} and strict convergence criteria of at most $10^{-7}$ Ha. For an all-electron basis set, the multiplicative factors ($P$, $D$, or $F$) per shell should all be set equal to a value of $1$ as in the calculation for FCC Al shown in Figure \ref{figS1}a where we used a high quality all-electron basis set which has been optimized for the metallic phase of Al \cite{Ruiz2003}. For Au and W we used basis sets from Ref. \cite{Laun2021} with $P$, $D$, and $F$ SOC factors of $260$, $40$, and $10$ for Au, and $210$, $30$, and $10$ for W, whilst for Pb we used the basis set from \cite{Zagorac2011} with $P$, $D$, and $F$ SOC factors of $280$, $30$, and $10$. All basis sets required optimization of their outer diffuse GTOs in order to get the best empirical agreement with the reference method. We used the \texttt{billy} optimizer developed by Mike Towler for this purpose \cite{billy}. 

The implementation of SOC in our DFT electronic transport method is discussed in detail in other works\cite{DednamThesis2019,DednamSAIP2022}. We merely mention here that the code Atomistic NanoTransport~(\texttt{ANT.Gaussian})\cite{palacios2001fullerene,palacios2002transport}, interfacing with \texttt{Gaussian09}~\cite{GAUSSIAN09}, has been employed in this work. Out of the box, it allows performing scalar-relativistic, spin-unrestricted calculations. The SOC correction as implemented in \texttt{ANT.Gaussian} is freely available online\cite{ANTG}. 

It should be mentioned that the small polarizations obtained across a wide energy range about the Fermi level in Ref. \cite{Zollner2020b} in their Au--molecule--Au calculations using an all-electron basis set such as \texttt{x2c-SVPall-2c}, if not optimized for the solid phase of the pure metal, may be the result of SOC not having any real effect on the electronic structure of the metal nanofragments that are used as electrodes.  The bands without and with SOC in Figure \ref{figS2}a, b respectively, which were obtained from \texttt{CRYSTAL14} calculations using the unoptimized all-electron basis set  \texttt{x2c-SVPall-2c}.

\section{Transversal polarization and $n-$fold rotation symmetry}
Here we prove that the spin-polarization is vanishing along any transversal direction if there exists any rotation symmetry $C_{n,l}\in\pazocal{G}$ with $n\geq2$. 

Fixing $\hat{\bm{z}}$ as an arbitrary transversal direction, we may in principle assign any direction in the $xy-$plane to the longitudinal one, hence we write $\hat{\bm{e}}=(\sin(\varphi),\cos(\varphi),0)$, for any $\varphi\in[0,2\pi)$. Then by equation (4) of the main text:
\begin{equation*}
\pazocal{D}^{1/2}(C_{n}^{m})=\begin{pmatrix}\cos(\pi m/n)&\sin(\pi m/n)e^{i\varphi}\\-\sin(\pi m/n)e^{-i\varphi}&\cos(\pi m/n) \end{pmatrix}\;\;,\;\;\forall m=1,\dots,n-1
\end{equation*} 
and by equation (5);
\begin{equation*} \footnotesize \begin{aligned}
& G_{\uparrow\uparrow}=\frac{e^{2}}{h}\sum_{i=1}^{M_{A}}\sum_{j=1}^{M_{B}}\left|\cos(\pi m/n)^{2}S_{\uparrow\uparrow}^{\beta_{j}\alpha_{i}}+\sin(\pi m/n)^{2}S_{\downarrow\downarrow}^{\beta_{j}\alpha_{i}}+\cos(\pi m/n)\sin(\pi m/n)\left[e^{i\varphi}S_{\downarrow\uparrow}^{\beta_{j}\alpha_{i}}+e^{-i\varphi}S_{\uparrow\downarrow}^{\beta_{j}\alpha_{i}}\right]\right|^{2}\:,  \\
& G_{\uparrow\downarrow}=\frac{e^{2}}{h}\sum_{i=1}^{M_{A}}\sum_{j=1}^{M_{B}}\left|\cos(\pi m/n)\sin(\pi m/n)e^{i\varphi}\left[-S_{\uparrow\uparrow}^{\beta_{j}\alpha_{i}}+S_{\downarrow\downarrow}^{\beta_{j}\alpha_{i}}\right]-\sin(\pi m/n)^{2}e^{i2\varphi}S_{\downarrow\uparrow}^{\beta_{j}\alpha_{i}}+\cos(\pi m/n)^{2}S_{\uparrow\downarrow}^{\beta_{j}\alpha_{i}}\right|^{2}\:,  \\
& G_{\downarrow\uparrow}=\frac{e^{2}}{h}\sum_{i=1}^{M_{A}}\sum_{j=1}^{M_{B}}\left|\cos(\pi m/n)\sin(\pi m/n)e^{-i\varphi}\left[-S_{\uparrow\uparrow}^{\beta_{j}\alpha_{i}}+S_{\downarrow\downarrow}^{\beta_{j}\alpha_{i}}\right]+\cos(\pi m/n)^{2}S_{\downarrow\uparrow}^{\beta_{j}\alpha_{i}}-\sin(\pi m/n)^{2}e^{-i2\varphi}S_{\uparrow\downarrow}^{\beta_{j}\alpha_{i}})\right|^{2}\:,  \\
& G_{\downarrow\downarrow}=\frac{e^{2}}{h}\sum_{i=1}^{M_{A}}\sum_{j=1}^{M_{B}}\left|\sin(\pi m/n)^{2}S_{\uparrow\uparrow}^{\beta_{j}\alpha_{i}}+\cos(\pi m/n)^{2}S_{\downarrow\downarrow}^{\beta_{j}\alpha_{i}}-\cos(\pi m/n)\sin(\pi m/n)\left[e^{i\varphi}S_{\downarrow\uparrow}^{\beta_{j}\alpha_{i}}+e^{-i\varphi}S_{\uparrow\downarrow}^{\beta_{j}\alpha_{i}}\right]\right|^{2}  
\end{aligned}\end{equation*} 
Inserting these in equation (1) we obtain, after some algebra:
\begin{equation} \label{eqP0prev} \begin{aligned}
P_{z}=\cot(\pi m/n)\frac{e^{2}}{h}\sum_{i=1}^{M_{A}}\sum_{j=1}^{M_{B}}\left[e^{i\varphi}\left(S_{\uparrow\uparrow}^{\beta_{j}\alpha_{i}*}S_{\downarrow\uparrow}^{\beta_{j}\alpha_{i}}+S_{\uparrow\downarrow}^{\beta_{j}\alpha_{i}*}S_{\downarrow\downarrow}^{\beta_{j}\alpha_{i}}\right)+\text{c.c.}\right]=\\
\cot(\pi m/n)\operatorname{Re}\left[e^{i\varphi}(P_{x}+iP_{y})\right]
\;\;,\;\;\forall m=1,\dots,n-1
\end{aligned}\end{equation}
where we have introduced the perpendicular components of the spin-polarization ($\hat{\bm{z}}$ being the fixed quantization axis, along the chosen transversal direction in this case). Since $\cot(x)=-\cot(\pi-x)$, applying equation  \eqref{eqP0prev} for $m=1$ and $m=n-1$ (that is, for $C_{n,l}$ and its inverse $C_{n,l}^{n-1}$) we complete the proof.

This result was straightforward to obtain for $n=2$ since $\pazocal{D}^{1/2}(C_{2,l})$ has two null entries. In that case $C_{2,l}$ is its own inverse and $\cot(\pi/2)=0$ readily. For $n\geq3$, however, it must be $\operatorname{Re}[e^{i\varphi}(P_{x}+iP_{y})]=0$; which is consistent with the fact that in this case the polarization is a vector along the direction of transport, noting that $\varphi=0$ ($\pi/2$) if $\hat{\bm{y}}$ ($\hat{\bm{x}}$, respectively) corresponds to the longitudinal direction.

\section{On the carbon helix molecules}
In the calculations on chiral molecules in Figure 4 of the main manuscript, we have used the molecule in the supplementary material of Zöllner et al.\cite{Zollner2020b} and modified it in two ways. Technically, the original molecule lacks any unitary symmetry, but, in a way, it is not far from having a $2-$fold transversal rotation symmetry $C_{2,t}$ (perpendicular to the axis of the helix). In Figure 4a-f, the hydrogen atoms and two carbon atoms from one end have been removed in order to make the molecule ``more asymmetric''. In contrast, the molecule exhibits the $C_{2,t}$ symmetry in Figure 4g-i due to the removal of the hydrogen atoms from both ends. All inset structures in Figures 3 and 4 of the main text have been created with the Open Visualization Tool \texttt{Ovito} \cite{ovito}.

\section{Comment on previous results for achiral molecules}
Here we briefly comment the results of Guo \textit{et al}. \cite{Guo2016} in the context of the present text's formalism. In that reference, a finite longitudinal spin-polarization was obtained by employing a tight-binding model (where symmetries are implicitly present in the relationships between the numerical parameters) for an achiral hexagonal nanotube between chain electrodes, and displacing a single electrode so that longitudinal symmetry planes are broken. Note the results of Figures 2 and 4 in reference \cite{Guo2016} are perfectly consistent with our symmetry results for junctions with achiral molecules. In particular, for their $j=1,4$ connections, $\pazocal{G}=C_{2v},C_{2h}$ respectively, albeit with the rotation axis directed transversally, (so that one symmetry plane is longitudinal in both cases), they find $P_{l}=0$. Otherwise, mirror planes determine the reversal of the longitudinal polarization when changing the connection in that system according to longitudinal planes, which was also noted by the authors.

\section{Numerical calculations not included in the main text}

\begin{figure}[H]
\centering 
\includegraphics[width=0.45\textwidth]{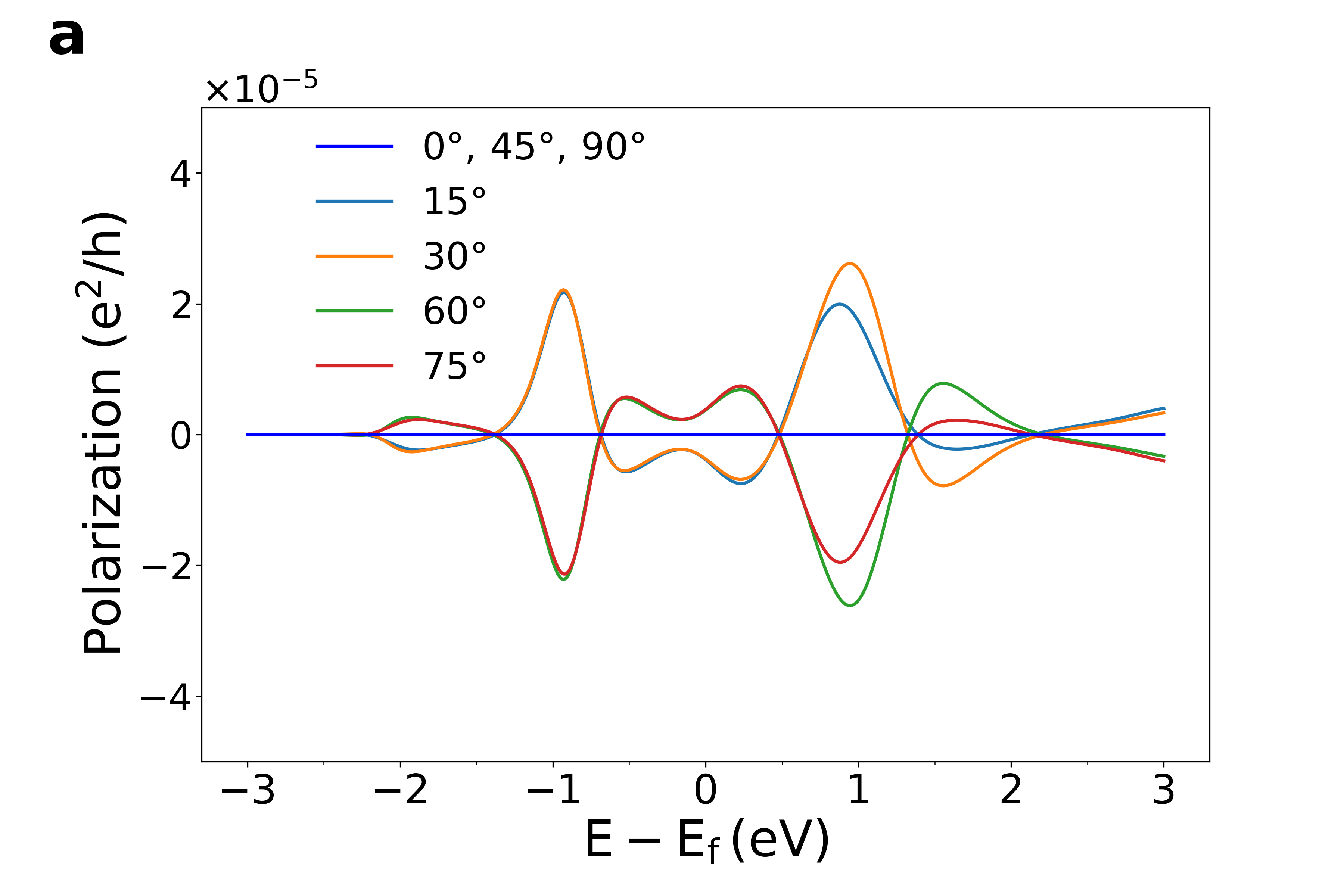}
\includegraphics[width=0.45\textwidth]{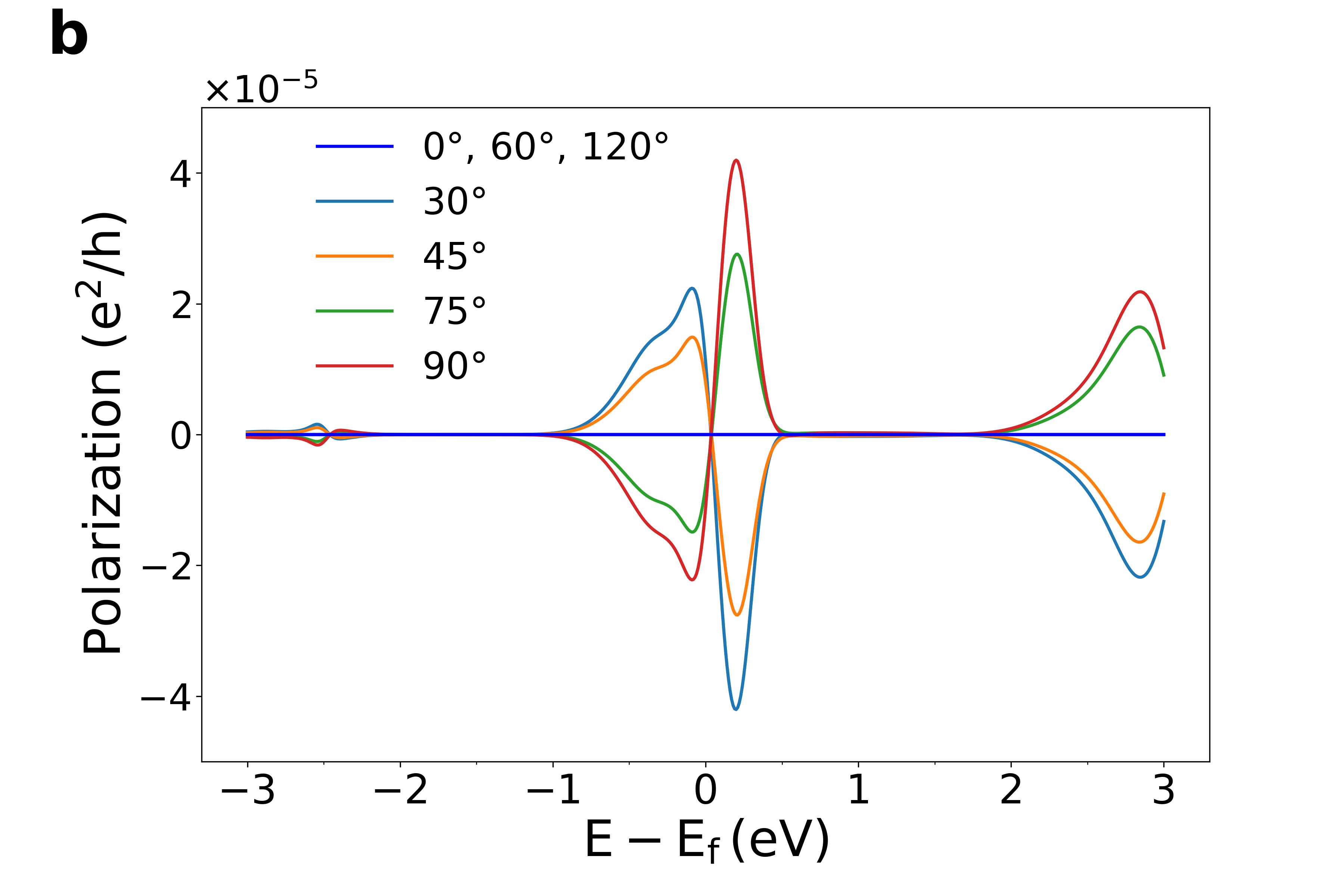}
\includegraphics[width=0.45\textwidth]{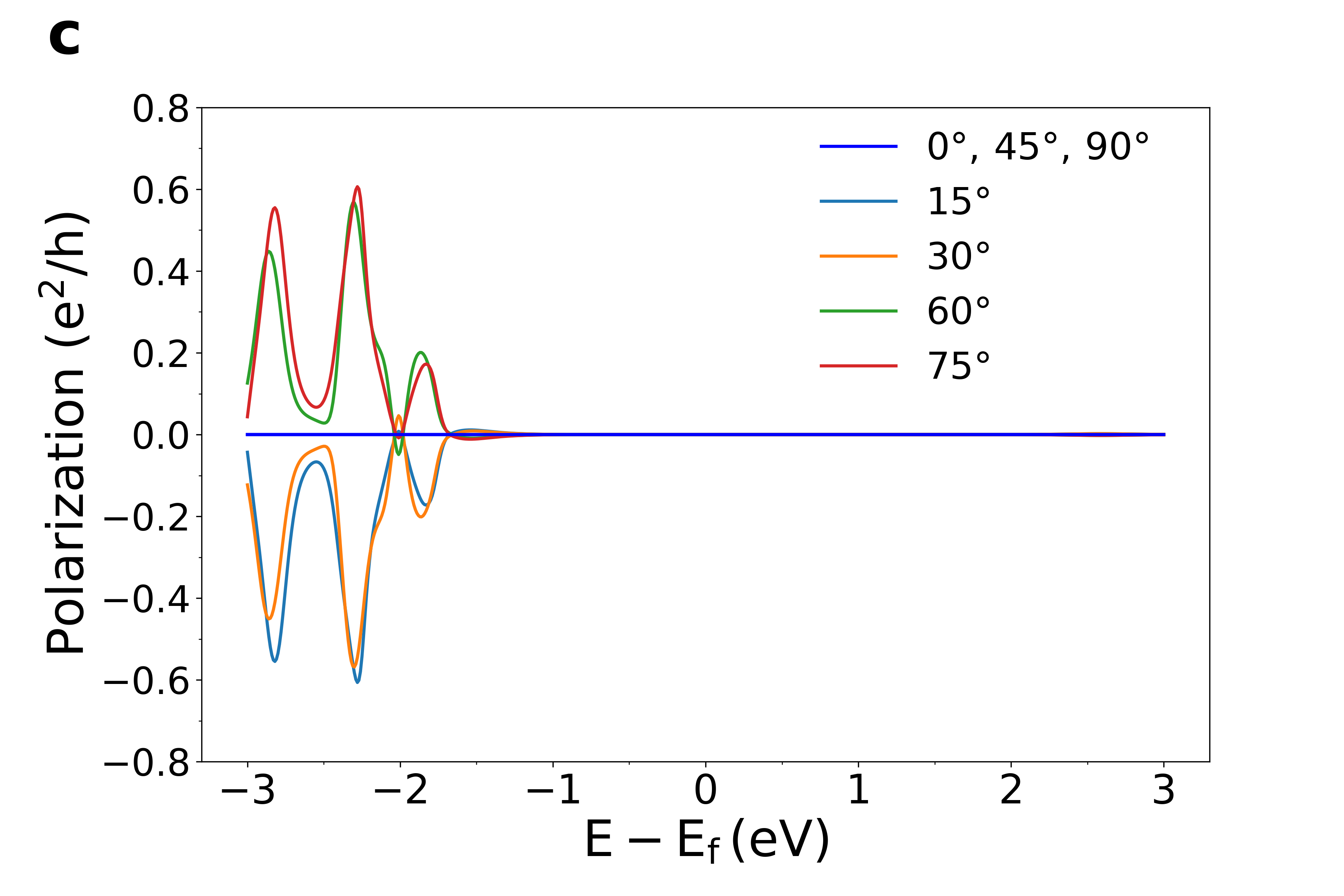}
\includegraphics[width=0.45\textwidth]{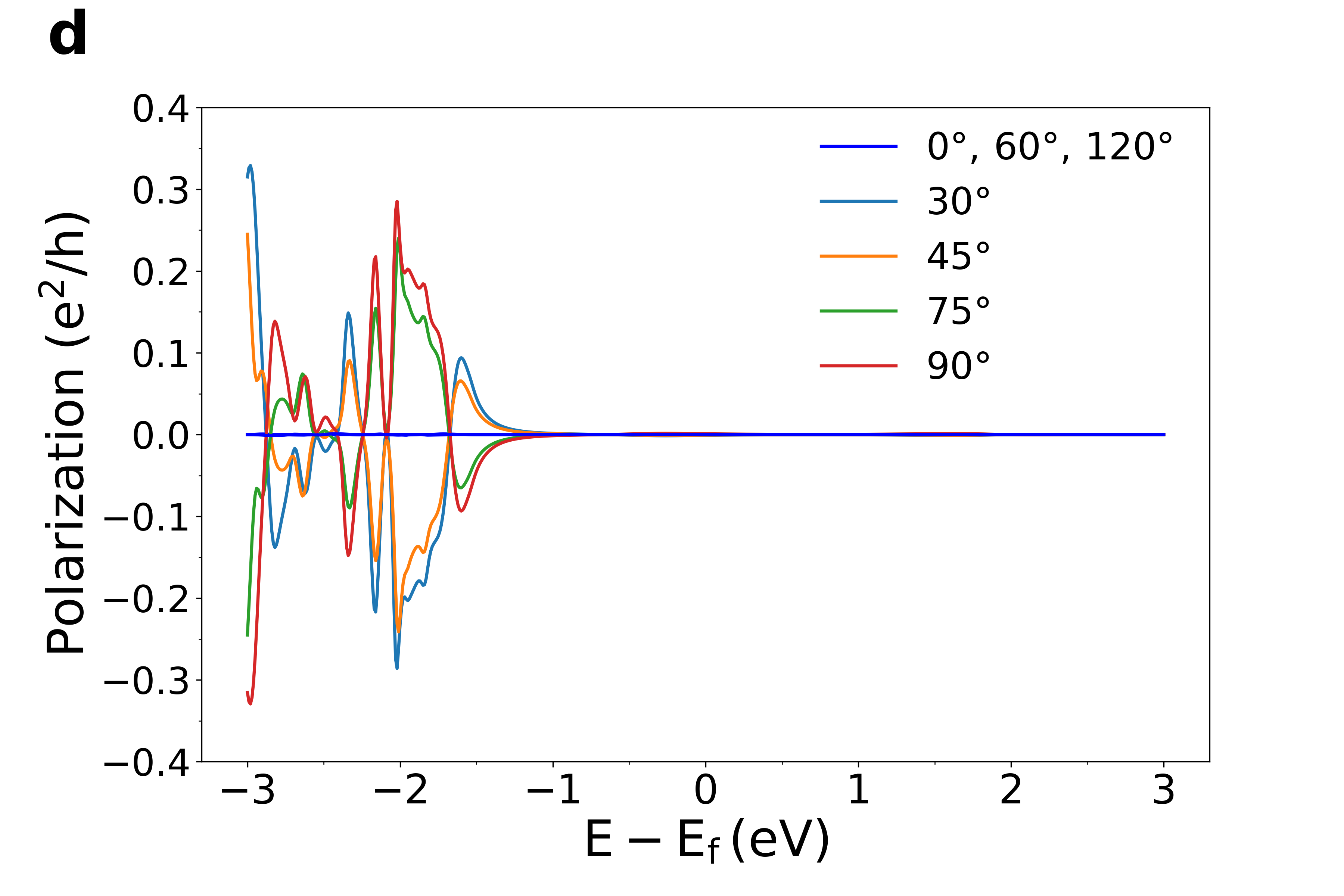}
\includegraphics[width=0.45\textwidth]{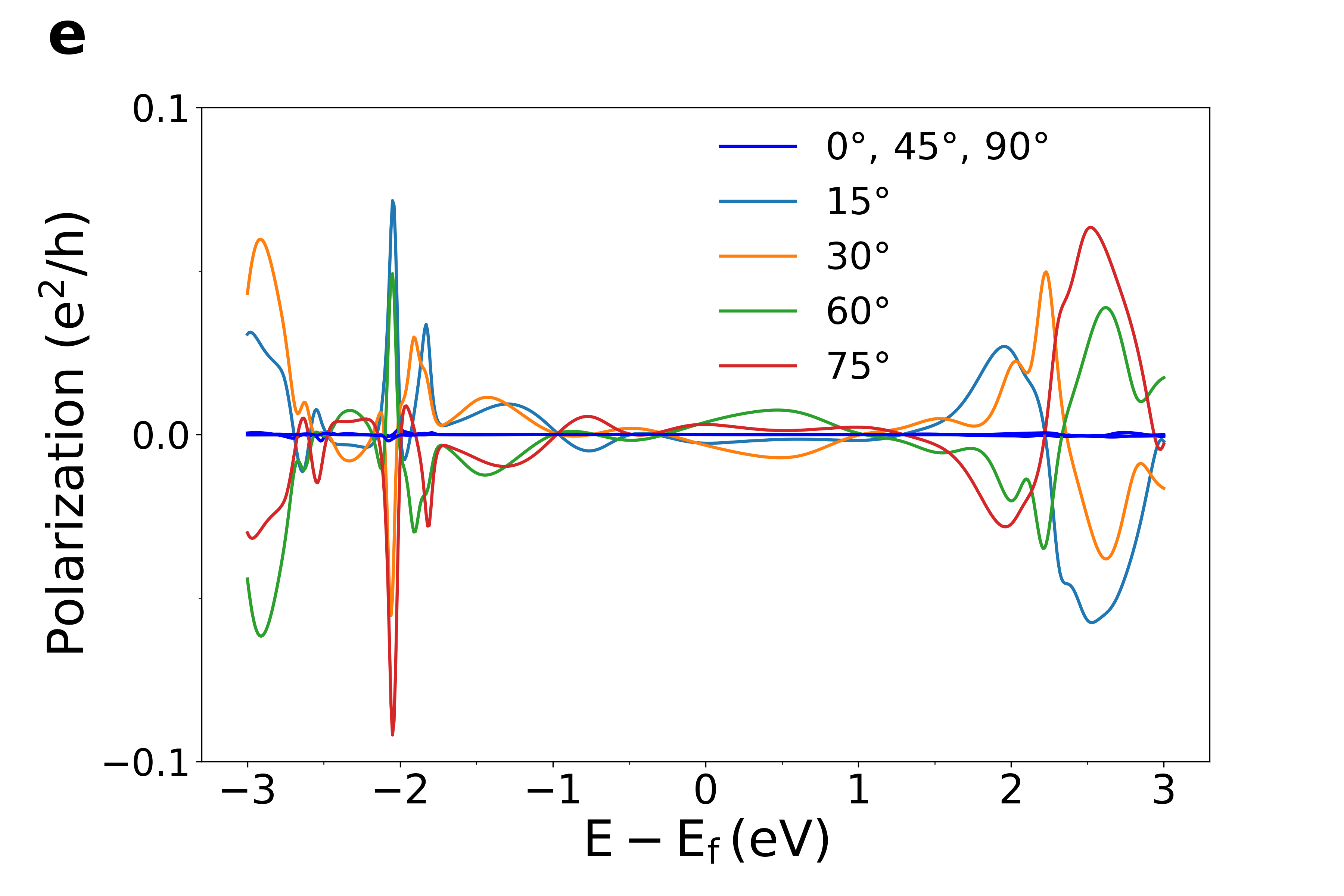}
\includegraphics[width=0.45\textwidth]{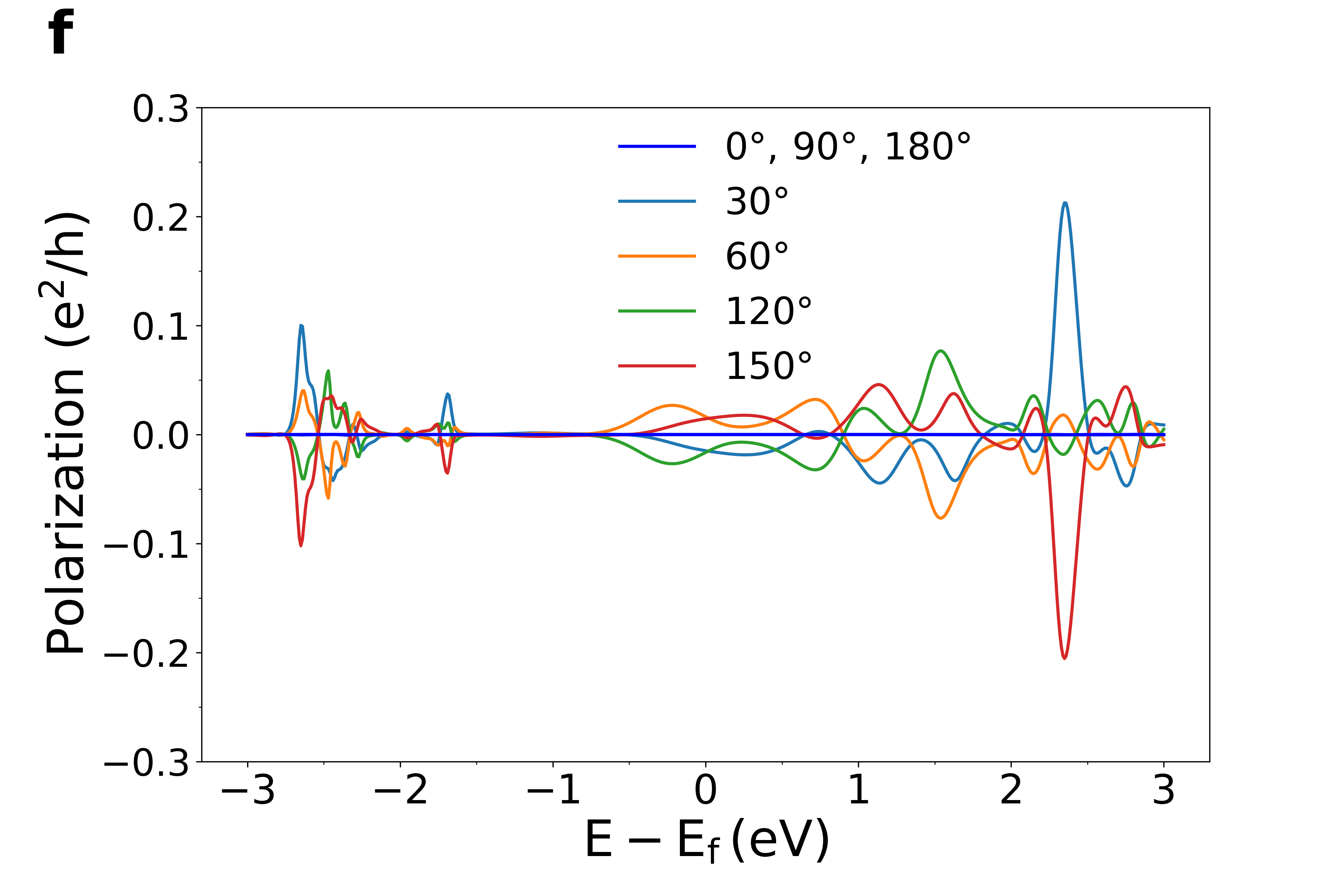}
\includegraphics[width=0.45\textwidth]{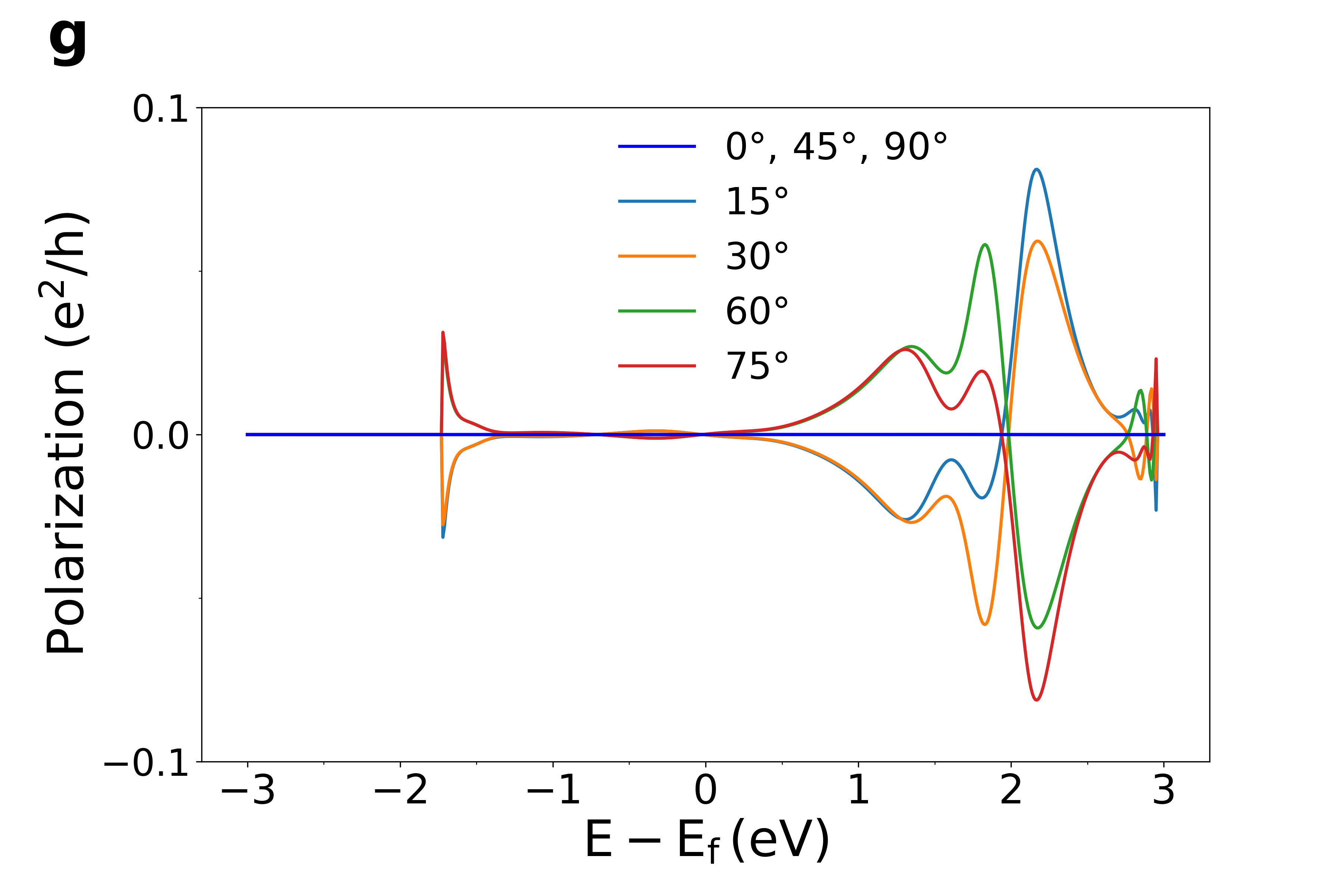}
\includegraphics[width=0.45\textwidth]{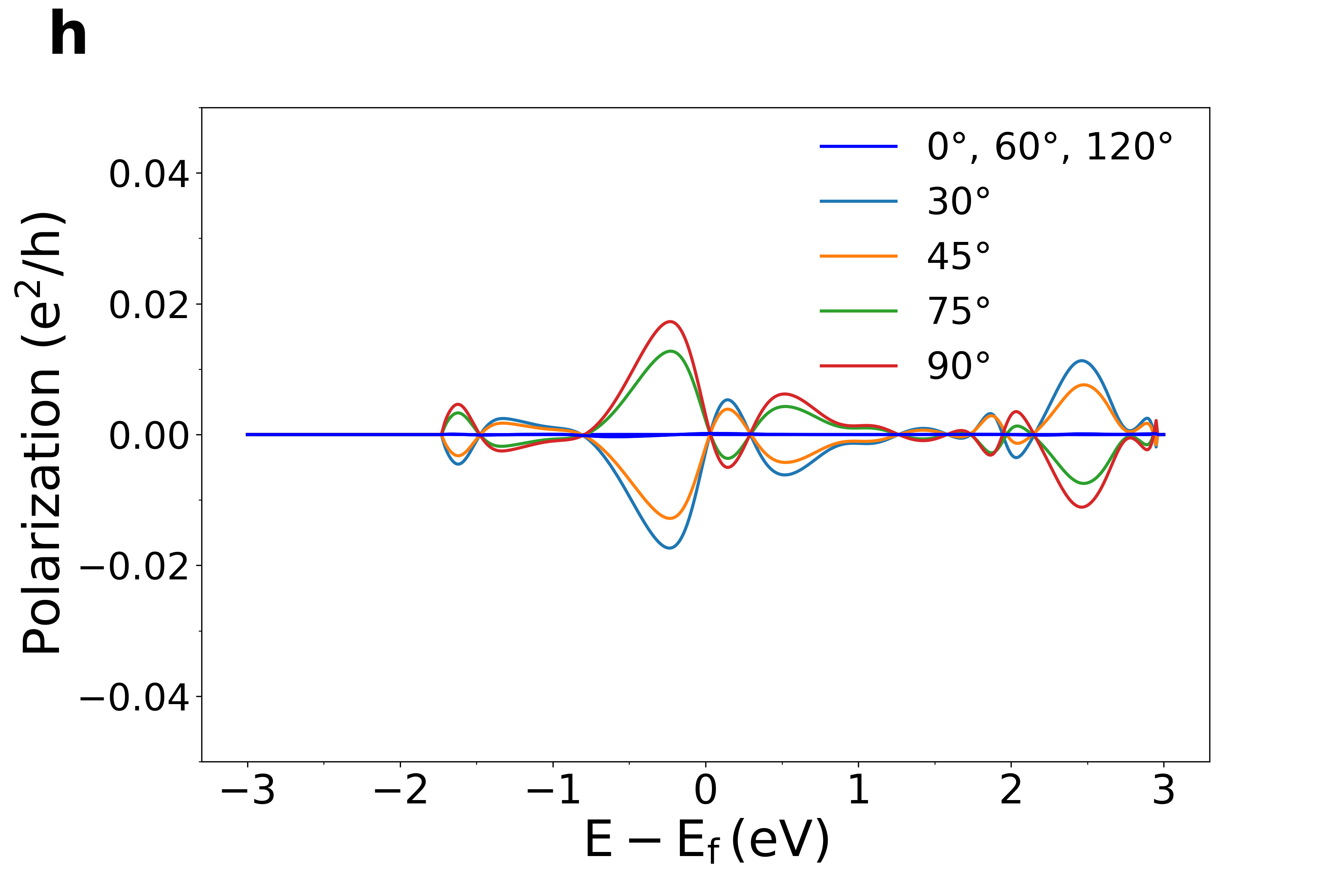}
\caption{
DFT-computed zero-bias spin-polarization along the direction of transport as a function of energy, for several angles of relative rotation between the electrodes (only the top half is rotated).
\textbf{a}, Al(001). \textbf{b}, Al(111). \textbf{c}, Au(001). \textbf{d}, Au(111). \textbf{e}, W(001). \textbf{f}, W(-110). \textbf{g}, Pb(001). \textbf{h}, Pb(111) stacked nanocontacts. Summarized in Figure 2 of the main text.}
\label{figS3}
\end{figure}

Here we present the full polarization results, over a range of energies centred on the Fermi level, and corresponding to Figure 2 of the main text. Figure \ref{figS3} shows the zero-bias spin-polarization along the direction of transport as a function of energy at various angles of relative rotation for \textbf{a} Al(001), \textbf{b} Al(111), \textbf{c} Au(001), \textbf{d} Au(111), \textbf{e} W(001), \textbf{f} W(-110), \textbf{g} Pb(001) and \textbf{h} Pb(111) nanocontacts.

\begin{figure}[H]
\centering 
\includegraphics[width=0.9\textwidth]{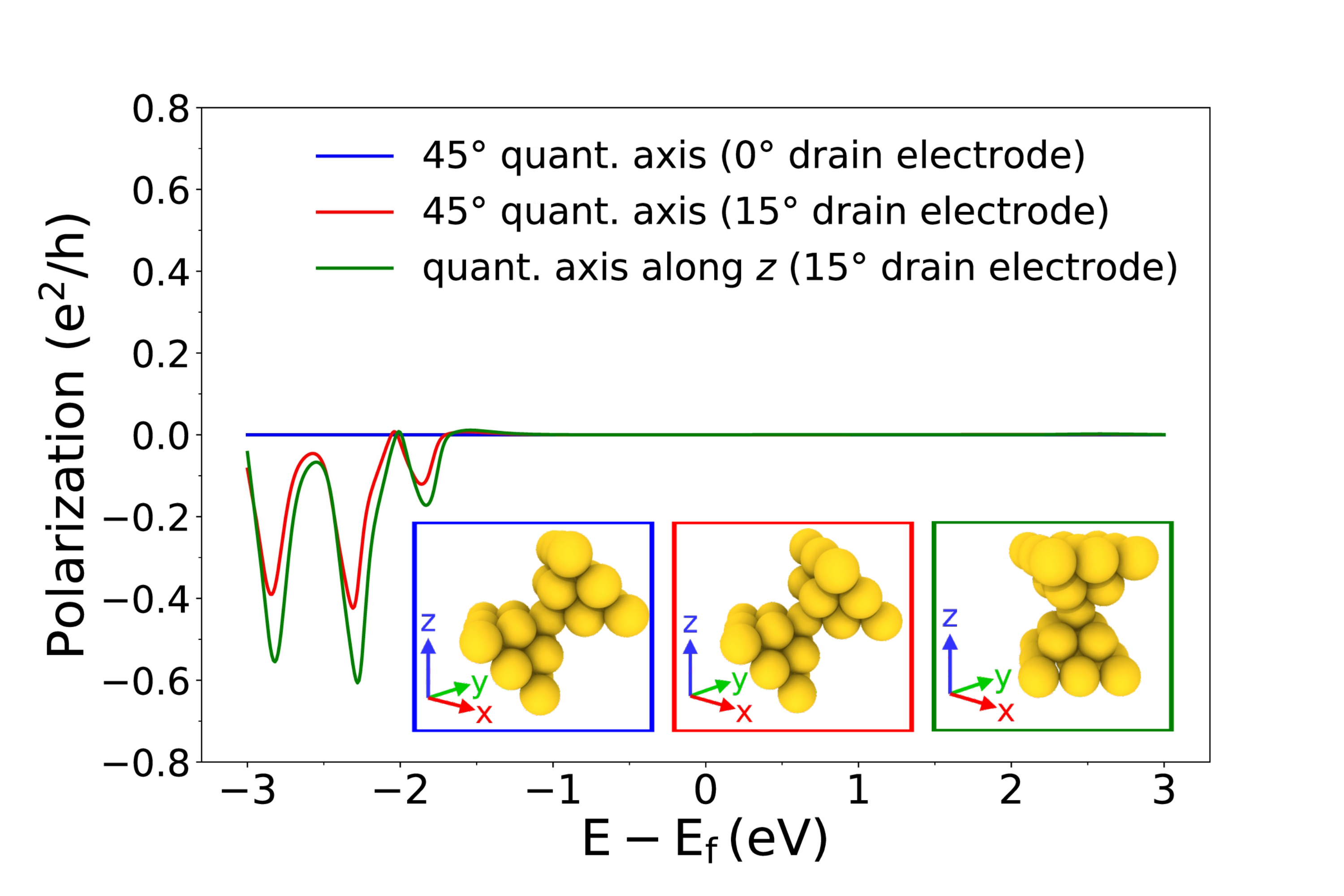}
\caption{DFT-computed zero-bias spin-polarization along $\hat{\bm{z}}$ as a function of energy, for Au(001) nanocontacts. The $15$\textdegree \space angles in the red and green inset boxes are obtained by rotating the drain electrode with respect to the source electrode.
}
\label{figS4}
\end{figure}

On the other hand, in Figure \ref{figS4} we explicitly show that when both longitudinal mirror and rotation symmetries are present in a structure the polarization is vanishing even when the quantization axis is neither parallel nor perpendicular to the direction of propagation (see blue line and inset in blue box in Figure \ref{figS4}), which corresponds to a neither longitudinal nor transversal spin-polarization component. However, when the drain electrode is rotated $15$\textdegree \space relative to the source electrode as in the red and green inset boxes, the longitudinal mirror is absent and only rotational symmetry remains, which as we saw in Figure 2a of the main text does not force the polarization to vanish if the spin quantization axis has a projection along the current. The magnitude of the polarization (red) in the structure tilted at $45$\textdegree \space with respect to the vertical quantization axis along $\hat{\bm{z}}$ is reduced relative to the structure oriented along $\hat{\bm{z}}$.  

\begin{figure}[H]
\centering 
\includegraphics[width=0.8\textwidth, height=16cm]{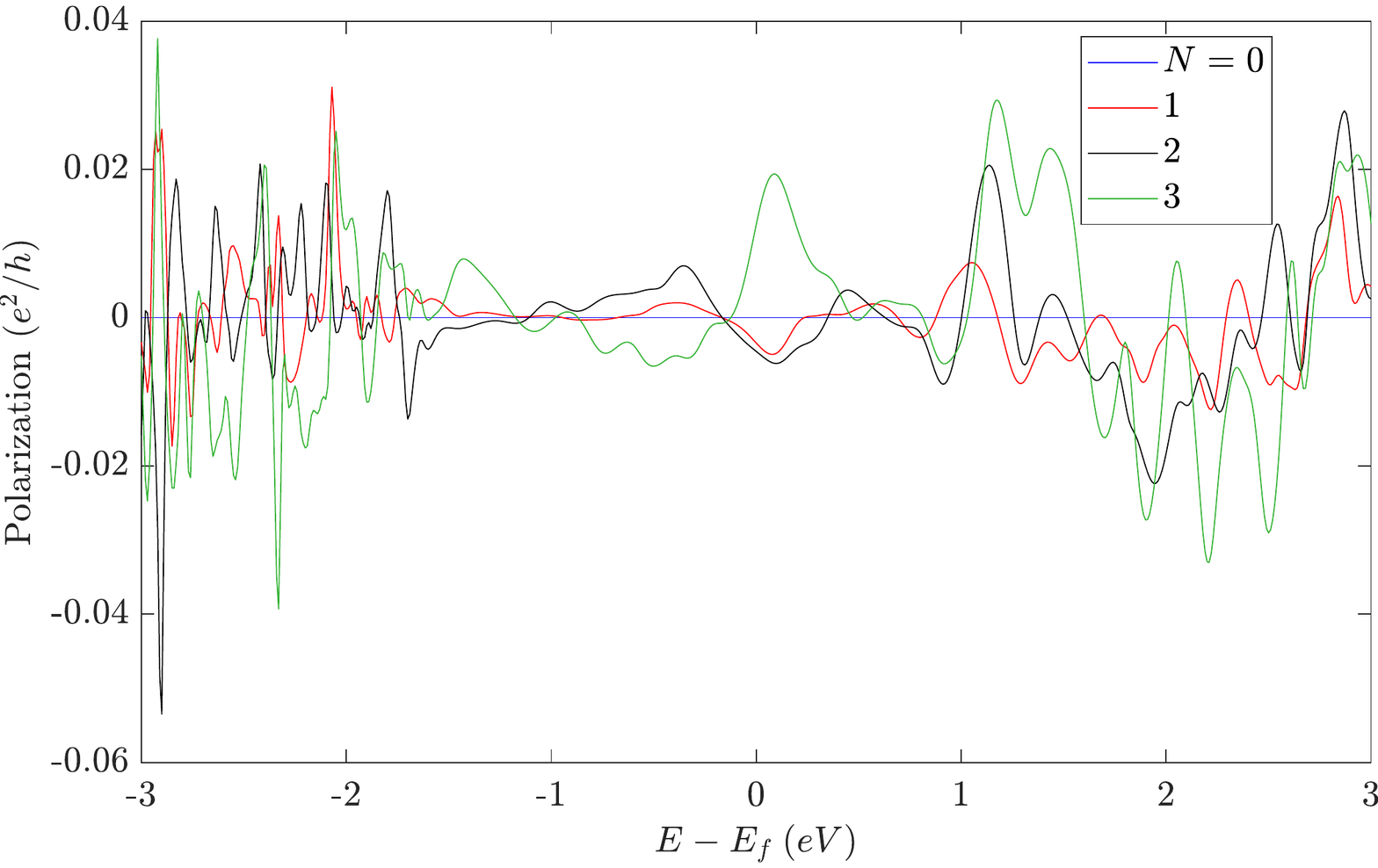}
\caption{DFT-computed zero-bias spin-polarization along the longitudinal direction as a function of energy, for W(110) nanocontacts (see first column of Figure 1c in the main text) with four atomic layers per contact in the scattering region. $N$ labels the number of atomic vacancies in the structure, which were randomly chosen within the source contact. The corresponding groups are $\pazocal{G}=D_{2h}$ for $N=0$, and $\pazocal{G}=C_{1}$ (trivial group) for the other cases.}
\label{figS5}
\end{figure}

Finally, in Figure \ref{figS5} we quantitatively exemplify the impact of defects, in particular atomic vacancies, on the spin-polarization. The original structure with no defects is a system of bare W(110) nanocontacts with point group $\pazocal{G}=D_{2h}$, so that $\bm{P}=0$ as stated in the main text. Upon introduction of random vacancies ($N$ indicating their number), all symmetries were removed at once, enabling a finite $\bm{P}$ whose component along the direction of transport we plot in the figure. The number of defects is sufficiently low in all cases, taking into account that the original scattering region is formed by 109 atoms, so that the symmetry is preserved at a course scale; but the selection rules for $\bm{P}$ are indeed determined by the symmetry at the local scale. It should be noted that not all defects will necessarily break all symmetries, or even any symmetry at all. Of course, the nature and position of the defects within a specific system may be important for numerical reasons.

\bibliography{Symmetry}